\PassOptionsToPackage{dvipsnames}{xcolor}
\documentclass[11pt]{article}

\usepackage[T1]{fontenc}
\usepackage[letterpaper,textwidth=6.5in,textheight=9in,headheight=14pt]{geometry}

\usepackage{amsmath}
\usepackage{amsthm}
\usepackage{mathtools}
\usepackage{lmodern}
\usepackage{amssymb}
\usepackage{bm}
\usepackage{bbm}
\usepackage{cancel}
\usepackage{microtype}

\usepackage{graphicx}
\usepackage{booktabs}
\usepackage{array}
\usepackage{multirow}
\usepackage{caption}
\usepackage{enumitem}
\usepackage{tikz}
\usetikzlibrary{arrows.meta,calc,positioning}
\usepackage{etoolbox}
\usepackage{placeins}

\usepackage{sectsty}
\sectionfont{\sffamily\bfseries\upshape\large}
\subsectionfont{\sffamily\bfseries\upshape\normalsize}
\subsubsectionfont{\sffamily\mdseries\upshape\normalsize}
\paragraphfont{\sffamily\bfseries\upshape}
\makeatletter
\renewcommand\@seccntformat[1]{\csname the#1\endcsname.\quad}
\makeatother

\usepackage{fancyhdr}
\newcommand{\runauthor}{J. E. Aguilar et al.}
\newcommand{\runtitle}{Dirichlet-Tree Allocation Priors}
\renewcommand{\headrulewidth}{0pt}

\usepackage[
  backend=biber,
  style=authoryear-comp,
  natbib=true,
  backref=true,
  maxcitenames=2,
  maxbibnames=99,
  giveninits=true,
  uniquename=false,
  sortcites=false,
  useprefix=true,
  uniquelist=false,
  dashed=false,
  date=year,
  isbn=false,
  eprint=false
]{biblatex}
\DeclareSourcemap{
  \maps[datatype=bibtex]{
    \map{\step[fieldset=issn, null]}
    \map{\step[fieldset=urldate, null]}
    \map{\step[fieldset=language, null]}
    \map{\step[fieldsource=note, match=\regexp{^\s*(Publisher|\\_eprint|_eprint)}, final]
         \step[fieldset=note, null]}
    \map{\step[fieldsource=url, match=\regexp{books\.google}, final]
         \step[fieldset=url, null]}
    \map{\step[fieldsource=doi, match=\regexp{^https?://(dx\.)?doi\.org/}, replace={}]}
    \map{\step[fieldsource=doi, final]
         \step[fieldset=url, null]}
  }
}

\DeclareFieldFormat[article,incollection,inproceedings,misc,unpublished]{title}{#1}
\DeclareFieldFormat{pages}{#1}
\renewbibmacro{in:}{\ifentrytype{article}{}{\printtext{\bibstring{in}\intitlepunct}}}
\DeclareNameAlias{sortname}{family-given}
\renewbibmacro*{volume+number+eid}{%
  \printfield{volume}%
  \iffieldundef{number}{}{\printtext[parens]{\printfield{number}}}%
  \setunit{\addcomma\space}\printfield{eid}}
\DeclareFieldFormat{doi}{\url{https://doi.org/#1}}
\DeclareFieldFormat{url}{\url{#1}}
\renewbibmacro*{pageref}{%
  \iflistundef{pageref}
    {}
    {\printtext[parens]{\printlist[pageref][-\value{listtotal}]{pageref}}}}

\usepackage{xcolor}
\definecolor{bablue}{rgb}{0,0,1}
\usepackage[
  colorlinks=true,
  linkcolor=bablue,
  citecolor=bablue,
  urlcolor=bablue,
  hyperfootnotes=false
]{hyperref}

\hypersetup{
  pdftitle={Dirichlet-Tree Allocation Priors for Bayesian Causal Forests},
  pdfauthor={Javier Enrique Aguilar, Daniel Klippert, Yusufhan Balci, Alexander Marx},
  pdfsubject={Statistics, Methodology},
  pdfkeywords={Bayesian additive regression trees, Bayesian causal forests,
    heterogeneous treatment effects, shrinkage priors, causal inference}}

\usepackage{etoc}

\newtheorem{proposition}{Proposition}

\newtheorem{theorem}{Theorem}

\newcommand{\R}{\mathbb R}
\newcommand{\E}{\mathbb E}

\newcommand{\Normal}{\operatorname{N}}
\newcommand{\Beta}{\operatorname{Beta}}
\newcommand{\Dir}{\operatorname{Dirichlet}}

\newcommand{\BP}{\operatorname{BetaPrime}}
\newcommand{\calL}{\mathcal L}
\newcommand{\calI}{\mathcal I}
\newcommand{\ind}{\mathbb I}
\newcommand{\bx}{\bm x}
\newcommand{\bh}{\bm h}
\newcommand{\bz}{\bm z}
\newcommand{\bphi}{\bm\phi}
\newcommand{\tsc}{\psi}
\newcommand{\method}{DT-BCF}
\newcommand{\Indep}{\mathop{\perp\!\!\!\perp}\nolimits}

\providecommand{\Indep}{\mathop{\perp\!\!\!\perp}\nolimits}

\newcommand{\tablesizes}{%
  \renewcommand{\scriptsize}{\fontsize{7}{8}\selectfont}%
  \renewcommand{\footnotesize}{\fontsize{8}{9.5}\selectfont}%
  \renewcommand{\small}{\fontsize{9}{11}\selectfont}%
  \setlength{\topsep}{0pt}\setlength{\partopsep}{0pt}}
\AtBeginEnvironment{table}{\tablesizes}
\AtBeginEnvironment{table*}{\tablesizes}

\newcommand{\maxupscale}{1.3}
\makeatletter
\newsavebox{\rb@box}
\newlength{\rb@w}
\def\rb@bang{!}
\let\rb@resizebox\resizebox
\newcommand{\rb@capped}[3]{%
  \def\rb@tmp{#2}%
  \ifx\rb@tmp\rb@bang
    \sbox{\rb@box}{#3}%
    \setlength{\rb@w}{#1}%
    \ifdim\maxupscale\wd\rb@box<\rb@w
      \setlength{\rb@w}{\maxupscale\wd\rb@box}%
    \fi
    \rb@resizebox{\rb@w}{!}{\usebox{\rb@box}}%
  \else
    \rb@resizebox{#1}{#2}{#3}%
  \fi}
\renewcommand{\resizebox}{\leavevmode\@ifstar{\rb@resizebox*}{\rb@capped}}
\def\maxwidth{\ifdim\Gin@nat@width>\linewidth\linewidth\else\Gin@nat@width\fi}
\makeatother
\setkeys{Gin}{width=\maxwidth,keepaspectratio}

\newcommand{\coderepo}{\url{https://github.com/causality-lab/bcf-dirichlet-trees}}

\title{\sffamily\bfseries Dirichlet-Tree Allocation Priors for\\ Bayesian Causal Forests}
\newcommand{\affmark}[1]{\textsuperscript{\textcolor{bablue}{#1}}}
\author{Javier Enrique Aguilar\affmark{1,2,*},
  Daniel Klippert\affmark{1,2},\\
  Yusufhan Balci\affmark{3,4}
  and Alexander Marx\affmark{1,2}}
\date{October 2026}

\newcommand{\affiliations}{%
  \begingroup
  \renewcommand{\thefootnote}{}%
  \footnotetext{\affmark{1}Department of Statistics, TU Dortmund University, Dortmund, Germany}%
  \footnotetext{\affmark{2}Research Center Trustworthy Data Science and Security, Research Alliance Ruhr, Dortmund, Germany}%
  \footnotetext{\affmark{3}Department of Epidemiology and Data Science, Amsterdam UMC, Amsterdam, The Netherlands}%
  \footnotetext{\affmark{4}Vrije Universiteit Amsterdam, Amsterdam, The Netherlands}%
  \footnotetext{\affmark{*}Corresponding author: \href{mailto:javier.aguilarr@icloud.com}{javier.aguilarr@icloud.com}}%
  \endgroup}

\begin{document}
\maketitle
\affiliations

\begin{abstract}
Bayesian Causal Forests (BCF) estimate heterogeneous treatment effects with a separate ensemble of treatment trees. The prior on the leaf values of the treatment trees determines whether sampling variability is attributed to the treatment effect. Independent shrinkage priors exert no control over how a tree allocates variability across its leaves, and their total prior variance grows as leaves are added. We propose the Dirichlet-Tree Bayesian Causal Forest, which jointly regularizes the leaf values of each treatment tree through a Dirichlet-Tree prior on their variance shares, generated by stick-breaking along the tree topology. We characterize the spike and tails of the marginal leaf prior, show that the prior adaptively bounds the number of regions of effect heterogeneity, and establish posterior consistency for the forest. In simulations, the model outperforms BCF when heterogeneity is confined to a few regions and reduces spurious heterogeneity elsewhere, and, unlike independent shrinkage priors, its accuracy does not deteriorate as the trees grow. We also evaluate the model on real data, including a pancreatic cancer survival study.


\end{abstract}

\etocdepthtag.toc{mainpart}

\section{Introduction}
\label{sec:introduction}

Estimating the effect of a policy or treatment is a key question in econometrics and biostatistics. Randomization is the gold standard to avoid confounding, but cannot always be attained due to ethical or practical concerns. In the observational setting, we need to account for treatment assignment and adjust for confounding to obtain valid inference \citep{rosenbaum1983central}. Besides the average treatment effect (ATE), we might also be interested in the conditional average treatment effect (CATE). The CATE indicates the expected benefit that an individual might have from a given treatment, given their observed covariates. It is especially of interest for making cost-effective policy choices and avoiding harm when treating a population. A medicine that might benefit the population on average might severely harm a certain subgroup \citep{kent2018personalized}.

Treatment heterogeneity is notoriously hard to estimate, as many randomized controlled trials (RCTs) are not powered to detect it \citep{hill2011causal,kent2018personalized}. Furthermore, treatment heterogeneity can be sparse or nonexistent, calling for methods that can deal with this setting. Causal machine learning methods like R-, T-, and S-learners can struggle with smaller sample sizes, depending on the structure of the treatment effect and the underlying learner, as learning relationships becomes harder in data sets where signal is hard to find \citep{nie2021quasi,kuenzel2019meta}. Bayesian Causal Forests (BCF) have been shown to perform well in settings with weak treatment signals, as they are flexible tree ensemble methods whose priors on the tree structure and leaf values act as regularizers \citep{hahn2020bcf}. The choice of these priors is therefore central to controlling the estimated treatment heterogeneity.

BCF use separate prognostic and treatment-effect ensembles, which helps prevent the prognostic signal from distorting regularization of the treatment effect \citep{hahn2018regularization,hahn2020bcf}. The leaf prior in the treatment ensemble then determines how strongly regional effects are regularized. The BART prior assigns the same variance to every leaf, while the recently proposed Horseshoe Forests consider independent local scales \citep{jacobs2026horseshoe}. These choices fix the relative variances to be equal or leave them unrelated, hence, they do not model how treatment-effect variation is distributed across the regions of a tree.

Joint variance allocation has proved useful in high-dimensional regression, where a total prior variance is distributed across coefficients to shrink negligible signals while retaining substantial ones \citep{bhattacharya2015dirichlet,zhang2022r2d2,aguilar2023r2d2m2, yanchenko2024r2d2priorgeneralizedlinear}. We adapt this principle to treatment trees through the Dirichlet-Tree Bayesian Causal Forest, \method{}. Our contributions are as follows:

\begin{itemize}[noitemsep,topsep=0pt,parsep=2pt,partopsep=0pt,leftmargin=*]
\item We construct a topology-indexed family of priors that separates the total leaf variance of each treatment tree from its allocation across regions. The family contains equal division as a limiting allocation and Dirichlet-tree priors with random unequal shares. 
\item We characterize the marginal shrinkage of the leaf values and derive a theoretical limit on how many regions can express substantial treatment-effect heterogeneity. We also establish posterior consistency.
\item We benchmark \method{} against Bayesian tree priors and external learners in controlled simulations, a causal-inference benchmark, and a real-data application. \method{} performs comparably to the Bayesian alternatives across a broad range of settings, with the largest gains when heterogeneity is concentrated in a few regions. 
\end{itemize}

\section{Related work}
\label{sec:related-work}

Bayesian tree models have been used for causal inference and treatment effect estimation \citep{hill2011causal,green2012heterogeneous}. BCF separates the prognostic and treatment-effect functions so that each can receive its own regularization and includes the estimated propensity score in the prognostic function to mitigate regularization-induced confounding \citep{hahn2018regularization,hahn2020bcf}. Later extensions modify different parts of this construction. Targeted smooth BCF imposes smoothness along a designated effect modifier \citep{starling2021targeted}. Shrinkage BCF places sparsity-inducing priors on the splitting probabilities to identify covariates that drive prognosis and treatment heterogeneity \citep{caron2022shrinkage,linero2018dart}. Horseshoe Forests consider using the Horseshoe (HS) prior on the leaf nodes to allow different amounts of shrinkage across the covariate space \citep{jacobs2026horseshoe}. \method{} changes the dependence among the leaf variances within each treatment tree, jointly regularizing regional contributions.

The closest prior constructions arise in joint shrinkage and variance decomposition. The Dirichlet--Laplace prior distributes a global scale across regression coefficients, while R2 based priors allocate explained variance across model terms \citep{bhattacharya2015dirichlet,zhang2022r2d2,aguilar2023r2d2m2, aguilar_generalized_2025, yanchenko2024r2d2priorgeneralizedlinear,kohns_arr2_2025}. Hierarchical decomposition priors allocate variance along a tree of additive model components specified by the analyst \citep{fuglstad_intuitive_2020}. These approaches separate the overall amount of variation from its relative allocation among a fixed collection of model components. In \method{}, we allocate the variance by using the discovered topology of the tree recursively.

Outside Bayesian tree ensembles, the R-, X-learners, causal forests, multi-task Gaussian processes, and neural networks provide alternative estimators of heterogeneous effects \citep{nie2021quasi,kuenzel2019meta,kennedy2023drlearner,athey2019grf,alaa2017bayesian,shalit2017estimating}. Our experiments emphasize comparisons among Bayesian tree priors and use the X- and R-learners as external references.

\section{Methods}
\label{sec:model}

We first review the BCF treatment ensemble and identify the role of the leaf prior. We then show how to specify a tree prior which separates the total leaf variance of tree from its allocation across regions. To construct the allocation, we make use of the topology of tree. We also study the prior's properties. Proofs are shown in Appendix~\ref{app:proofs}.

\subsection{Bayesian Causal Forests}
\label{sec:bcf}

For unit $i$, let $A_i\in\{0,1\}$ denote treatment, $\bx_i\in\mathbb{R}^p$
the observed covariates, and $y_i(a)$ the potential outcome under treatment $a\in\{0,1\}$. The observed response is $y_i=y_i(A_i)$. We assume consistency and no interference, conditional exchangeability $\{y_i(0),y_i(1)\}\Indep A_i\mid\bx_i$, and positivity $0<e(\bx)<1$ on the covariate support, where $e(\bx)=\Pr(A_i=1\mid\bx_i=\bx)$ \citep{imbens2015causal}. Under these conditions, the Conditional Average Treatment Effect (CATE)
\begin{equation}
\label{eq:cate}
\begin{aligned}
\tau(\bx)
  &= \E\{y_i(1)-y_i(0)\mid\bx_i=\bx\} \\
  &= \E(y_i\mid\bx_i=\bx,A_i=1)
     -\E(y_i\mid\bx_i=\bx,A_i=0)
\end{aligned}
\end{equation}
is identified by the difference between the treated and control outcome
regressions.

Bayesian Causal Forests (BCF) model the observed outcome as
\begin{equation}
y_i=\mu(\bx_i,\widehat e_i)
    +A_i\{\alpha+g_\tau(\bx_i)\}+\varepsilon_i,
\qquad \varepsilon_i\sim\Normal(0,\sigma^2),
\label{eq:bcf-model}
\end{equation}
where $\widehat e_i$ estimates $e(\bx_i)$, $\mu$ is the prognostic function, and $\tau(\bx)=\alpha+g_\tau(\bx)$ is the treatment-effect function \citep{hahn2020bcf}. Including the estimated propensity score in $\mu$ helps protect the treatment-effect estimate from regularization-induced confounding \citep{hahn2018regularization}. We retain the BCF prior for $\mu$ and modify only the leaf prior in the treatment ensemble $g_\tau$.


To define leaf priors we make use of the representation of a tree by its leaves. Let treatment tree $T_j$ partition the covariate space into $L_j$ leaves $\calL(T_j)$. Let $\bz_j(\bx)\in\{0,1\}^{L_j}$ be the one-hot vector indicating the leaf that contains $\bx$, and let $\bh_j=(h_{j\ell})_{\ell\in\calL(T_j)}$ collect the leaf values. Tree $j$ contributes
\begin{equation}
g(\bx;T_j,\bh_j)=\bz_j(\bx)^\top\bh_j,
\label{eq:tree-representation}
\end{equation}
and $g_\tau(\bx)=\sum_{j=1}^{m_\tau}g(\bx;T_j,\bh_j)$. 
We consider \citet{chipman2010bart}'s prior on the topology of a tree, in which a node at depth $d$ splits with probability $\nu(1+d)^{-\eta}$. We select splitting variables 
using a discrete uniform distributions. Given $T_j$, BART assigns independent $\Normal(0,\sigma_h^2)$ distributions to the elements of $\bh_j$ \citep{chipman2010bart}.  This leaves the choice of scale prior to determine how variation is distributed across the leaves. We consider common and leaf-specific scale priors next.

\subsection{Problem Statement}
\label{sec:joint}

Under BART, given that $\bx$ belongs to exactly one leaf, the prior variance of $g(\bx;T_j,\bh_j)$ is $\sigma_h^2$ . While considering further splits leaves the variance unchanged, the total variance the leaves represent increases. 
For a treatment tree, we seek two properties. 1) Leaf variances should be able to differ across covariate space, so that a leaf is shrunk strongly toward zero where the data supports no heterogeneity. 2) Second, the sum of the leaf variances of a tree should have a prior that does not depend on $L_j$, so that a split redistributes prior variance within a tree (i.e the total variability of a tree) instead of adding to it.

Neither property holds under the BART leaf prior. Priors on the leafs with conditionally independent local scales, such as the Horseshoe \citep{carvalho2010horseshoe,piironen2017sparsity,pas_theoretical_2021}, meet the first but not the second requirement. Under any such prior, the sum of the leaf variances is increasing in $L_j$, implying that the second property cannot be satisfied. Scaling BART leaf variances by $L_j$ could endow the prior with the second property, but every leaf would receive the same fixed variance, offering no adaptability. We meet both by letting the law of the leaf variances depend on the topology of the tree.



To this end, we propose a prior that separates the total variance of a tree from its allocation across leaves. Tree $j$ receives a positive tree scale $\tsc_j^2$ and an allocation vector $\bphi_j=(\phi_{j\ell})_{\ell\in\calL(T_j)}$ satisfying $\phi_{j\ell} >0$ and $\sum_\ell\phi_{j\ell}=1$. The prior on $\tsc_j^2$ is independent of the tree, while the law of $\bphi_j$ is constructed using information from the topology. Given $T_j$, $\tsc_j^2$ and $\bphi_j$, the leaf values are independent with $h_{j\ell} \sim \Normal(0,V_{j\ell})$, with $V_{j\ell}=V_j\phi_{j\ell}$ and $V_j=\sigma^2\omega\tsc_j^2$,
where $\omega>0$ is a shared constant of the forest. Since, $\sum_{\ell\in\calL(T_j)}V_{j\ell}=V_j$ for every tree $j$ and every value of $L_j$, we immediately satisfy the second property we seek. We are therefore able to jointly control the total amount of signal that the tree can represent. 
The compositional nature of $V_{j\ell}$ implies that a larger share for one leaf leaves less total variance $V_j$ for the others. This is precisely the joint regularization of leaf values we aim to achieve. To achieve this we need to specify prior distributions for $\tsc_j^2$ and $\bphi_j$. 

\subsection{Dirichlet-Tree Allocation}
\label{sec:dirichlet-tree}

We now specify the law of the shares $\bphi_j$ given the topology of the tree. The random topology of $T_j$ requires a law for the shares on every possible tree. A law chosen separately for each topology would not be able to express how the shares of a tree relate to those of a tree obtained by splitting one of its leaves. Therefore, we define the shares $\bphi_j$ through the same binary recursion that constructs the partition. The main idea is that each internal node divides the share of total variance $V_j$ it receives between its two children, while the rest of the tree remains unchanged.

Let $\calI(T_j)$ be the internal nodes of $T_j$, and $\phi_{jv}$ the share reaching node $v$, with $\phi_{jv}=1$ at the root. Let the branch fraction $U_{jv} \in (0,1)$ at node $v\in\calI(T_j)$ send
\begin{equation}
\phi_{jv_L}=U_{jv}\phi_{jv}, \qquad \phi_{jv_R}=(1-U_{jv})\phi_{jv}
\label{eq:recursive-split}
\end{equation}
to the left and right children $v_L$ and $v_R$. Therefore,  each leaf share is the product of the branch fractions along its root-to-leaf path, using $U_{jv}$ on a left branch and $1-U_{jv}$ on a right branch. 

If the branch fractions $U_{jv}$ are independent with
\begin{equation}
U_{jv}\mid T_j\sim\Beta\{c_L(v),c_R(v)\},
\qquad v\in\calI(T_j),
\label{eq:general-branch}
\end{equation}
the shares $\bphi_j$ follow a Dirichlet-Tree distribution \citep{dennis1991hyperdirichlet,minka2004dirichlettree}. The name refers to the recursive beta construction, and the terminal shares need not to have an  ordinary Dirichlet distribution. For example, the symmetric choice  $c_L(v)=c_R(v)=a$ divides shares evenly on average. A leaf at depth $d_\ell$ has mean share $2^{-d_\ell}$ for every value of $a$, so the shares cannot be exchangeable unless all leaves have the same depth. The share is a product of $d_\ell$ independent $\Beta(a,a)$ variables, which is not beta distributed for $d_\ell\ge2$. 

Instead, we aim to control how $V_j$ is allocated across the leaves by specifying the mean shares directly. For this purpose, the terminal law should separate two aspects of the allocation. Relative leaf weights should determine the mean shares, while a total concentration should determine how strongly the realized shares can depart from those means. We obtain this separation by choosing the branch parameters so that the induced terminal law is an ordinary Dirichlet distribution. Hence, each share then has a beta marginal whose mean and concentration are explicit in the parameters. The Dirichlet aggregation property allows us to compare the allocation laws before and after a leaf splits \citep{aitchison_statistical_1986,lin_dirichlet_2016}. Within the Dirichlet-Tree construction, the desired leaf law results from setting the beta parameters at each internal node as the sums of the leaf parameters below its two children \citep{minka2004dirichlettree}. We call the result weighted allocation. 

Specifically, we derive the branch parameters from positive leaf weights $(w_\ell)_{\ell\in\calL(T_j)}$ and a concentration $a>0$. If $W_v^L$ and $W_v^R$ are the total weights of the leaves below $v^L$ and $v^R$, we set
\begin{equation}
c_L(v)=aW_v^L, \qquad c_R(v)=aW_v^R.
\label{eq:weighted-branch}
\end{equation}
The shares are then $\bphi_j\mid T_j\sim\Dir\{(aw_\ell)_{\ell\in\calL(T_j)}\}$. Hence, 
\begin{equation}
\phi_{j\ell}\mid T_j \sim \Beta\{aw_\ell,a(W_j-w_\ell)\}, \ \ \E(\phi_{j\ell}\mid T_j)=\frac{w_\ell}{W_j},
\end{equation}
where $W_j=\sum_{\ell} w_{\ell}$. The weights $w_\ell$ set the mean allocation, while the total concentration $aW_j$ controls dispersion around it. 

Since splits can be considered local refinements of variance allocation, 
we impose the requirement that a split on a share between children will not change the law of the other shares. 
Let $T'$ be obtained from $T$ by splitting leaf $\ell$ into $\ell_L$ and $\ell_R$, and index the other leaves by $1,..., L-1$ under both trees. We call the allocation consistent under splits when adding the two shares of the two new leaves recovers the law of the shares before the split. Namely, $(\phi_1',\ldots,\phi_{L-1}',\phi_{\ell_L}'+\phi_{\ell_R}')\overset{d}{=}(\phi_1,\ldots,\phi_{L-1},\phi_\ell)$. 
By the aggregation property of the Dirichlet distribution, consistency is a condition on the weights. Hence we set the depth weights $w_\ell=\rho^{d_\ell}$, where $\rho>0$ and $d_\ell$ is the depth of leaf $\ell$. The value of $\rho$ determines how a leaf's contribution changes when it is replaced by its children.

\begin{proposition}[Consistency under splits]
\label{prop:split-consistency}
The weighted allocation is consistent under a split of leaf $\ell$ if and only if $w_{\ell_L}+w_{\ell_R}=w_\ell.$
For the depth weights $w_\ell=\rho^{d_\ell}$, consistency under every split holds if and only if $\rho=1/2$. Additionally, the symmetric choice is consistent under every split. 
\end{proposition}

At $\rho=1/2$, each child inherits half of its parent's weight, so $W_j=1$ for every topology. Hence, the Dirichlet concentration is always $a$ regardless of the topology, making it comparable across the different trees of the forest. In this case, a leaf at depth $d_\ell$ has mean share $2^{-d_\ell}$, and each child receives, on average, half of its parent's share. At $\rho=1$, all weights equal one. Thus, the shares have a symmetric Dirichlet distribution with parameter $a$, and every leaf has mean share $1/L_j$ irrespective of its depth.

\subsection{Tree Scale Prior}
\label{sec:tree-scale}

The tree scale $\tsc_j^2$ controls all leaf values in tree $j$ through their total variance $V_j=\sigma^2\omega\tsc_j^2$. When $\tsc_j^2$ is small, every leaf value is concentrated near zero and the tree contributes little throughout the covariate space. As $\tsc_j^2$ increases, the tree can make progressively larger regional contributions. Its prior should therefore concentrate near zero and decay slowly enough to keep large scales plausible.  To achieve both behaviors we consider the scaled beta-prime family
\begin{equation}
\tsc_j^2\sim\BP(a_\tsc,b_\tsc,s_\tsc),
\label{eq:beta-prime}
\end{equation}
with density proportional to $(t/s_\tsc)^{a_\tsc-1}(1+t/s_\tsc)^{-(a_\tsc+b_\tsc)}$ for $t>0$. The tree scales are independent across trees and independent of the topologies and shares. Since $\sum_{\ell\in\calL(T_j)}V_{j\ell}=V_j=\sigma^2\omega\tsc_j^2$, the induced prior on the total leaf variance is the same for every tree size and topology. 

We set $a_\tsc=b_\tsc=1/2$ and $s_\tsc=1$, for which $\tsc_j$ is standard half-Cauchy. Conditioned on $\bphi_j$, integrating over $\tsc_j$ implies that $h_{j\ell}/(\sigma\sqrt{\omega\phi_{j\ell}})$ has a Horseshoe marginal, with a logarithmic spike at zero and $|h|^{-2}$ tails \citep{pas_horseshoe_2014, pas_adaptive_2017}. It remains to integrate over the Dirichlet-Tree prior on $\bphi_j$. Proposition \ref{prop:marginal-behavior} shows that, under the weighted allocation, this integration preserves the spike and the horseshoe tail order, and can sharpen the spike from logarithmic to polynomial.

\begin{proposition}[Marginal behavior of a leaf value]
\label{prop:marginal-behavior}
Let $T_j$ have at least two leaves, and let $\underline c_\ell$ be the smallest branch parameter on the path to leaf $\ell$, taking $c_L(v)$ at a left branch and $c_R(v)$ at a right branch. Under the general Dirichlet-Tree distribution, conditionally on $\tsc_j^2$, the density of $h_{j\ell}$ is bounded at zero if and only if $\underline c_\ell>1/2$.

Under the weighted allocation and Equation~\eqref{eq:beta-prime}, let $c_\ell^\star=\min(a_\tsc,aw_\ell)$ and let $p$ be the density of $h_{j\ell}\mid T_j$. Then $p$ is bounded at zero if and only if $c_\ell^\star>1/2$. If $c_\ell^\star<1/2$ and $a_\tsc\ne aw_\ell$, then $p(h)\asymp |h|^{2c_\ell^\star-1}$ as $h\to0$. In all cases, $p(h)\asymp  |h|^{-(2b_\tsc+1)}$ as $|h|\to\infty$.
\end{proposition}

With $a=\rho=a_\tsc=b_\tsc=1/2$, a leaf at depth $d\ge1$ has $aw_\ell=2^{-d-1}$, so its density is of order $|h|^{2^{-d}-1}$ near zero and $|h|^{-2}$ in the tails. This polynomial spike sharpens with depth and is stronger than the logarithmic spike of the horseshoe prior. The Dirichlet-Tree allocation therefore concentrates deeper leaf values more strongly near zero while preserving the horseshoe tail order available to large values. This is attractive when most regional contributions should be negligible and a few may be large.

\subsection{Hyperparameter Specification}
\label{sec:hyperparameters}

We specify the remaining components through interpretable prior quantities. We begin with $\omega$, which rescales every leaf variance and hence the prior of $g_\tau$. We express its effect through $S_\tau=\operatorname{sd} \{g_\tau(\bx_i)\}$  the across-unit standard deviation of the treatment effect at the observed covariates. Following a prior-predictive approach to hyperparameter specification, \citep{gelman2020bayesianworkflow,manderson_translating_2026}, we choose $\omega$ through the induced prior for $S_\tau$. Since $V_j$ is proportional to $\sigma^2$, the prior of $S_\tau/ \sigma $ does not depend on $\sigma$. We set $\omega$ so that the prior median of $S_\tau/\sigma$ targets $\varsigma = 1/2$.

Once $\omega$ sets the overall spread, $a$ controls how $V_j$ is distributed among the leaves. We measure the concentration of this allocation by $\sum_\ell\phi_{j\ell}^2$, which equals one when a single leaf receives all of $V_j$ and $1/L_j$ under equal division. Since $\bphi_j$ is random, we define the prior effective allocation size as $A_{\mathrm{eff}}(T_j)=\{\E(\sum_\ell\phi_{j\ell}^2\mid T_j)\}^{-1}$. If $\rho = 1/2$, then $(1+a)/(1+a/2)\le A_{\mathrm{eff}}(T_j)<1+a$ (See Appendix~\ref{app:effective-size}). Thus, $1+a$ provides an interpretable upper bound for choosing $a$. We set $a=1/2$, giving $A_{\mathrm{eff}}(T_j)\in[1.2,1.5)$, thus favoring allocations in which a few leaves receive most of $V_j$.

\subsection{Theoretical Properties}
\label{sec:theory}
\label{sec:retention}

By construction, the Dirichlet-Tree prior meets the two requirements of Section~\ref{sec:joint}. We now show the implications for the shrinkage of the leaf values and state posterior consistency.

\paragraph{Within-tree Shrinkage}
Conditional on the rest of the model, let $n_{j\ell}$ be the number of treated observations in leaf $\ell$ and $\bar R_{j\ell}$ their mean partial residual, the local estimate of tree $j$'s contribution. Given $\tsc_j^2$ and $\bphi_j$, the conditional posterior mean of $h_{j\ell}$ is
$(1-\kappa_{j\ell})\bar R_{j\ell}$, where
\begin{equation}
\kappa_{j\ell}
=\bigl(1+n_{j\ell}\omega\tsc_j^2\phi_{j\ell}\bigr)^{-1}
\label{eq:shrinkage-factor}
\end{equation}
is its shrinkage factor \citep{piironen2017sparsity,bai_large-scale_2019,aguilar2025dependencyawareshrinkagepriorshigh}. For $q\in(0,1)$, define $N_{j,q}=\#\{\ell:\kappa_{j\ell}\leq 1-q\}$, the number of leaves that retain at least the fraction $q$ of their local mean.

\begin{proposition}
\label{prop:weak-shrinkage}
Let every leaf of $T_j$ contain between $1$ and $n$ treated observations. Under the Dirichlet-Tree prior,
\begin{equation}
\label{eq:upperbound}
    N_{j,q}\leq \frac{(1-q)n \omega\tsc_j^2}{q}
\end{equation}
for every $\tsc_j^2>0$ and $\bphi_j$. If the shares in
\eqref{eq:shrinkage-factor} are instead i.i.d. local scales, independent of $\tsc_j^2$, whose distribution has unbounded support independent of $L_j$, then $\E(N_{j,q}\mid\tsc_j^2)\geq p_qL_j$, where $p_q>0$. 
\end{proposition}

The proposition bounds the number of regions in which a tree can report a substantial estimated effect. With independent local scales, the expected count instead grows at least linearly in $L_j$ (See Appendix~\ref{app:effective-size}). Under the BART leaf prior, the shrinkage factor $\kappa_{j\ell}$ depends only on $n_{j\ell}$, so the prior cannot reallocate variance between regions.

\paragraph{Treatment Forest}
At the forest level, different trees may assign large allocations of variance to different regions. We can extend the bound in Equation~\eqref{eq:upperbound} to the forest of $m_\tau$ trees, showing that under our prior the forest can consider at most $(1-q)n\omega\sum_j\tsc_j^2/q$ regions of substantial effect.

We can additionally consider the size of the effects the forest can attain. Conditioned on the topologies $T_j$ and $\sigma^2$, for a fixed $\bx$ we have
\begin{equation}
    \Pr\{g_\tau(\bx)>t\}\sim\sum_{j=1}^{m_\tau}\Pr\{g(\bx;T_j,\bh_j)>t\}
\end{equation}
as $t \to \infty$, since the tree contributions $g(\bx;T_j,\bh_j)$ are independent and the tail of each tree is $t^{-2b_\tsc}$ asymptotically \citep{feller1971introduction}. Therefore our prior allows for a large effect at $\bx$ to arise from at least one large tree contribution, while allowing the others to remain small. This shows that the forest prior limits the number of regions with a substantial effect and still allows a large effect in any one of them.


\paragraph{Posterior Consistency}
Concentrating a tree's variance in a few regions through the joint prior raises the concern that we may rule out treatment effects that vary throughout the covariate space. Consistency results for BART cannot settle this, since they assume independent priors on the leaf values \citep{rockova2019theory}. Let $P_0$ denote the data-generating law of $(\bx,A,y)$ and let $P_{\mu,\tau,\sigma}$ denote the law induced by the fitted regression surfaces and residual scale.

\begin{theorem}[Weak consistency]
\label{thm:consistency}
Under Conditions~(i)--(iv) of Appendix~\ref{app:consistency}, the prior of \method{} assigns positive probability to every Kullback--Leibler neighborhood of $P_0$. For every weak neighborhood $\mathcal W$ of $P_0$,
\begin{equation}
 \Pi\!\left(P_{\mu,\tau,\sigma}\notin\mathcal W \mid\mathcal D_n\right)\longrightarrow 0 \qquad P_0^\infty\text{-almost surely}.
 \label{eq:weak-consistency}
\end{equation}
If $P_{\mu,\tau,\sigma}=P_0$, then $\sigma=\sigma_0$, $\mu\{\bx,e_0(\bx)\}=\mu_0(\bx)$, and $\tau(\bx)=\tau_0(\bx)$ for $Q$-almost every $\bx$.
\end{theorem}

In practical terms, Theorem~\ref{thm:consistency} shows that even if \method{} induces a preference for few regions of heterogeneity, this does not exclude effects that vary throughout the covariate space. With enough data, the posterior concentrates around the true distribution of the data, $P_0$, which determines the treatment effect. This holds when the true surfaces $\mu_0$ and $\tau_0$ are continuous, treated and control units overlap, and the propensity score is known. 

At first sight, Proposition~\ref{prop:weak-shrinkage} seems to say otherwise, since it bounds the number of regions in which a tree can report a substantial estimated effect. That bound, however, grows with the tree scale $\tsc_j^2$ and with the number $n$ of treated observations in a leaf, so it loosens as the data support larger tree scales and as leaves gather more observations. Hence, the prior shrinks spurious heterogeneity in small samples without ruling out widespread heterogeneity in large ones.


\section{Computation} \label{sec:computation}

Computation follows standard BART/BCF backfitting proposals \citep{hastie_tibshirani_2000bayesian,hahn2020bcf,pratola2016efficient}, with the full sampler described in Appendix~\ref{app:sampler}. A treatment-tree grow move additionally draws a branch fraction $U_{jv}$ from its beta prior, while the reverse prune move removes it. The Metropolis--Hastings ratio includes the corresponding Dirichlet-Tree density and topology-induced leaf variances.

We sample the continuous parameters of a treatment tree in one of two ways. The first integrates the leaf values out and updates each branch fraction and the tree scale by univariate slice sampling \citep{neal2003slice}, for any $a$  and $\rho$. The second applies when $s_\tsc=1$ and $a_\tsc=aW_j$, which holds at our default $\rho=1/2$. Writing $B_{j\ell}=\tsc_j^2\phi_{j\ell}$, we use the hierarchical gamma representation of \citet{zhang2022r2d2}, which makes their full conditionals given the leafs generalized inverse Gaussian. We proceed to draw and normalize them, which gives a blocked Gibbs update of the tree scale and the shares \citep{aguilar2023r2d2m2}. The intercept and residual variance retain their standard full conditional updates.

\section{Experiments}
\label{sec:experiments}

We now present experiments designed to evaluate the empirical performance of \method{} across different patterns of treatment-effect heterogeneity and covariate dimensions, relative to alternative priors and external learners.

\subsection{Models and Evaluation}
\label{sec:exp-setup}

We compare \textbf{\method{}} with three groups of models, namely variants of \textbf{\method{}a} (\textbf{\method{} + DART}, \textbf{\method{} ($\rho=1$)}, and \textbf{Equal}), priors with independent leaf variances (\textbf{Horseshoe HS} and \textbf{Fixed}), and external references (\textbf{BCF}, the \textbf{X-learner} and the \textbf{R-learner}). All Bayesian models are the BCF model \eqref{eq:bcf-model} with the prognostic forest of Section~\ref{sec:bcf}.

\textbf{\method{}} uses the weighted allocation with $a=\rho=1/2$ and the tree scale $\tsc_j^2\sim\BP(1/2,1/2,1)$. \textbf{\method{} + DART} adds, in both forests, the DART prior on the probability that a split uses each covariate \citep{linero2018dart}. \textbf{\method{} ($\rho=1$)} and \textbf{Equal} keep the tree scale and change the allocation. \method{} ($\rho=1$) the shares are symmetric Dirichlet with mean $1/L_j$, and \textbf{Equal}, the limit as $a\to\infty$, gives every leaf the variance $V_j/L_j$. \textbf{HS} is the horseshoe prior, which considers shrinkage but no joint regularization of the scales, within each tree \citep{carvalho_handling_2009}.  \textbf{Fixed} is the BART prior with constant variance $\sigma_h^2=\sigma^2\omega$. In these six models, $\omega$ is set so that the prior median of $S_\tau/\sigma$ is $\varsigma=1/2$,  which makes them comparable at the same prior spread of the effect. In the application, HS is instead the Horseshoe Forest of \citet{jacobs2026horseshoe}, fitted with its authors' package at its defaults.
\begin{figure}[t]
\centering
\includegraphics[width=\columnwidth]{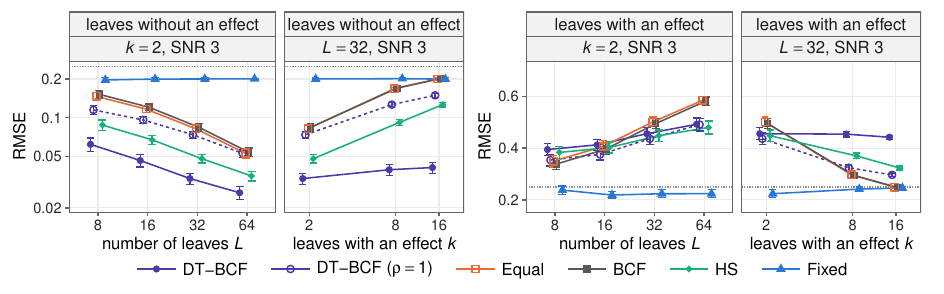}
\caption{Fixed Tree. log-RMSE of the posterior mean segmented over  leaves without (top) and with (bottom) an effect, mean and two standard errors over 200 data sets. Left, $k=2$, against the number of leaves $L$. Right, $L=32$, against the number $k$ of leaves with an effect. Dotted, the RMSE $\sigma/m^{1/2}$ of the leaf means.}
\label{fig:partition}
\end{figure}

\begin{table}[b]
\caption{Tree-Size Study. $p=50$ over 40 data sets ($\max$ s.e $\leq$ 0.06). Bold, the best RMSE, FPR and coverage. BCF and the X-learner do not depend on $\eta$.}
\label{tab:forest}
\begin{center}
\scriptsize
\setlength{\tabcolsep}{2pt}
\renewcommand{\arraystretch}{0.9}
\resizebox{\columnwidth}{!}{%
\begin{tabular}{lcccccccc}
\toprule
& \multicolumn{4}{c}{2 Regions} & \multicolumn{4}{c}{4 Regions Graded}\\
\cmidrule(lr){2-5}\cmidrule(lr){6-9}
Model & RMSE & Cov. & FPR & TPR & RMSE & Cov. & FPR & TPR\\
\midrule
\multicolumn{9}{l}{\emph{$\eta=2$}}\\
\method{} & 0.70 & 0.75 & 0.17 & 0.69 & 0.92 & \textbf{0.73} & 0.24 & 0.63\\
\method{} + DART & \textbf{0.60} & \textbf{0.78} & \textbf{0.16} & 0.77 & \textbf{0.70} & 0.72 & \textbf{0.22} & 0.71\\
BCF & 0.77 & 0.52 & 0.33 & 0.74 & 0.96 & \textbf{0.73} & 0.23 & 0.67\\
HS & 0.75 & 0.68 & 0.22 & 0.70 & 1.11 & 0.65 & 0.32 & 0.64\\
Fixed & 0.86 & 0.12 & 0.60 & 0.83 & 1.45 & 0.16 & 0.94 & 0.98\\
X-learner & 0.67 & & & & 0.84 & & & \\
\multicolumn{9}{l}{\emph{$\eta=0.25$}}\\
\method{} & 0.69 & \textbf{0.77} & \textbf{0.15} & 0.69 & 0.91 & \textbf{0.76} & 0.20 & 0.61\\
\method{} + DART & \textbf{0.62} & \textbf{0.77} & \textbf{0.15} & 0.73 & \textbf{0.76} & 0.75 & \textbf{0.19} & 0.68\\
HS & 0.79 & 0.46 & 0.34 & 0.74 & 1.13 & 0.57 & 0.41 & 0.65\\
Fixed & 0.88 & 0.00 & 0.87 & 0.92 & 1.52 & 0.14 & 1.00 & 1.00\\
\bottomrule
\end{tabular}}
\end{center}
\end{table}
Finally, \textbf{BCF} \citep{hahn2020bcf}, the baseline, is fitted at its published defaults, without matching $\varsigma$. 
We fit BCF with the authors' package, and with our sampler in the designs only with $p\ge n$, where the package fails. The \textbf{X-learner} and the \textbf{R-learner}, both with gradient-boosted trees provide loss-based references \citep{chen2016xgboost,kuenzel2019meta,nie2021quasi}. Bayesian models run 4 Markov chains, with the first half of each discarded as warmup \citep{vehtari_rank-normalization_2021}.

We measure accuracy by the root mean squared error (RMSE), over the $n$ units, of the estimate of the centered effect $\delta_i=\tau(\bx_i)-n^{-1}\sum_{i'}\tau(\bx_{i'})$, which is the posterior mean for the Bayesian models. Centering removes the error in the average effect, so the comparison focuses on the variation of the effect across units, which the leaf prior regularizes. For the Bayesian models we also report the coverage (Cov.) and length of the 90\% intervals for $\delta_i$. In designs where some units have no effect, the false positive rate (FPR) and the true positive rate (TPR) are the proportions of units with $\tau(\bx_i)=0$ and with $\tau(\bx_i)\neq0$ whose 90\% interval for $\tau(\bx_i)$ excludes zero. The log predictive density of held-out data (lpd) and  complete tables are in Appendix~\ref{app:experiments}.

\subsection{Leaf Priors}
\label{sec:exp-partition}

\begin{figure*}[t]
\centering
\includegraphics[width=\textwidth]{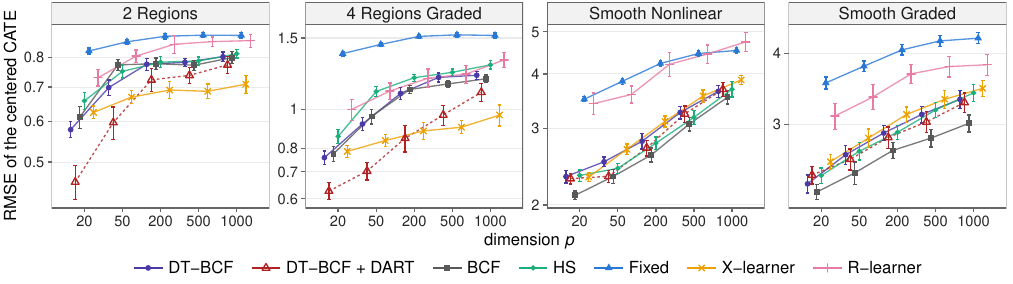}
\caption{Dimension study ($n=200$, $\eta=2$), RMSE of the centered effect against the number of covariates $p$, mean and two standard errors over 40 data sets, on a log scale.}
\label{fig:experiments}
\end{figure*}

To isolate how the leaf priors allocate variance once a partition has been found, we fix a balanced tree with $L$ leaves. Each leaf holds $m=16$ observations, summarized by their mean $\bar R_\ell\mid h_\ell\sim\Normal(h_\ell,\sigma^2/m)$ with $\sigma=1$. Of the $L$ leaves, $k$ have effects $\pm\Delta$, with signal-to-noise ratio (SNR) $\Delta m^{1/2}/\sigma=3$, and the others have none (null leaves). We take $k=2$ with $L\in\{8,16,32,64\}$ and $k\in\{8,16\}$ with $L=32$, with 200 data sets per condition and chains of 8,000 iterations. We match $\omega$ at every $L$ for every model, including the leaf prior of BCF, so that all share the prior median spread of the leaf values.

On null leaves, \method{} has the smallest RMSE in every condition (Figure~\ref{fig:partition}). Equal, BCF and Fixed give every leaf a common variance. Equal and BCF differ only in its prior and nearly coincide. Keeping the null leaves near zero then requires a small common variance, also shrinking the non-null leaves. Fixed has the smallest RMSE on leaves with an effect and the largest on the others. This is expected since its matched variance is proportional to $\sigma^2/m$, thus shrinking every leaf mean only slightly. \method{} instead can give most of $V_j$ to the two leaves with large $|\bar R_\ell|$. It is less accurate than BCF on leaves with an effect at $L\le16$ and more accurate than BCF and Equal at $L\ge32$, where their common variance is informed mainly by the leaves without an effect. Within the Dirichlet-Tree family, the RMSE on null leaves is smallest for \method{} and largest for Equal, whose shares are the least concentrated. HS also shrinks null leaves less than \method{}. With $k=8$ or $16$, \method{} keeps the smallest RMSE on null leaves but has the largest on leaves with an effect. With $a=1/2$, the prior effective allocation size satisfies $1.2\le A_{\mathrm{eff}}(T_j)<1.5$, so the prior concentrates $V$ on few leaves. This behavior suits effects confined to a few regions of the covariate space.

\subsection{Forest Simulations}
\label{sec:exp-forest}

We next consider causal forest simulations, in which the tree search must also find the partition. Each data set has $n=200$ units with covariates $\bx_i\sim\mathrm{U}(0,1)^p$ and a propensity score that depends only on $x_1$ to $x_5$. We consider four scenarios that differ in how the treatment effect is distributed over the covariate space. In \emph{2 Regions}, the effect equals 2 on two rectangles defined by $x_1$ to $x_5$ and zero elsewhere. In \emph{4 Regions Graded}, it equals $4\cdot0.6^{k-1}$ on rectangle $k=1,\dots,4$ and zero elsewhere, so that the regions carry effects of different sizes. 
In \emph{Smooth Nonlinear}, every unit has an effect, given by the Friedman function $\tau_0(\bx)=10\sin(\pi x_1x_2)+20(x_3-1/2)^2+10x_4+5x_5$. \emph{Smooth Graded} multiplies $\tau_0$ by $1.65^{-q(\bx)}$, where $q(\bx)\in\{0,1,2,3\}$ indexes the quadrant of $(x_6,x_7)$, so that the amplitude of the effect varies across regions. Smooth scenarios have $\sigma=\sqrt3$.

We perform two studies: In the tree-size study we fix $p=50$ and vary $\eta\in\{0.25,0.5,1,2\}$ in the split probability $0.95(1+d)^{-\eta}$ of the treatment trees, where smaller $\eta$ favors larger trees. In the dimension study we fix $\eta=2$ and vary $p\in\{20,50,200,500,1000\}$, of which at most seven covariates modify the effect. Each case has 40 data sets, and every chain runs 4,000 iterations.

\paragraph{Tree-size Prior.} On the region surfaces, \method{} has lower RMSE than BCF and HS at every $\eta$, with a smaller margin over BCF on 4 Regions Graded; \method{} + DART has the lowest RMSE in Table~\ref{tab:forest}.  The gain comes from null regions, where leaves receive small shares of a fixed total variance $V_j$. On 2 Regions, \method{} reports an effect for about half as many null units as BCF; on both surfaces its FPR is lower than that of HS, at the cost of lower TPR. Its intervals also cover more often than those of HS and, on 2 Regions, of BCF. As $\eta$ decreases, the RMSE and FPR of \method{} remain nearly constant. Fixed and HS deteriorate on both region surfaces. This pattern is consistent with the fixed total variance $V_j$. A new leaf receives part of $V_j$ under \method{}, while the independent priors in HS and Fixed introduce another leaf-specific source of variance. On the smooth surfaces, where every unit has an effect, BCF is more accurate than \method{}. The X-learner can be more accurate than \method{} on the region surfaces but less accurate than \method{} + DART.

\paragraph{Dimension}
The Bayesian models with uniform splitting probabilities lose accuracy as $p$ grows because most proposed splits use covariates that do not modify the effect (Figure~\ref{fig:experiments}). Up to $p=50$, \method{} is the most accurate of these models on the region surfaces, or is as accurate as BCF. At larger $p$, it remains close to BCF on 2 Regions and ahead of HS on 4 Regions Graded, although BCF has lower RMSE there when $p\ge500$. The leaf prior regularizes a partition after it has been formed, so it cannot recover an effect region that the tree search misses. DART addresses this search problem by concentrating splits on effect modifiers. With DART, \method{} has lower RMSE than BCF on both region surfaces at every $p$. Holding DART fixed also gives \method{} + DART lower RMSE than HS + DART at every $p$, while Equal + DART performs similarly. These comparisons provide more consistent evidence for a tree-level total variance than for an additional benefit from unequal allocation in these designs.

\begin{table}[t]
\centering
\caption{ACIC 2016 Benchmark. 19 settings, 10 replicates each. Geometric-mean RMSE ratio to BCF with a two-s.e interval, number of settings with a ratio below one, coverage and length of 90\% intervals. }
\label{tab:acic}
\begin{center}
\scriptsize
\setlength{\tabcolsep}{2pt}
\renewcommand{\arraystretch}{0.9}
\resizebox{\columnwidth}{!}{%
\begin{tabular}{lcccc}
\toprule
Model & RMSE ratio & Number of Settings & Cov. & Length\\
\midrule
\method{} & 0.99 (0.95, 1.04) & 10 & 0.80 & 0.20\\
\method{} + DART & \textbf{0.88} (0.83, 0.92) & \textbf{18} & 0.80 & 0.17\\
BCF & 1 & & 0.77 & 0.21\\
HS & 0.99 (0.95, 1.04) & 10 & 0.79 & 0.20\\
Fixed & 1.06 (0.99, 1.13) & 5 & \textbf{0.83} & 0.19\\
X-learner & 1.25 (1.19, 1.30) & 0 & & \\
R-learner & 1.90 (1.74, 2.08) & 0 & & \\
\bottomrule
\end{tabular}}
\end{center}
\end{table}

\subsection{ACIC 2016 Benchmark}
\label{sec:exp-acic}

We next evaluate the models on the data analysis competition of the 2016 Atlantic Causal Inference Conference (ACIC 2016), a widely used benchmark for heterogeneous effects \citep{dorie2019automated}. BART-based methods were among the most accurate, so BCF is a strong reference. The benchmark simulates treatment and outcome from the 58 correlated, mixed-type covariates of 4,802 units of a perinatal study.

The competition defines 77 settings, which differ in how treatment and outcome are generated. Before fitting any model, we drew 20 of them at random so that they cover the competition's range of heterogeneity (Appendix~\ref{app:exp-acic}). One has a constant effect and is reported there. In the other 19, the effect varies across nearly all units, as in the smooth scenarios.

 Models are fitted to the first 10 of the 100 data sets (replicates) of every setting, with chains of 20,000 iterations. Since the RMSE ranges over more than an order of magnitude across data sets, we report it relative to BCF on the same dataset in Table~\ref{tab:acic}. \method{} and HS have a geometric-mean RMSE close to that of BCF. \method{} + DART has the lowest, with an interval below one and a ratio below one in 18 of the 19 settings (Table~\ref{tab:acic}), consistent with DART concentrating the splits on the covariates that modify the effect. The intervals of \method{} cover more often than those of BCF and HS, at similar mean length.

\subsection{Application to Pancreatic Cancer Survival}
\label{sec:application}

We also study how our models perform in real life data in the context of censored survival outcomes. We consider the pancreatic ductal adenocarcinoma cohort of The Cancer Genome Atlas \citep{tcga2017pdac, jacobs2026horseshoe} dataset. Of the 130 patients, 36 received adjuvant radiotherapy after resection. We model overall survival on the log scale with an accelerated failure time model that imputes censored survival times and uses 3,027 clinical and gene-expression covariates. Each model fit 4 chains of 40,000 iterations. See Appendix~\ref{app:exp-pdac} for more details.

\begin{table}[b]
\centering
\caption{Pancreatic Cohort. Effect of adjuvant radiotherapy on log survival time. ATE, posterior mean with 95\% CI. SD across patients, prior median and posterior median with 95\% interval. PSIS-LOO elpd (s.e).}
\label{tab:pdac}
\begin{center}
\scriptsize
\setlength{\tabcolsep}{2pt}
\renewcommand{\arraystretch}{0.9}
\resizebox{\columnwidth}{!}{%
\begin{tabular}{@{}lcccc@{}}
\toprule
 & & \multicolumn{2}{c}{SD across patients} & \\
\cmidrule(lr){3-4}
Model & ATE & prior & posterior & elpd \\
\midrule
\method{} & 1.00 (0.45, 1.70) & 0.50 & 0.31 (0.10, 1.87) & $-140.7$ (9.9) \\
\method{} + DART & 1.04 (0.47, 1.87) & 0.45 & 0.31 (0.10, 1.95) & $-142.8$ (10.2) \\
BCF & 0.99 (0.45, 1.64) & 0.96 & 0.34 (0.02, 1.17) & $-139.7$ (10.0) \\
HS & 0.95 (0.34, 1.56) & -- & 0.78 (0.46, 1.24) & $-167.0$ (12.8) \\
Fixed & 0.97 (0.44, 1.51) & 0.47 & 0.42 (0.25, 0.61) & $-141.2$ (10.0) \\
\bottomrule
\end{tabular}}
\end{center}
\end{table}

\begin{figure}[t]
\centering
\includegraphics{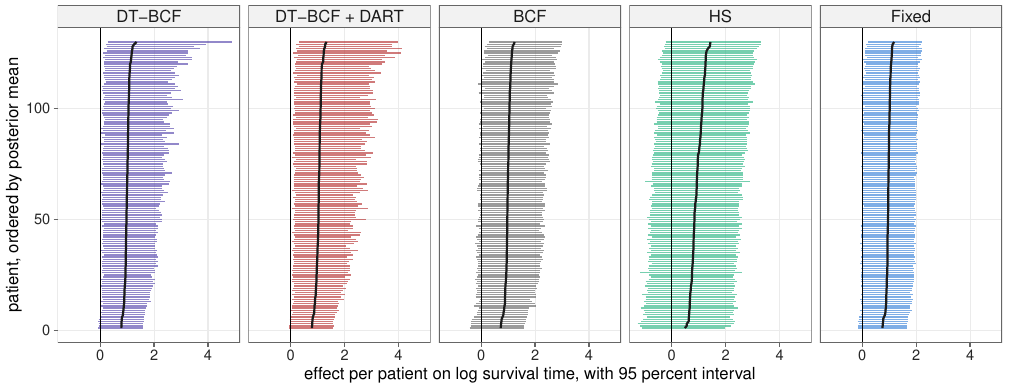}
\caption{Pancreatic cohort. Effect $\tau(\bx_i)$ of adjuvant radiotherapy on log survival time for each patient under the models of Table~\ref{tab:pdac}. The horizontal lines are the 95\% intervals. The black line connects the posterior means, with the patients ordered by posterior mean within each model. The vertical line marks zero.}
\label{fig:app-pdac-cate}
\end{figure}
 
\method{} and BCF are in accordance (Table~\ref{tab:pdac}). Every model estimates a positive ATE, with posterior means between 0.95 and 1.04 on the log survival scale and 95 $\%$ intervals above zero. The estimated leave-one-out predictive densities of \method{} and BCF are similar, and that of HS is lower. To judge whether the cohort is informative about heterogeneity, we compare the prior and posterior of the standard deviation of the effect across patients. Under the matched priors, its posterior median stays at or below the prior median, with wide intervals. Hence, the cohort carries little information about how much the effect varies across patients. This is also reflected in the the 95\% CI of the departure $\tau(\bx_i)-\bar\tau$ from the average effect which contains zero for every model and patient, meaning that the data does not identify patients who benefit more or less than average. Under every model, the estimated effects of the 130 patients lie close together with their wide 95\% intervals all overlapping (Figure~\ref{fig:app-pdac-cate}). From a practical perspective, the estimates give no basis for choosing which patients receive radiotherapy based on their clinical or genomic profile.

We show in Figure~\ref{fig:app-pdac-km} that every model predicts higher survival with radiotherapy than without it throughout follow-up. At two years, the predicted survival is about 0.7 with radiotherapy and about 0.4 without it. \method{}, \method{} + DART, BCF and Fixed give nearly identical curves, while HS gives a somewhat smaller difference between the two curves. For this dataset, the estimated benefit of radiotherapy therefore depends little on the choice of prior.
 
\begin{figure}[bp]
\centering
\includegraphics{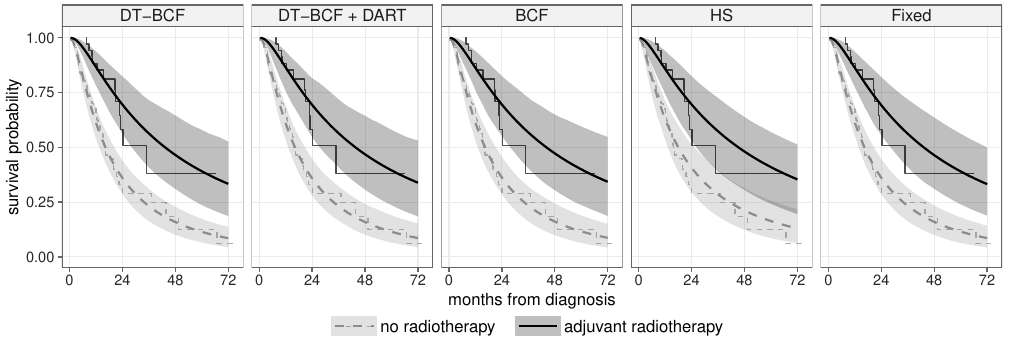}
\caption{Pancreatic cohort. Predicted survival of the 130 patients under each model of Table~\ref{tab:pdac} if every patient received adjuvant radiotherapy (solid) or none did (dashed). The bands show pointwise 95\% credible intervals and the steps show Kaplan--Meier curves of the 36 treated and the 94 untreated patients.}
\label{fig:app-pdac-km}
\end{figure}
\FloatBarrier
\section{Discussion}
\label{sec:discussion}

We have introduced \method{}, which gives the leaves of each treatment tree a joint variance prior informed by the tree topology. Our construction separates the overall contribution of a tree from its distribution across regions and combines strong shrinkage of negligible regional effects with support for substantial ones. Since standard leaf priors set the variance of each region independently, the number of regions in which a tree reports a substantial effect is allowed to grow with the size of the tree. We have shown that our prior controls this by sharing a total variance among the regions of each tree, which compete for this variance. Regions whose data support an effect claim a large share of it, leaving the others strongly shrunk toward zero. Joint shrinkage priors in regression apply the same principle, distributing a total variance or global scale over model components fixed in advance \citep{fuglstad_constructing_2019,zhang2022r2d2,aguilar2023r2d2m2}. In \method{}, the components are the regions of a random tree, so the prior allocates variance over a partition that is itself learned from the data.

Across our empirical studies, \method{} performs competitively over diverse effect surfaces, providing its clearest gains when heterogeneity is concentrated in a few regions.  In high dimensions, performance also depends on discovering which covariates are important, since the leaf prior regularizes a partition only after the search has formed it. Adding DART produces consistent gains in high dimensions and in the benchmarks, indicating that joint leaf allocation and adaptive splitting address complementary regularization problems. 

The Dirichlet-Tree construction restricts the dependence structures available among the leaf variances. More flexible compositional laws, including logistic-normal allocations \citep{aitchison_logistic-normal_1980,aguilar_generalized_2025}, could represent richer relationships among regions, although they would require new posterior computation. Our prior also acts within each tree, while the trees of the treatment forest remain independent a priori. Allocating a forest-level total variance across the trees, and then across the leaves of each tree, would extend the joint regularization to the whole ensemble. This approach would reflect the spirit of grouped shrinkage priors that regularize both groups and their members \citep{boss_group_2023,agarwal_bayesian_2025,graspgroupedregressionadaptive2025,aguilar_r2_2025}.  The survival application shows that the method extends to censored outcomes. Adapting it to other BART variants, such as targeted smooth BCF and shrinkage BCF \citep{starling2021targeted,caron2022shrinkage}, and to other outcome models is another direction for future work.

To adjust for confounding, we build on the work of \citet{hahn2020bcf}. The estimated propensity score enters the prognostic function, which guards against regularization-induced confounding \citep{hahn2018regularization}, while identification rests on the observed covariates being sufficient to adjust for treatment assignment. Because \method{} changes only the leaf prior of the treatment trees, this adjustment remains unchanged. Our prior could also be paired with other adjustments such as Robinson's orthogonalization. This approach underlies the R-learner \citep{kuenzel2019meta}, separating the estimation of the treatment effect from that of nuisance functions \citep{robinson1988root,nie2021quasi}. Within such a decomposition, the Dirichlet-Tree prior could be used to regularize the effect function. When unmeasured confounding is a concern, instrumental variables can identify local average treatment effects under different assumptions \citep{angrist1996identification}. We can apply our joint leaf prior to the tree-based models of that effect.

\section*{Code availability}
An implementation of \method{} and the code to reproduce the experiments are available at \coderepo.

\clearpage
\printbibliography

@article{bhattacharya2015dirichlet,
  author  = {Bhattacharya, Anirban and Pati, Debdeep and Pillai, Natesh S. and Dunson, David B.},
  title   = {Dirichlet--Laplace Priors for Optimal Shrinkage},
  journal = {Journal of the American Statistical Association},
  volume  = {110},
  number  = {512},
  pages   = {1479--1490},
  year    = {2015},
  doi     = {10.1080/01621459.2014.960967}
}

@article{aguilar2023r2d2m2,
  author  = {Aguilar, Javier Enrique and B{\"u}rkner, Paul-Christian},
  title   = {Intuitive Joint Priors for Bayesian Linear Multilevel Models: The {R2D2M2} Prior},
  journal = {Electronic Journal of Statistics},
  volume  = {17},
  number  = {1},
  pages   = {1711--1767},
  year    = {2023},
  doi     = {10.1214/23-EJS2136}
}

@article{caron2022shrinkage,
  author  = {Caron, Alberto and Baio, Gianluca and Manolopoulou, Ioanna},
  title   = {Shrinkage Bayesian Causal Forests for Heterogeneous Treatment Effects Estimation},
  journal = {Journal of Computational and Graphical Statistics},
  volume  = {31},
  number  = {4},
  pages   = {1202--1214},
  year    = {2022},
  doi     = {10.1080/10618600.2022.2067549}
}

@article{chipman2010bart,
  author  = {Chipman, Hugh A. and George, Edward I. and McCulloch, Robert E.},
  title   = {{BART}: Bayesian Additive Regression Trees},
  journal = {The Annals of Applied Statistics},
  volume  = {4},
  number  = {1},
  pages   = {266--298},
  year    = {2010},
  doi     = {10.1214/09-AOAS285}
}

@article{dennis1991hyperdirichlet,
  author  = {Dennis, III, Samuel Y.},
  title   = {On the Hyper-{D}irichlet Type 1 and Hyper-{L}iouville Distributions},
  journal = {Communications in Statistics---Theory and Methods},
  volume  = {20},
  number  = {12},
  pages   = {4069--4081},
  year    = {1991},
  doi     = {10.1080/03610929108830757}
}

@article{hastie_tibshirani_2000bayesian,
  title={Bayesian backfitting (with comments and a rejoinder by the authors},
  author={Hastie, Trevor and Tibshirani, Robert},
  journal={Statistical Science},
  volume={15},
  number={3},
  pages={196--223},
  year={2000},
  publisher={Institute of Mathematical Statistics}
}

@article{hahn2018regularization,
  author  = {Hahn, P. Richard and Carvalho, Carlos M. and Puelz, David and He, Jingyu},
  title   = {Regularization and Confounding in Linear Regression for Treatment Effect Estimation},
  journal = {Bayesian Analysis},
  volume  = {13},
  number  = {1},
  pages   = {163--182},
  year    = {2018},
  doi     = {10.1214/16-BA1044}
}

@article{hahn2020bcf,
  author  = {Hahn, P. Richard and Murray, Jared S. and Carvalho, Carlos M.},
  title   = {Bayesian Regression Tree Models for Causal Inference: Regularization, Confounding, and Heterogeneous Effects},
  journal = {Bayesian Analysis},
  volume  = {15},
  number  = {3},
  pages   = {965--1056},
  year    = {2020},
  doi     = {10.1214/19-BA1195}
}

@article{green2012heterogeneous,
  author  = {Green, Donald P. and Kern, Holger L.},
  title   = {Modeling Heterogeneous Treatment Effects in Survey Experiments with Bayesian Additive Regression Trees},
  journal = {Public Opinion Quarterly},
  volume  = {76},
  number  = {3},
  pages   = {491--511},
  year    = {2012},
  doi     = {10.1093/poq/nfs036}
}

@article{hill2011causal,
  author  = {Hill, Jennifer L.},
  title   = {Bayesian Nonparametric Modeling for Causal Inference},
  journal = {Journal of Computational and Graphical Statistics},
  volume  = {20},
  number  = {1},
  pages   = {217--240},
  year    = {2011},
  doi     = {10.1198/jcgs.2010.08162}
}

@article{dorie2019automated,
  author  = {Dorie, Vincent and Hill, Jennifer and Shalit, Uri and Scott, Marc and Cervone, Dan},
  title   = {Automated versus Do-It-Yourself Methods for Causal Inference: Lessons Learned from a Data Analysis Competition},
  journal = {Statistical Science},
  volume  = {34},
  number  = {1},
  pages   = {43--68},
  year    = {2019},
  doi     = {10.1214/18-STS667}
}

@article{jacobs2026horseshoe,
  author  = {Jacobs, Tijn and van Wieringen, Wessel N. and Pas, St{\'e}phanie L. van der},
  title   = {Horseshoe Forests for High-Dimensional Causal Survival Analysis},
  journal = {Bayesian Analysis},
  year    = {2026},
  note    = {Advance publication},
  doi     = {10.1214/26-BA1603}
}

@article{linero2018dart,
  author  = {Linero, Antonio R.},
  title   = {Bayesian Regression Trees for High-Dimensional Prediction and Variable Selection},
  journal = {Journal of the American Statistical Association},
  volume  = {113},
  number  = {522},
  pages   = {626--636},
  year    = {2018},
  doi     = {10.1080/01621459.2016.1264957}
}

@techreport{minka2004dirichlettree,
  author      = {Minka, Thomas P.},
  title       = {The {D}irichlet-tree Distribution},
  institution = {Justsystem Pittsburgh Research Center},
  year        = {2004},
  note        = {Originally circulated in 1999, revised October 2004},
  url         = {https://tminka.github.io/papers/dirichlet/minka-dirtree.pdf}
}

@article{piironen2017sparsity,
  author  = {Piironen, Juho and Vehtari, Aki},
  title   = {Sparsity Information and Regularization in the Horseshoe and Other Shrinkage Priors},
  journal = {Electronic Journal of Statistics},
  volume  = {11},
  number  = {2},
  pages   = {5018--5051},
  year    = {2017},
  doi     = {10.1214/17-EJS1337SI}
}

@article{pratola2016efficient,
  author  = {Pratola, Matthew T.},
  title   = {Efficient Metropolis--Hastings Proposal Mechanisms for Bayesian Regression Tree Models},
  journal = {Bayesian Analysis},
  volume  = {11},
  number  = {3},
  pages   = {885--911},
  year    = {2016},
  doi     = {10.1214/16-BA999}
}

@inproceedings{rockova2019theory,
  author    = {Ro\v{c}kov{\'a}, Veronika and Saha, Enakshi},
  title     = {On Theory for {BART}},
  booktitle = {Proceedings of the Twenty-Second International Conference on Artificial Intelligence and Statistics},
  series    = {Proceedings of Machine Learning Research},
  volume    = {89},
  pages     = {2839--2848},
  year      = {2019},
  url       = {https://proceedings.mlr.press/v89/rockova19a.html}
}

@article{schwartz1965bayes,
  author  = {Schwartz, Lorraine},
  title   = {On Bayes Procedures},
  journal = {Zeitschrift f{\"u}r Wahrscheinlichkeitstheorie und Verwandte Gebiete},
  volume  = {4},
  number  = {1},
  pages   = {10--26},
  year    = {1965},
  doi     = {10.1007/BF00535479}
}

@article{starling2021targeted,
  author  = {Starling, Jennifer E. and Murray, Jared S. and Lohr, Patricia A. and Aiken, Abigail R. A. and Carvalho, Carlos M. and Scott, James G.},
  title   = {Targeted Smooth Bayesian Causal Forests: An Analysis of Heterogeneous Treatment Effects for Simultaneous versus Interval Medical Abortion Regimens over Gestation},
  journal = {The Annals of Applied Statistics},
  volume  = {15},
  number  = {3},
  pages   = {1194--1219},
  year    = {2021},
  doi     = {10.1214/20-AOAS1438}
}

@article{zhang2022r2d2,
  author  = {Zhang, Yan Dora and Naughton, Brian P. and Bondell, Howard D. and Reich, Brian J.},
  title   = {Bayesian Regression Using a Prior on the Model Fit: The {R2-D2} Shrinkage Prior},
  journal = {Journal of the American Statistical Association},
  volume  = {117},
  number  = {538},
  pages   = {862--874},
  year    = {2022},
  doi     = {10.1080/01621459.2020.1825449}
}

@article{nie2021quasi,
  title   = {Quasi-oracle estimation of heterogeneous treatment effects},
  author  = {Nie, Xinkun and Wager, Stefan},
  journal = {Biometrika},
  volume  = {108},
  number  = {2},
  pages   = {299--319},
  year    = {2021}
}

@article{athey2019grf,
  title   = {Generalized random forests},
  author  = {Athey, Susan and Tibshirani, Julie and Wager, Stefan},
  journal = {The Annals of Statistics},
  volume  = {47},
  number  = {2},
  pages   = {1148--1178},
  year    = {2019}
}

@article{vehtari2017practical,
  title   = {Practical {B}ayesian model evaluation using leave-one-out cross-validation and {WAIC}},
  author  = {Vehtari, Aki and Gelman, Andrew and Gabry, Jonah},
  journal = {Statistics and Computing},
  volume  = {27},
  number  = {5},
  pages   = {1413--1432},
  year    = {2017}
}

@misc{yanchenko2024r2d2priorgeneralizedlinear,
      title={The R2D2 Prior for Generalized Linear Mixed Models}, 
      author={Eric Yanchenko and Howard D. Bondell and Brian J. Reich},
      year={2024},
      eprint={2111.10718},
      archivePrefix={arXiv},
      primaryClass={stat.ME},
      url={https://arxiv.org/abs/2111.10718}, 
}

@book{feller1971introduction,
  author    = {Feller, William},
  title     = {An Introduction to Probability Theory and Its Applications, Volume {II}},
  edition   = {Second},
  publisher = {John Wiley \& Sons},
  address   = {New York},
  year      = {1971}
}

@article{rosenbaum1983central,
  author  = {Rosenbaum, Paul R. and Rubin, Donald B.},
  title   = {The Central Role of the Propensity Score in Observational Studies for Causal Effects},
  journal = {Biometrika},
  volume  = {70},
  number  = {1},
  pages   = {41--55},
  year    = {1983}
}

@book{imbens2015causal,
  author    = {Imbens, Guido W. and Rubin, Donald B.},
  title     = {Causal Inference for Statistics, Social, and Biomedical Sciences: An Introduction},
  publisher = {Cambridge University Press},
  address   = {Cambridge},
  year      = {2015}
}

@article{kent2018personalized,
  author  = {Kent, David M. and Steyerberg, Ewout and van Klaveren, David},
  title   = {Personalized Evidence Based Medicine: Predictive Approaches to Heterogeneous Treatment Effects},
  journal = {BMJ},
  volume  = {363},
  pages   = {k4245},
  year    = {2018}
}

@inproceedings{alaa2017bayesian,
  author    = {Alaa, Ahmed M. and van der Schaar, Mihaela},
  title     = {Bayesian Inference of Individualized Treatment Effects Using Multi-Task {G}aussian Processes},
  booktitle = {Advances in Neural Information Processing Systems},
  volume    = {30},
  year      = {2017}
}

@inproceedings{shalit2017estimating,
  author    = {Shalit, Uri and Johansson, Fredrik D. and Sontag, David},
  title     = {Estimating Individual Treatment Effect: Generalization Bounds and Algorithms},
  booktitle = {Proceedings of the 34th International Conference on Machine Learning},
  series    = {Proceedings of Machine Learning Research},
  volume    = {70},
  pages     = {3076--3085},
  year      = {2017}
}

@article{kennedy2023drlearner,
  title={Towards optimal doubly robust estimation of heterogeneous causal effects},
  author={Kennedy, Edward H},
  journal={Electronic {J}ournal of {S}tatistics},
  volume={17},
  number={2},
  pages={3008--3049},
  year={2023},
  publisher={The Institute of Mathematical Statistics and the Bernoulli Society},
  doi = {10.1214/23-EJS2157}
}

@inproceedings{chen2016xgboost,
author = {Chen, Tianqi and Guestrin, Carlos},
title = {XGBoost: A Scalable Tree Boosting System},
year = {2016},
isbn = {9781450342322},
publisher = {Association for Computing Machinery},
address = {New York, NY, USA},
url = {https://doi.org/10.1145/2939672.2939785},
doi = {10.1145/2939672.2939785},
booktitle = {Proceedings of the 22nd ACM SIGKDD International Conference on Knowledge Discovery and Data Mining},
pages = {785–794},
location = {San Francisco, California, USA},
series = {KDD '16}
}

@article{carvalho2010horseshoe,
  author  = {Carvalho, Carlos M. and Polson, Nicholas G. and Scott, James G.},
  title   = {The Horseshoe Estimator for Sparse Signals},
  journal = {Biometrika},
  volume  = {97},
  number  = {2},
  pages   = {465--480},
  year    = {2010},
  doi     = {10.1093/biomet/asq017}
}

@article{vehtari_rank-normalization_2021,
	title = {Rank-{Normalization}, {Folding}, and {Localization}: {An} {Improved} {\textbackslash}{widehatR} for {Assessing} {Convergence} of {MCMC} (with {Discussion})},
	volume = {16},
	url = {https://doi.org/10.1214/20-BA1221},
	doi = {10.1214/20-BA1221},
	number = {2},
	journal = {Bayesian Analysis},
	publisher = {International Society for Bayesian Analysis},
	author = {Vehtari, Aki and Gelman, Andrew and Simpson, Daniel and Carpenter, Bob and Bürkner, Paul-Christian},
	year = {2021},
	pages = {667 -- 718},
}

@book{aitchison_statistical_1986,
	series = {Monographs on statistics and applied probability ({Series})},
	title = {The statistical analysis of compositional data},
	isbn = {0-412-28060-4},
	language = {eng},
	publisher = {Chapman and Hall},
	author = {Aitchison, John},
	year = {1986},
	lccn = {86002685},
}

@article{bai_large-scale_2019,
	title = {Large-scale multiple hypothesis testing with the normal-beta prime prior},
	volume = {53},
	url = {https://doi.org/10.1080/02331888.2019.1662017},
	doi = {10.1080/02331888.2019.1662017},
	number = {6},
	journal = {Statistics},
	author = {Bai, Ray and Ghosh, Malay},
	year = {2019},
	pages = {1210--1233},
}

@misc{gelman2020bayesianworkflow,
	title = {Bayesian {Workflow}},
	copyright = {Creative Commons Attribution 4.0 International},
	url = {https://arxiv.org/abs/2011.01808},
	doi = {10.48550/ARXIV.2011.01808},
	publisher = {arXiv},
	author = {Gelman, Andrew and Vehtari, Aki and Simpson, Daniel and Margossian, Charles C. and Carpenter, Bob and Yao, Yuling and Kennedy, Lauren and Gabry, Jonah and Bürkner, Paul-Christian and Modrák, Martin},
	year = {2020},
    note = {arXiv:2011.01808},
}

@article{lin_dirichlet_2016,
	title={On the {D}irichlet distribution},
  author={Lin, Jiayu},
  journal={Department of Mathematics and Statistics, Queens University},
  volume={40}, 
  pages={46},
  year={2016}
}

@article{fuglstad_constructing_2019,
	title = {Constructing {Priors} that {Penalize} the {Complexity} of {Gaussian} {Random} {Fields}},
	volume = {114},
	url = {https://doi.org/10.1080/01621459.2017.1415907},
	doi = {10.1080/01621459.2017.1415907},
	number = {525},
	journal = {Journal of the American Statistical Association},
	author = {Fuglstad, Geir-Arne and Simpson, Daniel and Lindgren, Finn and Rue, Håvard},
	year = {2019},
	pages = {445--452},
}

@inproceedings{carvalho_handling_2009,
	series = {Proceedings of {Machine} {Learning} {Research}},
	title = {Handling {Sparsity} via the {Horseshoe}},
	volume = {5},
	url = {https://proceedings.mlr.press/v5/carvalho09a.html},
	booktitle = {Proceedings of the {Twelth} {International} {Conference} on {Artificial} {Intelligence} and {Statistics}},
	publisher = {PMLR},
	author = {Carvalho, Carlos M. and Polson, Nicholas G. and Scott, James G.},
	editor = {van Dyk, David and Welling, Max},
	month = apr,
	year = {2009},
	pages = {73--80},
}

@article{kruijer_adaptive_2010,
	title = {Adaptive {Bayesian} {Density} {Estimation} with {Location}-{Scale} {Mixtures}},
	volume = {4},
	doi = {10.1214/10-EJS584},
	journal = {Electronic Journal of Statistics},
	author = {Kruijer, Willem and Rousseau, Judith and van der Vaart, A. W.},
	month = jan,
	year = {2010},
}

@misc{vehtari_loo_2023,
	title = {loo: {Efficient} leave-one-out cross-validation and {WAIC} for {Bayesian} models},
	url = {https://mc-stan.org/loo/},
	author = {Vehtari, Aki and Gabry, Jonah and Magnusson, Mans and Yao, Yuling and Bürkner, Paul-Christian and Paananen, Topi and Gelman, Andrew},
	year = {2023},
}

@article{aitchison_logistic-normal_1980,
	title = {Logistic-{Normal} {Distributions}: {Some} {Properties} and {Uses}},
	volume = {67},
	issn = {00063444},
	url = {http://www.jstor.org/stable/2335470},
	number = {2},
	urldate = {2022-07-28},
	journal = {Biometrika},
	author = {Aitchison, J. and Shen, S. M.},
	year = {1980},
	pages = {261--272},
}

@article{pas_horseshoe_2014,
	title = {The horseshoe estimator: {Posterior} concentration around nearly black vectors},
	volume = {8},
	url = {https://doi.org/10.1214/14-EJS962},
	doi = {10.1214/14-EJS962},
	number = {2},
	journal = {Electronic Journal of Statistics},
	author = {Pas, S. L. van der and Kleijn, B. J. K. and Vaart, A. W. van der},
	year = {2014},
	pages = {2585 -- 2618},
}

@incollection{pas_theoretical_2021,
	title = {Theoretical guarantees for the horseshoe and other global-local shrinkage priors},
	booktitle = {Handbook of {Bayesian} {Variable} {Selection}},
	publisher = {Chapman and Hall/CRC},
	author = {Pas, S. L. van der},
	year = {2021},
	pages = {133--160},
}

@article{west_scale_1987,
	title = {On scale mixtures of normal distributions},
	volume = {74},
	issn = {0006-3444},
	url = {https://doi.org/10.1093/biomet/74.3.646},
	doi = {10.1093/biomet/74.3.646},
	number = {3},
	journal = {Biometrika},
	author = {West, Mike},
	month = sep,
	year = {1987},
	pages = {646--648},
}

@article{boss_group_2023,
	title = {Group {Inverse}-{Gamma} {Gamma} {Shrinkage} for {Sparse} {Linear} {Models} with {Block}-{Correlated} {Regressors}},
	url = {https://doi.org/10.1214/23-BA1371},
	doi = {10.1214/23-BA1371},
	journal = {Bayesian Analysis},
	author = {Boss, Jonathan and Datta, Jyotishka and Wang, Xin and Park, Sung Kyun and Kang, Jian and Mukherjee, Bhramar},
	year = {2023},
	pages = {1 -- 30},
}

@article{ghosal_convergence_2000,
	title = {Convergence rates of posterior distributions},
	volume = {28},
	url = {https://doi.org/10.1214/aos/1016218228},
	doi = {10.1214/aos/1016218228},
	number = {2},
	journal = {The Annals of Statistics},
	author = {Ghosal, Subhashis and Ghosh, Jayanta K. and van der Vaart, Aad W.},
	year = {2000},
	pages = {500 -- 531},
}

@article{aguilar_generalized_2025,
	title = {Generalized {Decomposition} {Priors} on {R2}},
	url = {https://doi.org/10.1214/25-BA1524},
	doi = {10.1214/25-BA1524},
	journal = {Bayesian Analysis},
	author = {Aguilar, Javier Enrique and Bürkner, Paul-Christian},
	year = {2025},
	pages = {1 -- 34},
}

@article{pas_adaptive_2017,
	title = {Adaptive posterior contraction rates for the horseshoe},
	volume = {11},
	issn = {1935-7524},
	url = {https://projecteuclid.org/journals/electronic-journal-of-statistics/volume-11/issue-2/Adaptive-posterior-contraction-rates-for-the-horseshoe/10.1214/17-EJS1316.full},
	doi = {10.1214/17-EJS1316},
	language = {en},
	number = {2},
	urldate = {2026-01-29},
	journal = {Electronic Journal of Statistics},
	author = {Pas, S. L. van der and Szabó, Botond and Vaart, A. W. van der},
	month = jan,
	year = {2017},
}

@article{fuglstad_intuitive_2020,
	title = {Intuitive {Joint} {Priors} for {Variance} {Parameters}},
	volume = {15},
	url = {https://doi.org/10.1214/19-BA1185},
	doi = {10.1214/19-BA1185},
	number = {4},
	journal = {Bayesian Analysis},
	publisher = {International Society for Bayesian Analysis},
	author = {Fuglstad, Geir-Arne and Hem, Ingeborg Gullikstad and Knight, Alexander and Rue, Håvard and Riebler, Andrea},
	year = {2020},
	pages = {1109 -- 1137},
}

@article{manderson_translating_2026,
	title = {Translating predictive distributions into informative priors},
	volume = {0},
	url = {https://doi.org/10.1080/01621459.2026.2614034},
	doi = {10.1080/01621459.2026.2614034},
	number = {ja},
	journal = {Journal of the American Statistical Association},
	publisher = {Taylor \& Francis},
	author = {Manderson, Andrew A. and Goudie, Robert J. B.},
	year = {2026},
	note = {\_eprint: https://doi.org/10.1080/01621459.2026.2614034},
	pages = {1--18},
}

@article{linero_nonparametric_2024,
	title = {In {Nonparametric} and {High}-{Dimensional} {Models}, {Bayesian} {Ignorability} is an {Informative} {Prior}},
	volume = {119},
	issn = {0162-1459, 1537-274X},
	doi = {10.1080/01621459.2023.2278202},
	language = {en},
	number = {548},
	urldate = {2025-11-13},
	journal = {Journal of the American Statistical Association},
	author = {Linero, Antonio R.},
	month = oct,
	year = {2024},
	pages = {2785--2798},
}

@article{aguilar_r2_2025,
	title = {R2 priors for {Grouped} {Variance} {Decomposition} in {High}-dimensional {Regression}},
	journal = {arXiv preprint arXiv:2507.11833},
	author = {Aguilar, Javier Enrique and Kohns, David and Vehtari, Aki and Bürkner, Paul-Christian},
	year = {2025},
}

@article{kohns_arr2_2025,
	title = {The {ARR2} prior: flexible predictive prior definition for {Bayesian} auto-regressions},
	volume = {21},
	number = {3},
	journal = {Bayesian Analysis},
	publisher = {International Society for Bayesian Analysis},
	author = {Kohns, David and Kallioinen, Noa and McLatchie, Yann and Vehtari, Aki},
	year = {2026},
	pages = {1131--1162},
}

@misc{aguilar2025dependencyawareshrinkagepriorshigh,
      title={Dependency-Aware Shrinkage Priors for High Dimensional Regression}, 
      author={Javier Enrique Aguilar and Paul-Christian Bürkner},
      year={2025},
      eprint={2505.10715},
      archivePrefix={arXiv},
      primaryClass={stat.ME},
      url={https://arxiv.org/abs/2505.10715}, 
}

@article{andrews_scale_1974,
  author  = {Andrews, D. F. and Mallows, C. L.},
  title   = {Scale Mixtures of Normal Distributions},
  journal = {Journal of the Royal Statistical Society:
             Series B (Methodological)},
  year    = {1974},
  volume  = {36},
  number  = {1},
  pages   = {99--102},
  doi     = {10.1111/j.2517-6161.1974.tb00989.x}
}

@article{tcga2017pdac,
  author    = {{Cancer Genome Atlas Research Network}},
  title     = {Integrated Genomic Characterization of Pancreatic Ductal Adenocarcinoma},
  journal   = {Cancer Cell},
  volume    = {32},
  number    = {2},
  pages     = {185--203.e13},
  year      = {2017},
  doi       = {10.1016/j.ccell.2017.07.007},
  publisher = {Elsevier}
}

@article{kuenzel2019meta,
   title={Metalearners for estimating heterogeneous treatment effects using machine learning},
   volume={116},
   ISSN={1091-6490},
   url={http://dx.doi.org/10.1073/pnas.1804597116},
   DOI={10.1073/pnas.1804597116},
   number={10},
   journal={Proceedings of the National Academy of Sciences},
   publisher={National Academy of Sciences},
   author={Künzel, Sören R. and Sekhon, Jasjeet S. and Bickel, Peter J. and Yu, Bin},
   year={2019},
   month=Feb, pages={4156–4165} }

@misc{klippert2026guide,
      title={A Guide to Estimating Conditional Average Treatment Effects in Competing Risks Settings}, 
      author={Daniel Klippert and Sarah Friedrich and Markus Pauly},
      year={2026},
      eprint={2606.18281},
      archivePrefix={arXiv},
      primaryClass={stat.AP},
      url={https://arxiv.org/abs/2606.18281}, 
      note={arXiv:2606.18281},
}

@book{kalbfleisch2002statistical,
  title={The Statistical Analysis of Failure Time Data},
  author={Kalbfleisch, John D. and Prentice, Ross L.},
  volume={460},
  edition={2nd},
  year={2002},
  publisher={John Wiley \& Sons},
  address={Hoboken, NJ},
  isbn={978-0-471-36357-6}
}

@article{alberto2022ITEnonparam,
  title={Estimating individual treatment effects using non-parametric regression models: A review},
  author={Caron, Alberto and Baio, Gianluca and Manolopoulou, Ioanna},
  journal={Journal of the Royal Statistical Society Series A: Statistics in Society},
  volume={185},
  number={3},
  pages={1115--1149},
  year={2022},
  publisher={Oxford University Press},
  doi={10.1111/rssa.12824}
}

@incollection{xu2023handbook,
  title={Treatment Heterogeneity with Survival Outcomes},
  author={Xu, Yizhe and Ignatiadis, Nikolaos and Sverdrup, Erik and Fleming, Scott and Wager, Stefan and Shah, Nigam},
  booktitle={Handbook of Matching and Weighting Adjustments for Causal Inference},
  edition={1st},
  pages={38},
  year={2023},
  publisher={Chapman and Hall/CRC},
  doi={10.1201/9781003102670-21}
}

@inproceedings{grinsztajn2022tree,
  title={Why do tree-based models still outperform deep learning on typical tabular data?},
  author={Grinsztajn, L{\'e}o and Oyallon, Edouard and Varoquaux, Ga{\"e}l},
  booktitle={Advances in Neural Information Processing Systems},
  volume={35},
  pages={507--520},
  year={2022},
  url={https://proceedings.neurips.cc/paper_files/paper/2022/hash/0378c7692da36807bdec87ab043cdadc-Abstract-Datasets_and_Benchmarks.html}
}

@article{friedman2010glmnet,
  author  = {Friedman, Jerome and Hastie, Trevor and Tibshirani, Robert},
  title   = {Regularization Paths for Generalized Linear Models via Coordinate Descent},
  journal = {Journal of Statistical Software},
  volume  = {33},
  number  = {1},
  pages   = {1--22},
  year    = {2010},
  doi     = {10.18637/jss.v033.i01}
}

@article{eddelbuettel2011rcpp,
  author  = {Eddelbuettel, Dirk and Fran{\c{c}}ois, Romain},
  title   = {{Rcpp}: Seamless {R} and {C++} Integration},
  journal = {Journal of Statistical Software},
  volume  = {40},
  number  = {8},
  pages   = {1--18},
  year    = {2011},
  doi     = {10.18637/jss.v040.i08}
}

@manual{rcoreteam2026,
  author       = {{R Core Team}},
  title        = {R: A Language and Environment for Statistical Computing},
  organization = {R Foundation for Statistical Computing},
  address      = {Vienna, Austria},
  year         = {2026},
  url          = {https://www.R-project.org/}
}

@manual{therneau2024survival,
  author = {Therneau, Terry M.},
  title  = {A Package for Survival Analysis in {R}},
  year   = {2024},
  note   = {R package version 3.8-12},
  url    = {https://CRAN.R-project.org/package=survival}
}

@article{torrehealy2023pdacr,
  author  = {Torre-Healy, Luke A. and Kawalerski, Ryan R. and Oh, Ki and Chrastecka, Lucie and Peng, Xianlu L. and Aguirre, Andrew J. and Rashid, Naim U. and Yeh, Jen Jen and Moffitt, Richard A.},
  title   = {Open-Source Curation of a Pancreatic Ductal Adenocarcinoma Gene Expression Analysis Platform ({pdacR}) Supports a Two-Subtype Model},
  journal = {Communications Biology},
  volume  = {6},
  pages   = {163},
  year    = {2023},
  doi     = {10.1038/s42003-023-04461-6}
}

@article{vehtari2024pareto,
  title   = {Pareto Smoothed Importance Sampling},
  author  = {Vehtari, Aki and Simpson, Daniel and Gelman, Andrew and Yao, Yuling and Gabry, Jonah},
  journal = {Journal of Machine Learning Research},
  volume  = {25},
  number  = {72},
  pages   = {1--58},
  year    = {2024}
}

@misc{nci2014tcgapolicies,
  title        = {The Cancer Genome Atlas Program: Human Subjects Protection and Data Access Policies},
  author       = {{National Cancer Institute}},
  year         = {2014},
  note         = {Revision of 16 January 2014},
  howpublished = {\url{https://www.cancer.gov/about-nci/organization/ccg/research/structural-genomics/tcga/history/policies/tcga-human-subjects-data-policies.pdf}}
}

@article{neal2003slice,
  title   = {Slice Sampling},
  author  = {Neal, Radford M.},
  journal = {The Annals of Statistics},
  volume  = {31},
  number  = {3},
  pages   = {705--767},
  year    = {2003}
}

@article{robert1995simulation,
  title   = {Simulation of Truncated Normal Variables},
  author  = {Robert, Christian P.},
  journal = {Statistics and Computing},
  volume  = {5},
  number  = {2},
  pages   = {121--125},
  year    = {1995}
}

@article{angrist1996identification,
author = {Joshua D. Angrist and Guido W. Imbens and Donald B. Rubin},
title = {Identification of Causal Effects Using Instrumental Variables},
journal = {Journal of the American Statistical Association},
volume = {91},
number = {434},
pages = {444--455},
year = {1996},
publisher = {Taylor \& Francis},
doi = {10.1080/01621459.1996.10476902},
URL = { 
https://www.tandfonline.com/doi/abs/10.1080/01621459.1996.10476902
},
eprint = {     https://www.tandfonline.com/doi/pdf/10.1080/01621459.1996.10476902
}
}

@article{robinson1988root,
 ISSN = {00129682, 14680262},
 URL = {http://www.jstor.org/stable/1912705},
 author = {P. M. Robinson},
 journal = {Econometrica},
 number = {4},
 pages = {931--954},
 publisher = {[Wiley, Econometric Society]},
 title = {Root-N-Consistent Semiparametric Regression},
 urldate = {2026-10-19},
 volume = {56},
 year = {1988}
}

@article{agarwal_bayesian_2025,
	title = {Bayesian {Variable} {Selection} {Under} {High}-{Dimensional} {Settings} with {Grouped} {Covariates}},
	url = {https://doi.org/10.1214/25-BA1525},
	doi = {10.1214/25-BA1525},
	journal = {Bayesian Analysis},
	author = {Agarwal, Sumanta and Dutta, Somak and Mukhopadhyay, Minerva},
	year = {2025},
}

@misc{graspgroupedregressionadaptive2025,
      title={GRASP: Grouped Regression with Adaptive Shrinkage Priors}, 
      author={Shu Yu Tew and Daniel F. Schmidt and Mario Boley},
      year={2025},
      eprint={2506.18092},
      archivePrefix={arXiv},
      primaryClass={stat.ME},
      url={https://arxiv.org/abs/2506.18092}, 
}

\clearpage
\appendix
\normalsize
\etocdepthtag.toc{appendixpart}
\etocsettagdepth{mainpart}{none}
\etocsettagdepth{appendixpart}{subsection}
\etocsettocstyle{\section*{Appendix}}{}
\tableofcontents
\clearpage

\appendix
\label{sec:appendix}
\thispagestyle{empty}

\section{Bayesian Causal Forest recap}
\label{sec:bcf-recap}

This section writes down the BART and BCF specification that the rest of the paper starts from. The tree prior, the prognostic forest, and the backfitting update are the usual ones. What changes, in Section~\ref{sec:dirichlet-tree}, is the prior on the leaf values of the treatment trees.

\subsection{BART}
\label{app:bart}

Bayesian additive regression trees (BART) model a continuous outcome by a sum of regression trees \citep{chipman2010bart}. For observation $i$,
\begin{equation}
y_i = f(\bx_i) + \varepsilon_i,
\qquad
f(\bx_i) = \sum_{j=1}^{m} g(\bx_i; T_j, \bh_j),
\qquad
\varepsilon_i \sim \Normal(0,\sigma^2).
\label{eq:appendix-bart}
\end{equation}
The tree $T_j$ is a binary partition of the covariate space, and $\bh_j=(h_{j\ell})_{\ell\in\calL(T_j)}$ holds the value in each leaf. Let $\bz_j(\bx)$ indicate the leaf that contains $\bx$. The contribution of the tree is then the piecewise constant function $g(\bx;T_j,\bh_j)=\bz_j(\bx)^\top\bh_j$. Regularizing the depth and the leaf values prevents overfitting. The sum allows the forest to model nonlinearities.

We put a priors on how a tree is grown and on the leaf values. Following \citet{chipman2010bart}, a node at depth $d$, counting the root as depth zero, splits with probability
\begin{equation}
p_{\mathrm{split}}(d)=\nu(1+d)^{-\eta},
\qquad 0<\nu<1,\qquad \eta\ge 0.
\label{eq:appendix-bart-split}
\end{equation}
The parameters are set as $\nu=0.95$ and $\eta=2$ as defaults. This equals a splitting probability of $0.95$ at the root and $0.95/4$ at the first split, so the prior prefers shallow trees and places less mass on deep partitions. When a split is drawn, one variable is drawn uniformly from the covariates, and the cut-point is drawn uniformly from the cut-points that the observations reaching that node still permit.

Conditional on the tree structure, the leaf values have independent conjugate normal priors,
\begin{equation}
h_{j\ell}\mid T_j,\sigma_h^2 \sim \Normal(0,\sigma_h^2),
\qquad \ell\in\calL(T_j).
\label{eq:appendix-bart-leaves}
\end{equation}
The zero mean pulls each trees contribution to zero. The scale is set from the number of trees. At a fixed $\bx$, each tree contributes exactly one leaf, so the prior variance of $f(\bx)$ given the partitions equals $m\sigma_h^2$. Taking $\sigma_h=\sigma_0/\sqrt{m}$ makes that variance $\sigma_0^2$ however many trees are used. \citet{chipman2010bart} first shift and rescale the observed $y_i$ into $[-0.5,0.5]$, and they set $\sigma_0=0.5/k$ with $k=2$. Then $f(\bx)$ lands outside the range of the rescaled outcome with small prior probability.

Each leaf is given its own variance $\sigma_h^2$ , and the sum of the leaf-parameter variances in the tree rises by $\sigma_h^2$. The prior on the tree depths means that the number of leafs, and the variance remains limited. Thus, the normal prior regularizes the leaf given the tree structure.

The noise variance has a conjugate inverse-gamma prior, $\sigma^2\sim\operatorname{InvGamma}(a_\sigma,b_\sigma)$, with density proportional to $(\sigma^2)^{-a_\sigma-1}\exp(-b_\sigma/\sigma^2)$. The scaled inverse-$\chi^2$ prior of \citet{chipman2010bart} is the special case $a_\sigma=\nu/2$ and $b_\sigma=\nu\lambda/2$. They take $\nu=3$ and choose $\lambda$ so that the $0.9$ quantile of the prior on $\sigma$ equals the residual standard deviation from a linear regression of the rescaled outcome.

\subsubsection{Sampling in BART}
\label{app:bart-sampling}

We draw the posterior by Bayesian backfitting \citep{hastie_tibshirani_2000bayesian,chipman2010bart}. One tree is updated at a time, on the part of the outcome the other trees do not already explain, and $\sigma^2$ is drawn after the trees. In one iteration the steps are the following.

\begin{enumerate}
    \item \textbf{Partial residual.} To update tree $j$, remove the other trees from the outcome,
    \begin{equation}
    R_{ij} = y_i - \sum_{k\neq j} g(\bx_i;T_k,\bh_k).
    \label{eq:appendix-bart-residual}
    \end{equation}
    A tree that has been updated in at this iteration is subtracted from the outcome. 

    \item \textbf{Tree structure.} Conditional on the other trees, and residual $R_{ij}$ updating tree $j$ simplifies to updating one regression tree. As \eqref{eq:appendix-bart-leaves} is conjugate normal, the leaf values integrate out of the likelihood of a proposed partition, and the Metropolis--Hastings step only has to compare partitions. The candidate is built with one of four moves \citep{chipman2010bart,pratola2016efficient}. The grow move picks a leaf and splits it. The variable and the cut-point are drawn uniformly from those that the observations in the leaf still allow. The prune move picks an internal node whose two children are both leaves and collapses them back into that node. The change replaces the splitting rule at an internal node. The swap move exchanges the rule at a node with the rule at one of its children. To ensure sparsity, the splits that would put fewer than five observations in a child are discarded. The move is accepted or rejected with the Metropolis--Hastings probability given by the marginal likelihood, the tree prior \eqref{eq:appendix-bart-split}, and the probability of proposing the move in each direction.

    \item \textbf{Leaf values.} Once a move has been selected, the leaves are drawn independently conditional on the tree structure. Let $\mathcal N_{j\ell}$ be the observations that fall in leaf $\ell$, let $n_{j\ell}=|\mathcal N_{j\ell}|$, and let $S_{j\ell}$ be the sum of $R_{ij}$ over those observations. Then
    \begin{equation}
    v_{j\ell}=\left(\frac{n_{j\ell}}{\sigma^2}+\frac{1}{\sigma_h^2}\right)^{-1},
    \qquad
    h_{j\ell}\mid -\sim\Normal\left(\frac{v_{j\ell}S_{j\ell}}{\sigma^2},\, v_{j\ell}\right).
    \label{eq:appendix-bart-leaf-posterior}
    \end{equation}
    The draw is pulled toward the residual mean in the leaf and toward zero. The two precisions are $n_{j\ell}/\sigma^2$ and $1/\sigma_h^2$. A leaf with either few observations, or a large $\sigma^2$, is the one that stays near zero.

    \item \textbf{Residual variance.} Once every tree has been updated, draw
    \begin{equation}
    \sigma^2\mid -\sim\operatorname{InvGamma}\!\left(
    a_\sigma+\frac{n}{2},\;
    b_\sigma+\frac12\sum_{i=1}^{n}\{y_i-f(\bx_i)\}^2
    \right),
    \label{eq:appendix-bart-noise-posterior}
    \end{equation}
    using the same parameterization as above.
\end{enumerate}

Draws of $f(\bx)$ from this sampler are posterior draws of the conditional mean, and the credible intervals are quantiles of those draws.

\subsection{Bayesian Causal Forests (BCF)}
\label{app:bcf}

Bayesian causal forests do not run a single BART of $y$ on $(\bx,A)$ \citep{hahn2020bcf}. Prognosis and treatment effect are given two forests, with different priors. The outcome model \eqref{eq:bcf-model} is
\begin{equation}
y_i=\mu(\bx_i,\widehat e_i)+A_i\,\tau(\bx_i)+\varepsilon_i,
\qquad \varepsilon_i\sim\Normal(0,\sigma^2),
\label{eq:appendix-bcf}
\end{equation}
where $A_i\in\{0,1\}$ and $\widehat e_i$ estimates the propensity score $e(\bx)=\Pr(A_i=1\mid\bx_i=\bx)$. Under the assumptions in Section~\ref{sec:bcf}, $\tau(\bx)$ equals the CATE. As in the main text we write
\begin{align}
\mu(\bx,\widehat e)
&= \sum_{j=1}^{m_\mu} g\bigl((\bx,\widehat e);T_j^{\mu},\bh_j^{\mu}\bigr),
\label{eq:appendix-bcf-mu}\\
\tau(\bx)
&= \alpha+\sum_{j=1}^{m_\tau} g(\bx;T_j^{\tau},\bh_j^{\tau})
= \alpha+g_\tau(\bx).
\label{eq:appendix-bcf-tau}
\end{align}

The two functions do not receive the same regularization. The prognostic function has to follow the baseline outcome, including non-linearities and interactions, so it gets a flexible BART prior. Treatment signal is often less pronounced than the prognostic signal. BCF therefore regularize $\tau$ more strongly than $\mu$. If the treatment trees never split, $g_\tau(\bx)$ is constant in $\bx$ and the CATE does not vary. In our parameterization that constant can also sit in $\alpha$, with $g_\tau$ shrunk toward zero.

For $\mu$, \cite{hahn2020bcf} use $m_\mu=200$ trees and splitting probability prior $0.95(1+d)^{-2}$. The leaf standard deviation is $\sigma_h=\sigma_0/\sqrt{m_\mu}$, so $\sigma_0$ is the prior standard deviation of $\mu(\bx)$ at a given $\bx$. They put a half-Cauchy prior on $\sigma_0$ with median twice the marginal standard deviation of $y$. For $\tau$ they use $m_\tau=50$ trees and splitting probability $0.25(1+d)^{-3}$, hence $0.25$ at the root, and a half-normal prior on the analogous standard deviation of $\tau(\bx)$, with median equal to the marginal standard deviation of $y$. Conditional on the scale, the leaf values remain independent $\Normal(0,\sigma_h^2)$ draws. A new leaf gets its own $\sigma_h^2$. It does not receive a share of a total that is held fixed when the tree grows. That is the leaf prior replaced in Section~\ref{sec:dirichlet-tree}.

Coding $A$ as $0/1$ or as $\pm 1/2$ changes how the prior variance is divided between the two conditional outcome means, and the BCF software estimates this coding \citep{hahn2020bcf}. Equation~\eqref{eq:bcf-model} keeps the $0/1$ coding fixed.

The propensity score is in $\mu$ to mitigate regularization-induced confounding \citep{hahn2018regularization,hahn2020bcf}. When the prior on $\mu$ shrinks too hard, some of the covariate--outcome association is left over, and the treatment contrast can absorb it. This is easiest to see under targeted selection, where the probability of treatment depends on the expected outcome under control, and $\mu(\bx)$ and $e(\bx)$ then describe much of the same variation. Approximating that surface with axis-aligned splits takes a deep tree, which \eqref{eq:appendix-bart-split} penalizes, while a split on $A$ is penalized less. Supplying $\widehat e$ gives the prognostic trees a single covariate that tracks this dependence. We estimate it once and keep it fixed while the outcome model is updated. \citet{hahn2020bcf} treat the estimated score as a covariate transformation, not as an extra unknown with its own posterior. A related point is made by \citet{linero_nonparametric_2024}: taking the outcome regression and the treatment assignment mechanism to be a priori independent is already an informative prior, and one way to break this independence is to let the propensity score enter the mean function.

Sampling alternates between the two forests on partial residuals 
using Bayesian backfitting \citep{hastie_tibshirani_2000bayesian}. With the latest draw of the other function we set
\begin{align}
R_i^{\mu} &= y_i-A_i\tau(\bx_i),
\label{eq:appendix-bcf-resid-mu}\\
R_i^{\tau} &= y_i-\mu(\bx_i,\widehat e_i),
\label{eq:appendix-bcf-resid-tau}
\end{align}
and we update the prognostic trees against $R_i^{\mu}$. The treatment update uses the conditional model $R_i^{\tau}=A_i\tau(\bx_i)+\varepsilon_i$. Under the fixed $0/1$ coding, a control observation has conditional mean $\mu(\bx_i,\widehat e_i)$ whatever $\tau$ is, and the treatment trees are updated from the treated observations. The prognostic scale and the treatment scale are then drawn from their full conditionals under the half-Cauchy and half-normal priors, given the current leaf values. The noise variance is drawn from its inverse-gamma full conditional given the residuals $y_i-\mu(\bx_i,\widehat e_i)-A_i\tau(\bx_i)$.
The stored draws of $\tau(\bx_i)$ are the posterior of the CATE. Appendix~\ref{app:sampler} gives the sampler for our model.

Other BCF variants change a different part of this model and still take the leaf values as independent given their scales. Targeted smooth BCF makes the treatment effect smooth along one chosen covariate \citep{starling2021targeted}. Shrinkage BCF learns the splitting probabilities rather than fixing \eqref{eq:appendix-bart-split} \citep{caron2022shrinkage,linero2018dart}. Horseshoe forests put a separate half-Cauchy scale on each leaf, in a survival model \citep{jacobs2026horseshoe}.  We use a BCF-type prior for $\mu$, with a half-Cauchy scale on the prognostic forest (Appendix~\ref{app:exp-setup}). On the treatment leaves we use the joint Dirichlet-Tree allocation of Section~\ref{sec:dirichlet-tree} instead of independent normal variances.

\clearpage
\section{Meta-Learners for Heterogeneous Treatment Effects}
\label{app:meta-learners}
Meta-learners are estimation strategies for heterogeneous treatment effects that utilize arbitrary supervised learning methods as so-called base learners. Prominent examples include the S-, T-, X-, R and DR-learner~\citep{kuenzel2019meta,nie2021quasi,kennedy2023drlearner}. The S- and T-learners constitute the most basic approaches. Under the identifiability assumptions (consistency, no interference, conditional exchangeability, positivity), the CATE can be written as \begin{align*}
    \tau(\bx) &  = \E\{y_i(1)-y_i(0)\mid\bx_i=\bx\} \\ 
  &  = \E\{y_i(a_i)\mid\bx_i=\bx, a_i=1\} - \E\{y_i(a_i)\mid\bx_i=\bx, a_i=0\}
\end{align*} The S-learner fits a single regression model $\widehat{m}$ to \begin{align*}
    m(\bx_i, a_i) = \E\{y_i(a_i)\mid\bx_i=\bx, a_i=a_i\}
\end{align*} and predicts the CATE as $\widehat{m}(\bx, 1)-\widehat{m}(\bx, 0)$. A limitation of this approach is that the treatment is treated as a regular covariate that can be overlooked by the model. 

In comparison, the T-learner improves upon this issue by fitting two separate regression models to \begin{align*}
    m_a(\bx_i) = \E\{y_i(a_i)\mid\bx_i=\bx, a_i=a\}, \quad a \in \{0,1\},
\end{align*} using only observations from the respective treatment arm. The T-learner estimates the CATE by taking the difference $\widehat{m}_1(\bx)- \widehat{m}_0(\bx)$. This approach has the disadvantage that it is susceptible to regularization bias~\citep{kuenzel2019meta}, particularly when the true CATE is close to zero. 

Since the true CATE is unobserved in real-data settings, simulation studies are a valuable tool to assist practitioners with choosing a suitable meta-learner. Based on previous simulation studies~\citep{alberto2022ITEnonparam,xu2023handbook,klippert2026guide}, we consider the X- and R-learners as strong choices and therefore include them as baselines within our experiments. For the nuisance functions learned within these algorithms, including the propensity score, we use XGBoost~\citep{chen2016xgboost} as the base learner which is reliable default choice on tabular data~\citep{grinsztajn2022tree}.

\subsection{X-Learner}
The X-learner~\citep{kuenzel2019meta} is a meta-learner that estimates the CATE by computing a weighted average of imputed treatment effects that requires an estimate of the propensity score. Analogously to the T-learner, the X-learner begins by learning a regression model for $m_1$ and $m_0$ stratified by treatment. Then, for each observation $i$, the imputed treatment effect constructed as  \begin{align}
    \widetilde{d}_i(1) &= y_i(1) - \widehat{m}_0(\bx_i) \quad \text{ if } a_i=1 \text{, or} \\
    \widetilde{d}_i(0) &= \widehat{m}_1(\bx_i) - y_i(0) \quad \text{ if } a_i=0.
\end{align}
Under $\widehat{m}_a = m_a$, we have that $\tau(\bx) = \E[\widetilde{d}_i(a)|\bx_i=\bx] =: \tau_a(\bx)$ for $a\in\{0,1\}$. In the second step, the X-learner fits regression models to $\tau_1(\bx)$ and $\tau_0(\bx)$ using the data from the respective treatment arm. We obtain the final CATE estimate as the propensity-score-weighted average of the two CATE models $$\widehat{\tau}(\bx) = \widehat{e}(\bx)\tau_0(\bx) + (1-\widehat{e}(\bx))\tau_1(\bx).$$ 
This construction is especially beneficial under unbalanced treatment designs. Without loss of generality, assume that group 0 is substantially smaller than group 1. Then, $\widehat{m}_1$ can be estimated accurately using the large treatment arm 1. Hence, the resulting imputed treatment effects $\widetilde{d}_i(0)$ in the small group can be considered highly informative. In the final estimate, the corresponding CATE estimate $\tau_0(\bx)$ receives more weight than the model $\tau_1(\bx)$ based on the less informative imputed treatment effects, given that treatment 1 is generally more common than treatment 0. In summary, the X-learner allows to construct accurate CATE estimates in unbalanced settings by efficiently using information from the larger treatment group. In comparison, small treatment groups can directly harm the T-learner as it requires reliable estimates of both $m_1$ and $m_0$ to obtain an accurate CATE model.

\subsection{R-Learner}
The R-learner~\citep{nie2021quasi} is a meta-learner that estimates the CATE by minimizing a loss function using residuals of the treatment $a$ and the conditional mean outcome $m^{*}(\bx) = \E[y_i|\bx_i = \bx]$. Specifically, the CATE function can be characterized by
\begin{align*}
    \tau(\cdot) & =  \text{argmin}_{\tilde{\tau}(\cdot)} \, \E \left[\left((y_i-m^*(\bx_i))-(a_i-e(\bx_i))\tilde{\tau}(\bx_i)\right)^2 \right]\\
& = \text{argmin}_{\tilde{\tau}(\cdot)}  \E \left[(a_i-e(\bx_i))^2\left(\dfrac{y_i-m^*(\bx_i)}{a_i-e(\bx_i)}-\tilde{\tau}(\bx_i)\right)^2 \right]. 
\end{align*}
Since $m^*$ and $e$ are generally unknown, these need to be replaced by estimates to obtain the CATE based on the above loss-based representation. 

The R-learner algorithm starts by dividing the training data into $q$ folds to obtain cross-fitted estimates of $\widehat{m^*}$ and $\widehat{e}$ for each observation. More precisely, for observation $i$, the estimate is obtained from a model fitted on the $q-1$ folds that do not contain observation $i$.
Let $\widehat{m^*}^{-q(i)}(\bx_i)$ and $\widehat{e}^{-q(i)}(\bx_i)$ denote the corresponding out-of-fold predictions. 
Using these estimates, the pseudo-outcomes $$h^{*}_i = \dfrac{y_i-\widehat{m^*}^{-q(i)}(\bx_i)}{a_i-\widehat{e}^{-q(i)}(\bx_i)}$$ are constructed. 
Then, the R-learner fits a weighted regression model of $h^{*}_i$ on $\bx$, using $(a_i-e^{-q(i)}(\bx_i))^2$ as regression weights, to obtain the CATE estimate $\widehat{\tau}(\bx)$.

An advantageous property of the R-learner is its quasi-oracle property. Specifically, \citet{nie2021quasi} show that, when penalized kernel regression is used to obtain the regression model for $\widehat{\tau}(\bx)$, the asymptotic error bounds are unaffected by replacing the ground-truth functions $m^*$ and $e$ by their estimated versions.

\section{Accelerated Failure Time Models}
\label{app:aft}
Under the accelerated failure time (AFT) model~\citep{kalbfleisch2002statistical}, the logarithm of the survival time is linked to covariates through an additive regression model. Under the CATE formulation in Eq.~\ref{eq:cate} used by the Bayesian causal forest~\citep{hahn2020bcf}, this can be written as
\begin{align*}
    \log(T(a)) = \mu(\bx,\widehat{e}) + a \tau(\bx) + \varepsilon,
\end{align*}
where $\varepsilon \thicksim N(0,\sigma^2)$ and $\tau(\bx)$ is the CATE on the log-survival-time scale. In our experiments in Section~\ref{sec:experiments}, we consider the right-censored survival data setting, where the survival time is observed only through $\min\{T,C\}$, with $C$ denoting the censoring time.
In addition to the causal identification assumptions (SUTVA and strong ignorability), identification from right-censored data requires a conditional non-informative censoring assumption. Specifically, we assume that $C(a)$ and $T(a)$ are independent given treatment $a\in{0,1}$ and covariates $\bx$, such that the distribution of $T(a)$ can be recovered based on right-censored data~\citep{jacobs2026horseshoe}.

\clearpage
\section{Posterior Computation}
\label{app:sampler}

We sample from the posterior by Bayesian backfitting \citep{chipman2010bart,hahn2020bcf}. The trees are updated one at a time, each conditional on its partial residuals. A structural proposal integrates out the leaf values. Conditional on the resulting tree, these values are drawn from their normal full conditionals. For a treatment tree, the branch fractions and tree scale are updated by one of two procedures. The collapsed procedure uses slice sampling with the leaf values integrated out and applies for all $a$ and $\rho$. When the tree-scale prior is coupled to the allocation, the blocked procedure jointly updates the tree scale and all terminal shares.

Sections~\ref{app:sampler-iteration} to~\ref{app:sampler-output} describe the sampler for the model of Section~\ref{sec:model} under the settings of Appendix~\ref{app:exp-setup}. Section~\ref{app:sampler-censoring} gives the additional update required for the accelerated failure time model with censored outcomes. The implementation uses R, with the leaf likelihoods and branch-fraction updates compiled in C++.

\subsection{One Iteration}
\label{app:sampler-iteration}

One iteration of the sampler performs the following steps in order, conditional on the most recent values of all remaining quantities.

\begin{enumerate}
\item \emph{Prognostic forest.} Each prognostic tree is updated by one structural move followed by a draw of its leaf values. For unit $i$, the partial residual is $y_i-A_i\{\alpha+g_\tau(\bx_i)\}$ minus the contribution of the other prognostic trees. All $n$ units contribute to this update. After the last tree, the forest scale $\tsc_\mu^2$ is drawn. With DART, the split probabilities of the prognostic forest are drawn next.
\item \emph{Intercept.} The intercept $\alpha$ is drawn from its normal full conditional.
\item \emph{Treatment forest.} Each treatment tree $j$ is updated by one structural move. Its branch fractions, tree scale $\tsc_j^2$ and leaf values are then updated in the order specified by the procedure in Appendix~\ref{app:sampler-scales}. For a treated unit $i$, the partial residual is $R_{ij}=y_i-\mu(\bx_i,\widehat e_i)-\alpha-\sum_{j'\neq j}g(\bx_i;T_{j'},\bh_{j'})$. Only treated units contribute to this update. With DART, the split probabilities of the treatment forest are drawn after the last tree.
\item \emph{Residual variance.} The variance $\sigma^2$ is drawn from its inverse-gamma full conditional.
\end{enumerate}

Under the $0/1$ coding of \eqref{eq:bcf-model}, the likelihood contribution of a control unit does not depend on $g_\tau$. The treatment trees are therefore fitted to the treated units, and their splitting rules are formed from the covariates of those units. The treatment effect $\tau(\bx_i)=\alpha+g_\tau(\bx_i)$ is nevertheless evaluated at every unit, including the control units.

\subsection{Leaf Values}
\label{app:sampler-leaf}

Consider one tree with partial residuals $R_i$. Let $b_\ell$ denote the prior variance of its leaf value $h_\ell$ divided by $\sigma^2$. For a treatment leaf, $b_\ell=\omega\tsc_j^2\phi_{j\ell}$, so its value depends on the terminal share $\phi_{j\ell}$. For a prognostic leaf, $b_\ell=\omega_\mu\tsc_\mu^2$ is common to all leaves of the forest. Let $n_\ell$ be the number of units in leaf $\ell$ and $S_\ell$ the sum of their partial residuals. Integrating out $h_\ell$ gives the leaf factor
\begin{equation}
\log M_\ell(b_\ell)=-\frac12\log(1+n_\ell b_\ell)
+\frac{S_\ell^2\,b_\ell}{2\sigma^2(1+n_\ell b_\ell)},
\label{eq:app-leaf-factor}
\end{equation}
up to terms that do not depend on the partition or on $b_\ell$. Whenever the leaf values are integrated out, the acceptance ratios and conditional densities below depend on products and ratios of these factors. Conditional on the tree and $b_\ell$, the leaf value is drawn from
\begin{equation*}
h_\ell\mid\cdot\sim\Normal\Bigl(\frac{S_\ell\,b_\ell}{1+n_\ell b_\ell},\,
\frac{\sigma^2 b_\ell}{1+n_\ell b_\ell}\Bigr).
\end{equation*}
The conditional mean can equivalently be written as $(1-\kappa_\ell)S_\ell/n_\ell$, where $\kappa_\ell=(1+n_\ell b_\ell)^{-1}$ is the shrinkage factor in \eqref{eq:shrinkage-factor}. 

\subsection{Splitting Rules and DART}
\label{app:sampler-rules}

A splitting rule at node $v$ consists of a covariate $k$ and a cutpoint $c$. Let $q$ be the number of covariates available to a forest. The treatment forest has $q=p$, whereas the prognostic forest has $q=p+1$ because it may also split on $\widehat e_i$. Given the units that reach $v$, covariate $k$ is selected with probability $s_k$. Conditional on $k$, the cutpoint is drawn uniformly from the $K_{vk}$ valid cutpoints, defined as the midpoints between consecutive distinct values of $x_k$ among these units that send at least one unit to each child. The resulting rule has probability $s_k/K_{vk}$. If the selected covariate has no valid cutpoint at $v$, the tree remains unchanged. These rule probabilities, together with the split probability in \eqref{eq:appendix-bart-split}, define the prior on the topology and its rules. We set $\nu=0.95$ and use the value of $\eta$ specified in Appendix~\ref{app:exp-setup}.

Without DART, $s_k=1/q$. Under DART \citep{linero2018dart}, the split probabilities $\bm s=(s_1,\dots,s_q)$ of a forest have the prior $\Dir(\theta/q,\dots,\theta/q)$ with $\theta=1$. After the last tree of the forest, they are drawn from $\bm s\mid\cdot\sim\Dir(\theta/q+c_1,\dots,\theta/q+c_q)$, where $c_k$ is the number of internal nodes that split on covariate $k$.

\subsection{Structural Moves}
\label{app:sampler-moves}

Each tree update proposes one of four moves, GROW, PRUNE, CHANGE and SWAP, with probabilities 0.3, 0.3, 0.2 and 0.2 \citep{chipman2010bart}. Its Metropolis--Hastings acceptance probability integrates out the leaf values through the factors in Equation~\eqref{eq:app-leaf-factor}.

\paragraph{GROW and PRUNE.}
GROW selects a terminal node $P$ uniformly from the $L$ leaves, draws a rule at $P$ as in Appendix~\ref{app:sampler-rules}, and splits $P$ into the children $P_L$ and $P_R$. In a treatment tree with a random allocation, GROW also draws the new branch fraction from its prior, $U_P\sim\Beta\{c_L(P),c_R(P)\}$ in \eqref{eq:weighted-branch}. The resulting child shares are $\phi_{P_L}=U_P\phi_P$ and $\phi_{P_R}=(1-U_P)\phi_P$. The proposal is accepted with probability $\min(1,r)$, where
\begin{equation*}
r=\frac{M_{P_L}M_{P_R}}{M_P}\cdot
\frac{p_{\mathrm{split}}(d_P)\{1-p_{\mathrm{split}}(d_P+1)\}^2}{1-p_{\mathrm{split}}(d_P)}\cdot
\frac{\pi(\bm U'\mid T')}{\pi(\bm U\mid T)\,\Beta\{U_P;c_L(P),c_R(P)\}}\cdot
\frac{L}{N'}.
\end{equation*}
Here $d_P$ is the depth of $P$, $N'$ is the number of internal nodes of the proposed tree $T'$ whose two children are terminal, and $\pi(\bm U\mid T)$ is the product of the beta densities of the branch fractions under $T$. The rule probability cancels because the rule is drawn from its prior, and the move probabilities cancel because GROW and PRUNE are proposed with the same probability. At $\rho=1/2$, splitting $P$ leaves the parameters $c_L(v)$ and $c_R(v)$ of every ancestor $v$ unchanged, so the third factor equals one. At $\rho=1$, the third factor accounts for the corresponding changes in the ancestor parameters. PRUNE selects uniformly one of the $N$ internal nodes whose two children are terminal and removes those children together with their branch fraction. Its acceptance probability uses the reciprocal of the corresponding GROW ratio. Prognostic trees and treatment-tree priors without random branch fractions omit the branch-fraction draw. Their factors $M_\ell$ use the leaf variances implied by the corresponding prior.

\paragraph{CHANGE and SWAP.}
CHANGE selects an internal node $v$ uniformly and draws a replacement rule conditional on the units that reach it. SWAP selects an internal node $v$ with at least one internal child, selects one such child uniformly, and exchanges the two rules. Neither proposal changes the topology or the branch fractions. Both proposals can redistribute units among the descendants of $v$, thereby changing the valid cutpoints and rule probabilities at those nodes. To preserve rule validity and reversibility, each internal node below $v$ retains its covariate and the rank of its cutpoint among the valid cutpoints for that covariate. After the proposal, its cutpoint is placed at the same rank among the cutpoints that are valid for the units that now reach the node. The proposal is rejected if any retained rank exceeds the number of available cutpoints. Otherwise, it is accepted with probability $\min(1,r)$. The ratio $r$ contains the ratio of the products of $M_\ell$ over the leaves below $v$ and the ratio of the products of the rule probabilities over the internal nodes below $v$. For SWAP, the latter product also includes $v$. For CHANGE, the rule probability at $v$ cancels with its proposal probability.

\subsection{Tree Scales and Branch fractions}
\label{app:sampler-scales}

After the structural move of treatment tree $j$, the sampler uses one of two procedures to update its branch fractions, tree scale and leaf values. Under the collapsed procedure, the branch fractions and tree scale are updated with the leaf values integrated out, after which the leaf values are redrawn. The blocked procedure first applies the collapsed branch updates and redraws the leaf values, then jointly updates the tree scale and terminal shares conditional on those values. The collapsed procedure applies for all $a$ and $\rho$. The blocked procedure requires the beta-prime prior on the tree scale to be coupled to the Dirichlet-Tree allocation through $s_\tsc=1$ and $a_\tsc=aW_j$ \citep{kruijer_adaptive_2010,zhang2022r2d2,aguilar2023r2d2m2}. This condition holds at the default $\rho=1/2$.

\paragraph{Collapsed Update.}
Each branch fraction is updated in turn with the leaf values integrated out. Write $\eta_v=\log\{U_{jv}/(1-U_{jv})\}$ and $\mathrm{expit}(x)=1/(1+e^{-x})$. Conditional on the remaining branch fractions, the tree scale and the partial residuals, the density of $\eta_v$ is
\begin{equation*}
p(\eta_v\mid\cdot)\propto
\mathrm{expit}(\eta_v)^{c_L(v)}\,\mathrm{expit}(-\eta_v)^{c_R(v)}
\prod_{\ell\in\calL_v}M_\ell(\omega\tsc_j^2\phi_{j\ell}),
\end{equation*}
where $\calL_v$ is the set of leaves below $v$, whose shares depend on $\eta_v$. The first two factors give the beta prior of $U_{jv}$ on the logit scale, including its Jacobian. We draw each $\eta_v$ by univariate slice sampling with stepping out \citep{neal2003slice}, using the initial width $\max\{1,1/\min(c_L(v),c_R(v))\}$ and at most 50 steps. The tree scale is then drawn by slice sampling with stepping out on $\log\tsc_j^2$. The initial width is one, and the target density is
\begin{equation*}
p(\tsc_j^2\mid\cdot)\propto\pi(\tsc_j^2)
\prod_{\ell\in\calL(T_j)}M_\ell(\omega\tsc_j^2\phi_{j\ell}),
\end{equation*}
where $\pi$ is the beta-prime density in \eqref{eq:beta-prime}. The leaf values are drawn last, conditional on the updated branch fractions and tree scale. Marginalizing the leaf values from the preceding updates addresses the posterior dependence induced by weak identification of a single tree. When the data contain little information about $\bh_j$, the leaf values and $\tsc_j^2$ are strongly dependent a posteriori. An update of $\tsc_j^2$ conditional on $\bh_j$ can consequently mix slowly, whereas the collapsed procedure updates the tree scale under its marginal conditional distribution.

\paragraph{Blocked Update.}
Define $B_{j\ell}=\tsc_j^2\phi_{j\ell}$, let $w_\ell$ be the weights in \eqref{eq:weighted-branch}, and write $W_j=\sum_\ell w_\ell$. When $s_\tsc=1$ and $a_\tsc=aW_j$, the prior admits the hierarchical gamma representation used in the R2D2 prior \citep{zhang2022r2d2}. Introduce the auxiliary variable $\xi\sim\operatorname{Gamma}(b_\tsc,1)$, where gamma distributions are parameterized by shape and rate. Conditional on $\xi$, the variables $B_{j\ell}$ are independently distributed as $\operatorname{Gamma}(aw_\ell,\xi)$. It follows that $\tsc_j^2=\sum_\ell B_{j\ell}\sim\BP(aW_j,b_\tsc,1)$ and that $\bphi_j=\bm B_j/\tsc_j^2\sim\Dir\{(aw_\ell)_\ell\}$ is independent of $\tsc_j^2$. Under the coupling condition, this representation recovers the prior of Section~\ref{sec:model}. Conditional on the leaf values, the $B_{j\ell}$ are independent with densities
\begin{equation*}
p(B_{j\ell}\mid\bh_j,\xi,\cdot)\propto
B_{j\ell}^{\,aw_\ell-3/2}\exp\Bigl\{-\xi B_{j\ell}
-\frac{h_{j\ell}^2}{2\sigma^2\omega B_{j\ell}}\Bigr\},
\end{equation*}
which are generalized inverse Gaussian densities. Conditional on the $B_{j\ell}$, the auxiliary variable is drawn from $\xi\mid\cdot\sim\operatorname{Gamma}(aW_j+b_\tsc,\,1+\tsc_j^2)$. Drawing and normalizing the $B_{j\ell}$ updates the tree scale and all terminal shares in one block, following the compositional strategy of \citet{aguilar2023r2d2m2}. The updated values are $\tsc_j^2=\sum_\ell B_{j\ell}$ and $\phi_{j\ell}=B_{j\ell}/\tsc_j^2$. Each branch fraction is reconstructed as the proportion of the share below its node that belongs to the left subtree.

At $\rho=1/2$, every topology has $W_j=1$, so the coupling condition holds with $a_\tsc=a=1/2$. At $\rho=1$, $W_j=L_j$ changes with the topology. No fixed value of $a_\tsc$ then satisfies the condition for every tree, and the blocked update would target a different prior. At $\rho=1/2$, the collapsed and blocked procedures target the same posterior. The reported fits use the collapsed update.

\subsection{Prognostic Forest, Intercept and Residual Variance}
\label{app:sampler-other}

All prognostic leaves share the variance $\sigma^2\omega_\mu\tsc_\mu^2$. The forest has one scale $\tsc_\mu^2\sim\BP(a_\mu,b_\mu,s_\mu)$, with $a_\mu=b_\mu=1/2$ and $s_\mu=1$. After the last prognostic tree, this scale is drawn through the gamma--inverse-gamma representation of the beta-prime distribution using an auxiliary variable $\zeta$,
\begin{equation*}
\tsc_\mu^2\mid\zeta,\cdot\sim\operatorname{InvGamma}\Bigl(b_\mu+\frac{L_\mu}2,\,
\zeta+\frac{\sum_\ell h_\ell^2}{2\sigma^2\omega_\mu}\Bigr),
\qquad
\zeta\mid\tsc_\mu^2\sim\operatorname{Gamma}\bigl(a_\mu+b_\mu,\,
1/\tsc_\mu^2+1/s_\mu\bigr),
\end{equation*}
where $L_\mu$ is the total number of leaves in the prognostic forest and the sum runs over those leaves.

The intercept has the prior $\alpha\sim\Normal(0,100\sigma^2)$. For each treated unit, define $r_i=y_i-\mu(\bx_i,\widehat e_i)-g_\tau(\bx_i)$, and let $n_1$ denote the number of treated units. Its full conditional is
\begin{equation*}
\alpha\mid\cdot\sim\Normal\Bigl(\frac{\sum_{i:A_i=1}r_i}{n_1+1/100},\,
\frac{\sigma^2}{n_1+1/100}\Bigr).
\end{equation*}
Because the prior variances of the leaf values and $\alpha$ are proportional to $\sigma^2$, their quadratic forms enter the full conditional of $\sigma^2$. Let $\mathrm{RSS}=\sum_i[y_i-\mu(\bx_i,\widehat e_i)-A_i\{\alpha+g_\tau(\bx_i)\}]^2$ be the residual sum of squares, let $L$ be the total number of leaves in both forests, and define $Q=\sum h_\ell^2/b_\ell$, where the sum runs over all leaves. Under the inverse-gamma prior with shape $a_\sigma$ and scale $b_\sigma$ given in Appendix~\ref{app:exp-setup}, the full conditional is
\begin{equation*}
\sigma^2\mid\cdot\sim\operatorname{InvGamma}\Bigl(a_\sigma+\frac{n+1+L}2,\,
b_\sigma+\frac{\mathrm{RSS}}2+\frac{\alpha^2}{200}+\frac Q2\Bigr).
\end{equation*}

\subsection{Initialization, Chains and Output}
\label{app:sampler-output}

Each chain is initialized with single-leaf trees, tree scales drawn from their priors, $\alpha=0$ and $\sigma^2$ equal to the sample variance of the centered outcome. The chains run independently under different seeds. Appendix~\ref{app:experiments} gives the numbers of chains, iterations and warmup iterations, together with the thinning interval, for each study. At every retained draw, the sampler records $\tau(\bx_i)=\alpha+g_\tau(\bx_i)$ for each unit, the average effect $n^{-1}\sum_i\tau(\bx_i)$ and the centered effects $\delta_i$ defined in Appendix~\ref{app:exp-setup}. Posterior summaries pool the retained draws across chains.

\subsection{Censored Outcomes}
\label{app:sampler-censoring}

Right censoring adds a latent-data update to the accelerated failure time model of Appendix~\ref{app:aft}. For patient $i$, $y_i$ is the log event time when the event is observed and the log censoring time otherwise. For each censored patient, introduce a latent log event time $z_i>y_i$. Conditional on the latent times, the model reduces to the Gaussian model in \eqref{eq:bcf-model}. The updates in Sections~\ref{app:sampler-iteration} to~\ref{app:sampler-other} therefore apply with $z_i$ in place of $y_i$. Each iteration additionally draws every latent time from
\begin{equation*}
z_i\mid\cdot\sim\Normal\{\mu(\bx_i,\widehat e_i)+A_i\tau(\bx_i),\sigma^2\}
\text{ truncated to }(y_i,\infty).
\end{equation*}
Each latent time is drawn by exact rejection sampling. When $y_i$ lies at or below the conditional mean, the proposal is the untruncated normal distribution. Otherwise, the sampler uses the exponential proposal of \citet{robert1995simulation}. The log event times of patients with an observed event remain fixed. A censored latent time is initialized at the median of the observed log event times that exceed $y_i$. If none exceeds $y_i$, it is initialized at $y_i$ plus half the standard deviation of the observed log event times.

For the leave-one-out estimates in Appendix~\ref{app:exp-pdac}, define $m_i=\mu(\bx_i,\widehat e_i)+A_i\tau(\bx_i)$ at each retained draw. For a patient with an observed event, the pointwise likelihood is the normal density with mean $m_i$ and variance $\sigma^2$ evaluated at $y_i$. For a censored patient, it is the survival probability $1-\Phi\{(y_i-m_i)/\sigma\}$, where $\Phi$ denotes the standard normal distribution function.
\clearpage
\section{Proofs}
\label{app:proofs}

\setcounter{proposition}{0}
\renewcommand{\theHproposition}{appendix.\arabic{proposition}}

\subsection{Split Consistency}
\label{app:split-consistency}

We begin by showing consistency under splits.

\begin{proposition}
For a tree with at least two leaves, and with the other leaf weights unchanged,
the weighted allocation is consistent under a split of leaf $\ell$ if and only
if $w_{\ell_L}+w_{\ell_R}=w_\ell$. For the depth weights
$w_\ell=\rho^{d_\ell}$, consistency under every split holds if and only if
$\rho=1/2$. Additionally, the symmetric choice is consistent under every
split.
\end{proposition}

\begin{proof}
For each $r\in\calL(T)$, let $G_r\sim\operatorname{Gamma}(aw_r,1)$
independently and define $\Phi_r=G_r/\sum_sG_s$. Then $\bm\Phi\sim\Dir\{(aw_r)_r\}$ \citep{kruijer_adaptive_2010,lin_dirichlet_2016}. At an internal node $v$, the sums of the $G_r$ below its left and right children are independent gamma variables with respective shapes $aW_v^L$ and $aW_v^R$. Their normalized ratio has distribution $\Beta(aW_v^L,aW_v^R)$ and is independent of the normalized compositions within the two subtrees. Recursive application of this gamma decomposition proves that Equation~\eqref{eq:weighted-branch} induces the terminal law $\Dir\{(aw_r)_r\}$.

Let $T'$ be obtained from a tree $T$ with at least two leaves by splitting
$\ell$, leaving every other leaf weight unchanged. Dirichlet aggregation
under $T'$ gives
\begin{equation}
 (\phi'_1,\ldots,\phi'_{L-1},
   \phi'_{\ell_L}+\phi'_{\ell_R})
 \sim
 \Dir\{aw_1,\ldots,aw_{L-1},a(w_{\ell_L}+w_{\ell_R})\}.
\end{equation}
whereas the corresponding vector under $T$ has distribution
$\Dir(aw_1,\ldots,aw_{L-1},aw_\ell)$. Dirichlet parameter vectors are
identifiable in dimension at least two \citep{lin_dirichlet_2016}. The two laws coincide precisely when $w_{\ell_L}+w_{\ell_R}=w_\ell$.

For $w_\ell=\rho^{d_\ell}$, the children of a leaf at depth $d_\ell$ have combined weight
\begin{equation}
 w_{\ell_L}+w_{\ell_R}
 =2\rho^{d_\ell+1}=2\rho w_\ell.
\end{equation}
The required equality holds for every split when $2\rho=1$.

For the symmetric recursion, retain the branch fractions at all internal nodes of $T$ and introduce an independent $\Beta(a,a)$ fraction at $\ell$. The shares of the unchanged leaves retain their values. The two new shares are $U\phi_\ell$ and $(1-U)\phi_\ell$, whose sum equals $\phi_\ell$. Aggregation under $T'$ recovers the terminal shares under $T$ for every realization of the branch fractions, and hence the two aggregated laws agree. 
\end{proof}

\subsection{Marginal Distributions}
\label{app:marginal-dist}

We now show the marginal behavior of the leaf values.

\begin{proposition}
Let $T_j$ have at least two leaves, and let $\underline c_\ell$ be the smallest
branch parameter on the path to leaf $\ell$, taking $c_L(v)$ at a left branch
and $c_R(v)$ at a right branch. Under the general Dirichlet-Tree distribution,
conditionally on $\tsc_j^2$, the density of $h_{j\ell}$ is bounded at zero if
and only if $\underline c_\ell>1/2$.

Under the weighted allocation and Equation~\eqref{eq:beta-prime}, let
$c_\ell^\star=\min(a_\tsc,aw_\ell)$ and let $p$ be the density of
$h_{j\ell}\mid T_j$. Then $p$ is bounded at zero if and only if
$c_\ell^\star>1/2$. If $c_\ell^\star<1/2$ and $a_\tsc\ne aw_\ell$, then
$p(h)\asymp |h|^{2c_\ell^\star-1}$ as $h\to0$. In all cases,
$p(h)\asymp |h|^{-(2b_\tsc+1)}$ as $|h|\to\infty$.
\end{proposition}

\begin{proof}
Write $\kappa=\sigma^2\omega$. For a node $v$ on the path to leaf $\ell$,
let $B_v$ denote the selected branch fraction, meaning $U_v$ on a left branch
and $1-U_v$ on a right branch. If $c_\ell(v)$ is the parameter on the selected
side and $\bar c_\ell(v)$ the parameter on the other side, then the $B_v$ are
independent with
$B_v\sim\Beta\{c_\ell(v),\bar c_\ell(v)\}$ and
$\phi_\ell=\prod_{v\prec\ell}B_v$.

Fix $\tsc_j^2=t>0$. We can write the conditional density of $h_{j\ell}$ as
\begin{equation}
 p_t(h)=\E\left[
 (2\pi\kappa t\phi_\ell)^{-1/2}
 \exp\left\{-\frac{h^2}{2\kappa t\phi_\ell}\right\}
 \middle|T_j\right].
\end{equation}
As $|h|\downarrow0$, the integrand increases to
$(2\pi\kappa t)^{-1/2}\phi_\ell^{-1/2}$. Monotone convergence and independence
along the path give
\begin{equation}
 p_t(0)=\frac{1}{\sqrt{2\pi\kappa t}}
 \prod_{v\prec\ell}
 \frac{B\{c_\ell(v)-1/2,\bar c_\ell(v)\}}
      {B\{c_\ell(v),\bar c_\ell(v)\}},
 \label{eq:conditional-origin-density}
\end{equation}
whenever every selected parameter exceeds $1/2$.

For each $v\prec\ell$, the moment $\E(B_v^{-1/2})$ is finite if and only if
$c_\ell(v)>1/2$. Hence, the product in
\eqref{eq:conditional-origin-density} is finite when this condition
holds at every node on the path to $\ell$, or equivalently,
\begin{equation}
 p_t(0)<\infty
 \quad\Longleftrightarrow\quad
 \underline c_\ell>1/2.
\end{equation}
Since $p_t(h)\uparrow p_t(0)$ as $|h|\downarrow0$, the same condition
characterizes boundedness of the conditional density at zero.

Under the weighted allocation, set
\begin{equation}
 \alpha=aw_\ell,\qquad
 \beta=a(W_j-w_\ell),\qquad
 T=\tsc_j^2,\qquad
 \Phi=\phi_{j\ell},\qquad
 S=T\Phi.
\end{equation}
Since $T_j$ has at least two leaves, $\beta>0$. Hence,  the terminal Dirichlet law is $\Phi\sim\Beta(\alpha,\beta)$, independently of
$T\sim\BP(a_\tsc,b_\tsc,s_\tsc)$. At the origin, monotone convergence gives
\begin{equation}
 p(0)=(2\pi\kappa)^{-1/2}\E(T^{-1/2})\E(\Phi^{-1/2}).
\end{equation}
The beta-prime moment $\E(T^r)$ is finite for $-a_\tsc<r<b_\tsc$, and the beta moment $\E(\Phi^r)$ is finite when $r>-\alpha$. The density $p$ is bounded at zero if and only if
$a_\tsc>1/2$ and $\alpha>1/2$, equivalently
$c_\ell^\star=\min(a_\tsc,aw_\ell)>1/2$.

Let $f_S$ denote the density of $S$, and suppose $a_\tsc\ne\alpha$. The
product-density identity and dominated convergence give
\begin{equation}
 f_S(s)=C_Ss^{c_\ell^\star-1}\{1+o(1)\},
 \qquad s\downarrow0,
 \label{eq:product-density-origin}
\end{equation}
where
\begin{equation}
 C_S=
 \begin{cases}
 \displaystyle
 \frac{s_\tsc^{-a_\tsc}}{B(a_\tsc,b_\tsc)}
 \frac{B(\alpha-a_\tsc,\beta)}{B(\alpha,\beta)},
 & a_\tsc<\alpha,\\[2.2ex]
 \displaystyle
 \frac{s_\tsc^{-\alpha}B(a_\tsc-\alpha,b_\tsc+\alpha)}
      {B(a_\tsc,b_\tsc)B(\alpha,\beta)},
 & \alpha<a_\tsc.
 \end{cases}
\end{equation}
When $a_\tsc<\alpha$, the identity
\begin{equation}
 s^{1-a_\tsc}f_S(s)
 =\frac{s_\tsc^{-a_\tsc}}{B(a_\tsc,b_\tsc)}
 \int_0^1 u^{-a_\tsc}
 \left(1+\frac{s}{s_\tsc u}\right)^{-(a_\tsc+b_\tsc)}
 f_\Phi(u)\,du
\end{equation}
converges by domination because $\E(\Phi^{-a_\tsc})<\infty$. The same
domination bounds $s^{1-a_\tsc}f_S(s)$ uniformly on a neighborhood of zero.

When $\alpha<a_\tsc$, conditioning on $T$ gives
\begin{equation}
 s^{1-\alpha}f_S(s)
 =\frac{1}{B(\alpha,\beta)}
 \int_s^\infty t^{-\alpha}
 \left(1-\frac{s}{t}\right)^{\beta-1}f_T(t)\,dt.
\end{equation}
Split the integral at $2s$. On $[2s,\infty)$, the factor involving $\beta$ is
bounded by $\max(1,2^{1-\beta})$, and
$\E(T^{-\alpha})<\infty$ because $\alpha<a_\tsc$. The contribution from this
interval converges to $\E(T^{-\alpha})/B(\alpha,\beta)$. On $[s,2s]$, the
bound $f_T(t)\leq C t^{a_\tsc-1}$ and the substitution $t=su$ give an upper
bound proportional to
\begin{equation}
 s^{a_\tsc-\alpha}
 \int_1^2 u^{a_\tsc-\alpha-1}(1-1/u)^{\beta-1}\,du.
\end{equation}
The last equation converges to zero because $a_\tsc>\alpha$ and $\beta>0$.
We can use these bounds to control $s^{1-\alpha}f_S(s)$ uniformly near zero. 

The beta-prime density also satisfies the global bound
\begin{equation}
 \sup_{t>0}t^{b_\tsc+1}f_T(t)
 \leq \frac{s_\tsc^{b_\tsc}}{B(a_\tsc,b_\tsc)}.
\end{equation}
The product-density formula gives
\begin{equation}
 f_S(s)\leq
 \frac{s_\tsc^{b_\tsc}\E(\Phi^{b_\tsc})}
      {B(a_\tsc,b_\tsc)}s^{-b_\tsc-1},
 \qquad s>0.
\end{equation}
Together with the small-$s$ estimates we present above, this bound yields a finite $M_0$ such
that
$f_S(s)\leq M_0s^{c_\ell^\star-1}$ for every $s>0$ when
$a_\tsc\ne\alpha$.

The normal scale-mixture density can be written as \citep{andrews_scale_1974, west_scale_1987}
\begin{equation}
 p(h)=|h|\int_0^\infty
 (2\pi\kappa r)^{-1/2}\exp\{-1/(2\kappa r)\}
 f_S(h^2r)\,dr.
 \label{eq:marginal-substitution-proof}
\end{equation}
If $c_\ell^\star<1/2$, substitution of Equation~\eqref{eq:product-density-origin} and dominated convergence yield
\begin{equation}
 \lim_{h\to0}|h|^{1-2c_\ell^\star}p(h)=C_{0,\ell},
 \qquad
 C_{0,\ell}=
 C_S(2\pi\kappa)^{-1/2}(2\kappa)^{1/2-c_\ell^\star}
 \Gamma(1/2-c_\ell^\star),
 \label{eq:leaf-origin-constant}
\end{equation}
and $C_{0,\ell}$ is positive and finite. The global bound above gives a
dominating function proportional to
$r^{c_\ell^\star-3/2}\exp\{-1/(2\kappa r)\}$, which is integrable when
$c_\ell^\star<1/2$. Set $b=b_\tsc$. The beta-prime density satisfies
\begin{equation}
 f_T(t)\sim\frac{s_\tsc^b}{B(a_\tsc,b)}t^{-b-1},
 \qquad t\to\infty.
\end{equation}
Since $0<\Phi<1$ and $\E(\Phi^b)<\infty$, dominated convergence in the product
density gives
\begin{equation}
 f_S(s)\sim Ds^{-b-1},
 \qquad
 D=\frac{s_\tsc^b}{B(a_\tsc,b)}\E(\Phi^b).
\end{equation}
After multiplication by $|h|^{2b+1}$, the global tail bound dominates the
integrand in Equation~\eqref{eq:marginal-substitution-proof} by a constant multiple of
$r^{-b-3/2}\exp\{-1/(2\kappa r)\}$, which is integrable for $b>0$. Dominated
convergence in Equation~\eqref{eq:marginal-substitution-proof} gives
\begin{equation}
 \lim_{|h|\to\infty}|h|^{2b+1}p(h)=K_\ell,
 \qquad
 K_\ell=
 \frac{(2\kappa s_\tsc)^b\Gamma(b+1/2)}
      {\sqrt{\pi}B(a_\tsc,b)}
 \frac{B(\alpha+b,\beta)}{B(\alpha,\beta)},
 \label{eq:leaf-tail-constant}
\end{equation}
where $K_\ell$ is positive and finite.
\end{proof}

\subsection{Heterogeneity Regions}

We proceed to bound the number of regions that express heterogeneity. 

\begin{proposition}[Number of regions expressing heterogeneity]
Let every leaf of $T_j$ contain between $1$ and $n$ treated observations.
Under the Dirichlet-Tree prior,
\begin{equation}
 N_{j,q}\leq \frac{(1-q)n\omega\tsc_j^2}{q}
\end{equation}
for every $\tsc_j^2>0$ and $\bphi_j$.
\end{proposition}

\begin{proof}
For each leaf counted by $N_{j,q}$,
\begin{equation}
 n_{j\ell}\omega\tsc_j^2\phi_{j\ell}
 \geq\frac{q}{1-q}.
\end{equation}
The bound $n_{j\ell}\leq n$ implies
\begin{equation}
 \phi_{j\ell}\geq
 c_q,
\end{equation}
where $c_q=\frac{q}{(1-q)n\omega\tsc_j^2}$. Since $\sum_\ell\phi_{j\ell}=1$,
\begin{equation}
  N_{j,q}c_q \leq
 \sum_{\ell:\,\kappa_{j\ell}\leq1-q}\phi_{j\ell}
 \leq 1.
\end{equation}
Rearranging the last inequality gives the claimed bound.
\end{proof}

\paragraph{Comparison with Independent Local Scales.}
Let $\lambda_{j\ell}\geq0$ be i.i.d. local scales, independent of $T_j$ and $\tsc_j^2$, whose common law has unbounded support and does not depend on $L_j$. Replace the Dirichlet-Tree variance $\sigma^2\omega\tsc_j^2\phi_{j\ell}$ by $\sigma^2\omega\tsc_j^2\lambda_{j\ell}^2$. Conditional on $T_j$ and $\tsc_j^2$, define
\begin{equation}
 \kappa_{j\ell}^{\mathrm{loc}} =\left(1+n_{j\ell}\omega\tsc_j^2\lambda_{j\ell}^2\right)^{-1}, \qquad N_{j,q}^{\mathrm{loc}} =\#\{\ell:\kappa_{j\ell}^{\mathrm{loc}}\leq1-q\}.
\end{equation}
Let
\begin{equation}
 p_q(\tsc_j^2)= \Pr\left\{\lambda_{j\ell}^2\geq \frac{q}{(1-q)\omega\tsc_j^2}\right\}.
\end{equation}
The unbounded support of the local-scale distribution implies
$p_q(\tsc_j^2)>0$. Moreover, since $n_{j\ell}\geq1$, we have 
\begin{equation}
 \left\{ \lambda_{j\ell}^2\geq \frac{q}{(1-q)\omega\tsc_j^2} \right\}
 \subseteq
 \left\{  \kappa_{j\ell}^{\mathrm{loc}}\leq1-q \right\}.
\end{equation}
Consequently,
\begin{equation}
\begin{split}
 \E\!\left( N_{j,q}^{\mathrm{loc}}\mid T_j,\tsc_j^2 \right)
 &= \sum_{\ell=1}^{L_j} \Pr\!\left( \kappa_{j\ell}^{\mathrm{loc}}\leq1-q \mid T_j,\tsc_j^2 \right) \geq L_jp_q(\tsc_j^2).
\end{split}
\end{equation}
Since the local-scale distribution does not depend on $L_j$, neither does
$p_q(\tsc_j^2)$. Thus, the expected number of regions satisfying the
criterion grows at least linearly with $L_j$. 

For example, consider standard half-Cauchy local scales, as in the horseshoe
prior \citep{carvalho2010horseshoe}. In this case,
\begin{equation}
 p_q(\tsc_j^2) = 1-\frac{2}{\pi}\arctan \left\{ \frac{q}{(1-q)\omega\tsc_j^2} \right\}^{1/2}.
\end{equation}
Substitution into the preceding bound gives
\begin{equation}
 \E\!\left( N_{j,q}^{\mathrm{loc}}\mid T_j,\tsc_j^2 \right) \geq L_j \left[ 1-\frac{2}{\pi}\arctan \left\{ \frac{q}{(1-q)\omega\tsc_j^2} \right\}^{1/2} \right].
\end{equation}
For fixed $q$, $\omega$, and $\tsc_j^2>0$, the term in brackets is strictly
positive and does not depend on $L_j$. Thus, theexpected number of regions satisfying the criterion grows at least linearly with the number of leaves. In contrast, Proposition~\ref{prop:weak-shrinkage} bounds the  count under the Dirichlet-Tree prior independently of $L_j$.

\subsection{Effective Allocation Size}
\label{app:effective-size}

We show how to derive the bounds presented in the main text for the effective allocation size $A_{\mathrm{eff}}(T_j)$. Under the weighted allocation,
\begin{equation}
 \E\left(\sum_\ell\phi_{j\ell}^2\middle|T_j\right) =\frac{a\sum_\ell w_\ell^2+W_j}{W_j(aW_j+1)}.
\end{equation}
For $\rho=1/2$, splitting a leaf of weight $2^{-d}$ replaces it by two leaves of weight $2^{-(d+1)}$, so $W_j=1$ for every finite binary tree. If the tree has at least two leaves, then $0<\sum_\ell w_\ell^2\leq(\max_\ell w_\ell)\sum_\ell w_\ell\leq1/2$. The preceding bounds give
\begin{equation}
 \frac{1+a}{1+a/2} \leq A_{\mathrm{eff}}(T_j) =\frac{1+a}{1+a\sum_\ell w_\ell^2} <1+a.
\end{equation}
The lower bound is attained by the two-leaf tree. The upper bound is not
attained by a finite tree, but complete trees approach it as their depth tends
to infinity.

\subsection{Posterior Consistency}
\label{app:consistency}

The proof follows the classical Kullback--Leibler support argument of \citet{schwartz1965bayes,ghosal_convergence_2000}; see also \citet{ghosal_convergence_2000} for a broader treatment of posterior convergence. We first establish that the prior assigns positive probability to uniform neighborhoods of the true prognostic and treatment-effect surfaces. We then show that these neighborhoods are contained in arbitrarily small Kullback--Leibler neighborhoods of the true observable law. Weak posterior consistency follows from Schwartz's theorem.

Consider a fixed number $p$ of covariates, $\bx\sim Q$ on a compact rectangle $\mathcal X\subset\R^p$, $A\mid\bx\sim\operatorname{Bernoulli}\{e_0(\bx)\}$, and
\begin{equation}
 y\mid\bx,A\sim \Normal\left(\mu_0(\bx)+A\tau_0(\bx),\sigma_0^2\right).
\end{equation}
The fitted model uses the true propensity $e_0$, and $\tau(\bx)=\alpha+g_\tau(\bx)$ denotes its complete treatment-effect surface.

\emph{Conditions.} Assume the following:
\begin{enumerate}[label=(\roman*),leftmargin=*,itemsep=2pt,topsep=3pt]
\item The surfaces $\mu_0$ and $\tau_0$ are continuous on $\mathcal X$, $\sigma_0>0$, and $\underline e\leq e_0(\bx)\leq1-\underline e$ for some $\underline e\in(0,1/2]$ and $Q$-almost every $\bx$.
\item Each forest contains a fixed positive number of trees. The topology prior assigns positive mass to every finite binary topology, every covariate has positive splitting probability, and each conditional cutpoint law has full support on the interior of its coordinate range.
\item Treatment-tree branch fractions are independent beta variables with positive parameters. The tree scale is independent of the branch fractions and has a positive density on $(0,\infty)$, and $\omega>0$. Conditional on a prognostic topology and any $\sigma>0$, the prognostic terminal-value density is positive everywhere on its Euclidean space.
\item Conditional on $\sigma$, the tree-specific states have product form across the fixed collection of trees. The conditional density of $\alpha$ is  positive in a neighborhood of zero, the prior for $\sigma$ assigns positive mass to every neighborhood of $\sigma_0$, and the induced joint densities are measurable.
\end{enumerate}

\setcounter{theorem}{0}
\renewcommand{\theHtheorem}{appendix.\arabic{theorem}}
\begin{theorem}[Weak consistency]
Under these conditions, the prior of \method{} assigns positive probability to every Kullback--Leibler neighborhood of the law $P_0$ of $(\bx,A,y)$. Moreover, for every weak neighborhood $\mathcal W$ of $P_0$,
\begin{equation}
 \Pi\!\left(P_{\mu,\tau,\sigma}\notin\mathcal W \mid\mathcal D_n\right)\longrightarrow0 \qquad P_0^\infty\text{-almost surely}.
\end{equation}
If $P_{\mu,\tau,\sigma}=P_0$, then $\sigma=\sigma_0$, $\mu\{\bx,e_0(\bx)\}=\mu_0(\bx)$, and $\tau(\bx)=\tau_0(\bx)$ for $Q$-almost every $\bx$.
\end{theorem}

\begin{proof}
In the following let $\widetilde\mu(\bx)=\mu\{\bx,e_0(\bx)\}$. Consider a finite treatment tree $T$ with $L$ leaves. For $t=\tsc_j^2>0$, $\sigma>0$, and a realized allocation $\bphi$, the covariance matrix of its terminal values is
\begin{equation}
 \Sigma_T(t,\bphi,\sigma) =\sigma^2\omega t\operatorname{diag}(\bphi).
\end{equation}
For $L\geq2$, positivity of the beta branch parameters gives $\phi_\ell\in(0,1)$ almost surely for every leaf. When $L=1$, $\phi_1=1$. In either case, $\Sigma_T(t,\bphi,\sigma)$ is positive definite.

Let $\bh=(h_1,\ldots,h_L)^\top$ denote the vector of terminal values. Conditional on $t$, $\bphi$, and $\sigma$, $\bh\sim\Normal_L\{\bm 0,\Sigma_T(t,\bphi,\sigma)\}$. Since $\Sigma_T(t,\bphi,\sigma)$ is positive definite, the conditional density of $\bh$ is strictly positive on $\R^L$.  Hence, for every $\bh_0\in\R^L$ and $\eta>0$,
\begin{equation}
 \Pi\!\left(\|\bh-\bh_0\|_\infty<\eta\mid T,\sigma\right)>0.
 \label{eq:fixed-tree-support}
\end{equation}
Assumption~(iii) gives the same conclusion for a prognostic tree.

Fix $f\in C(\mathcal X)$ and $\eta>0$. Since $\mathcal X$ is compact,
$f$ is uniformly continuous. Hence, there exists $\delta>0$ such that
\begin{equation}
 \|\bx-\bx'\|_\infty<\delta \quad\Longrightarrow\quad |f(\bx)-f(\bx')|<\frac{\eta}{2}.
\end{equation}
Choose a finite axis-aligned recursive partition $\{R_k\}_{k=1}^K$ whose cells have diameter smaller than $\delta/2$, and select $\bx_k\in R_k$ for each cell. Let $T_f$ be a binary tree representing this partition, with terminal value $f(\bx_k)$ assigned to leaf $R_k$.

Since $T_f$ contains finitely many cutpoints, we can chose sufficiently small open neighborhoods of these cutpoints such that every perturbed partition has corresponding leaves $\{R_k'\}_{k=1}^K$ satisfying
\begin{equation}
 \sup_{\bx\in R_k'}\|\bx-\bx_k\|_\infty<\delta, \qquad k=1,\ldots,K.
\end{equation}
It follows from the choice of $\delta$ that
\begin{equation}
 \sup_{\bx\in R_k'} |f(\bx)-f(\bx_k)| <\frac{\eta}{2}, \qquad k=1,\ldots,K.
\end{equation}
Assumption~(ii) assigns positive probability to the topology of $T_f$, its
split variables, and the specified cutpoint neighborhoods. Conditional on
this event, \eqref{eq:fixed-tree-support} assigns positive probability to
\begin{equation}
 \max_{1\leq k\leq K}|h_k-f(\bx_k)|<\frac{\eta}{2}.
\end{equation}
For every $\bx\in R_k'$, the triangle inequality gives
\begin{equation}
 |g(\bx)-f(\bx)| \leq |h_k-f(\bx_k)|+|f(\bx_k)-f(\bx)| <\eta.
\end{equation}
Consequently,
\begin{equation}
 \Pi\!\left(\|g-f\|_\infty<\eta\mid\sigma\right)>0.
\label{eq:single-tree-support}
\end{equation}

Apply Equation~\eqref{eq:single-tree-support} to one tree in each forest. In the prognostic forest, the approximating tree can split only on the original covariates, so continuity of $e_0$ is not required. Assign zero target values to the remaining trees and to $\alpha$. The product-support conditions in assumptions~(iii)--(iv), together with the prior support for $\sigma$, imply that every $\delta>0$ has positive prior probability of
the event
\begin{equation}
 \left\{
 \|\widetilde\mu-\mu_0\|_\infty<\delta,\quad
 \|\tau-\tau_0\|_\infty<\delta,\quad
 |\sigma-\sigma_0|<\delta
 \right\}.
 \label{eq:causal-supnorm-support}
\end{equation}

Relative to the product of $Q$, counting measure on $\{0,1\}$, and Lebesgue measure, the Kullback--Leibler divergence is
\begin{equation}
 \begin{split}
 K(P_0,P_{\mu,\tau,\sigma}) &=\log\frac{\sigma}{\sigma_0} +\frac{\sigma_0^2}{2\sigma^2}-\frac12\\
 &\quad+\frac{1}{2\sigma^2}\int_{\mathcal X} \left[ (1-e_0)(\widetilde\mu-\mu_0)^2 +e_0\{\widetilde\mu+\tau-\mu_0-\tau_0\}^2 \right]\,dQ ,
 \end{split}
 \label{eq:causal-kl}
\end{equation}
with every function in the integrand evaluated at $\bx$. On the event in Equation~\eqref{eq:causal-supnorm-support},
\begin{equation}
 (1-e_0)(\widetilde\mu-\mu_0)^2 +e_0(\widetilde\mu+\tau-\mu_0-\tau_0)^2 \leq 4\delta^2.
\end{equation}
For $\sigma\geq\sigma_0/2$, this yields
\begin{equation}
 K(P_0,P_{\mu,\tau,\sigma}) \leq r(\sigma)+\frac{8\delta^2}{\sigma_0^2}, \qquad
 r(\sigma)= \log\frac{\sigma}{\sigma_0} +\frac{\sigma_0^2}{2\sigma^2}-\frac12.
 \label{eq:causal-kl-bound}
\end{equation}
The function $r$ is continuous at $\sigma_0$ and $r(\sigma_0)=0$. Given $\varepsilon>0$, choose $\delta<\sigma_0/2$ so that $8\delta^2/\sigma_0^2<\varepsilon/2$, and restrict the $\sigma$-neighborhood further so that $r(\sigma)<\varepsilon/2$. Equations \eqref{eq:causal-supnorm-support} and \eqref{eq:causal-kl-bound} give
\begin{equation}
 \Pi\!\left(P:K(P_0,P)<\varepsilon\right)>0.
\end{equation}
The prior has Kullback--Leibler support at $P_0$. Schwartz's theorem \citep{schwartz1965bayes} establishes weak consistency.

It remains to verify the identification statement. If $P_{\mu,\tau,\sigma}=P_0$, overlap gives equality of the conditional Gaussian laws at both treatment levels for $Q$-almost every $\bx$. The equality at $A=0$ gives $\sigma=\sigma_0$ and $\widetilde\mu=\mu_0$. The equality at $A=1$ then gives $\widetilde\mu+\tau=\mu_0+\tau_0$, and hence $\tau=\tau_0$ $Q$-almost everywhere.
\end{proof}

\clearpage

\section{Experimental details and further results}
\label{app:experiments}

This appendix first gives the settings that the studies of
Section~\ref{sec:experiments} share. The design and the complete results of
each study follow, in the order of that section.

\subsection{Common setup}
\label{app:exp-setup}

\paragraph{Models.}
We consider the six models of Section~\ref{sec:exp-setup} whose $\omega$ is matched run in our sampler. They share the BCF model~\eqref{eq:bcf-model} and, apart from DART, differ only in the prior variance $V_{j\ell}$ of the leaf value $h_{j\ell}$ of a treatment tree. Section~\ref{sec:exp-setup} defines \method{}, its variants and Fixed. HS sets $V_{j\ell}=\sigma^2\omega\tsc_j^2\lambda_{j\ell}^2$, where the local scales $\lambda_{j\ell}$ are independent and half-Cauchy. Its tree scale has the prior $\tsc_j^2\sim\BP(1/2,1/2,1)$ of \method{}, under which $\tsc_j$ is also half-Cauchy. In the application, HS is instead the Horseshoe Forest of \citet{jacobs2026horseshoe}. It is fitted with the function \texttt{CausalHorseForest} of the authors' package at its defaults, including the value $k=1.5$ of the parameter that sets its prior width, with $\widehat e_i$ added to the covariates of its prognostic forest. We do not match its prior to $\varsigma$. The models marked + DART place the prior $\bm s\sim\Dir(\theta/q,\dots,\theta/q)$ of \citet{linero2018dart} on the split probabilities of each forest, where $q$ is the number of covariates of that forest and $\theta=1$ is held fixed. HS + DART and Equal + DART, which appear in Appendix~\ref{app:exp-forest}, are defined in the same way as \method{} + DART.

\paragraph{Prior Settings.}
We use $m_\tau=50$ treatment trees with split probability $0.95(1+d)^{-\eta}$, where $\eta=2$ outside the tree-size study, and $m_\mu=200$ prognostic trees with split probability $0.95(1+d)^{-2}$. The prognostic leaf values have variance $\sigma^2\omega_\mu\tsc_\mu^2$, with one
scale $\tsc_\mu^2\sim\BP(1/2,1/2,1)$ for the whole forest and $\omega_\mu=1/m_\mu$, so that given $\tsc_\mu^2$ the prior variance of $\mu(\bx,\widehat e)$ at any covariate value is $\sigma^2\tsc_\mu^2$. As in BCF, $\widehat e_i$ enters $\mu$ as a covariate. 

The outcome is centered at its sample mean. The intercept has the prior $\alpha\sim\Normal(0,100\sigma^2)$. The residual variance has an inverse-gamma prior with shape $3$ and scale $v$, so its prior mean is $v/2$. The value of $v$ is the sample variance of $y$ in the forest simulations, the residual variance of a lasso regression of $y$ on $(A_i,\bx_i)$ with a cross-validated penalty in ACIC 2016 \citep{friedman2010glmnet}, and the squared scale of an intercept-only Gaussian accelerated failure time model in the application \citep{therneau2024survival}. In every reported fit the leaf variances carry the factor $\sigma^2$, as in Section~\ref{sec:model}.

\paragraph{Specification of $\omega$.}
Section~\ref{sec:hyperparameters} sets $\omega$ so that the prior median of $S_\tau/\sigma$ equals $\varsigma=1/2$. We compute this value of $\omega$ by simulation with $\sigma=1$. Each of $1{,}000$ prior draws grows $m_\tau$ treatment trees from the tree prior, in which a node splits with the split
probability of its depth. The split variable is drawn uniformly, and the cut is the midpoint of the range of that variable among the units in the node. The shares, tree scales and leaf values are then drawn from their priors, which gives one draw of $S_\tau$ over the units of a reference design. Because $S_\tau$ is proportional to $\omega^{1/2}$, a run at $\omega_0=1/m_\tau$ whose median is $M_0$ gives $\omega=\omega_0(\varsigma/M_0)^2$. A second, independent run of $1{,}000$ draws at this $\omega$ checks the median. The reference design is a sample of 500 units from $\mathrm{U}(0,1)^5$ in the forest simulations, the $4{,}802$ units of the competition in ACIC 2016, and the 130 patients in the application. Each model is matched separately, and in the tree-size study separately at each $\eta$.

\paragraph{BCF.}
BCF is fitted with the \texttt{bcf} package of \citet{hahn2020bcf} when $p<n$, which covers the tree-size study, the dimension study at $p\in\{20,50\}$ and ACIC 2016. The package runs at its defaults, with the prior of Appendix~\ref{app:bcf}, and receives the same propensity score as the other models. Because the package does not run when $p\ge n$, BCF is then fitted in our sampler. This version keeps the numbers of trees and the split probabilities of the package, with a half-normal prior on the scale of the treatment forest and a half-Cauchy prior on the scale of the prognostic forest. In our sampler the treatment indicator enters with the fixed coding $0$ and $1$, without the coefficients $b_0$ and $b_1$ of Appendix~\ref{app:bcf}. Its $\omega$ is matched to $\varsigma=1$, which approximates the width of the package's default prior for the treatment forest. This is the only version of BCF whose prior is matched. On the fixed partition, BCF denotes its leaf prior on the given tree, with one half-normal scale shared by all leaves (Appendix~\ref{app:exp-partition}).

\paragraph{Loss-based Models.}
The X-learner \citep{kuenzel2019meta} and the R-learner \citep{nie2021quasi} are fitted with the functions \texttt{xboost} and \texttt{rboost} of the \texttt{rlearner} package, which estimate each regression of Appendix~\ref{app:meta-learners} by gradient-boosted trees \citep{chen2016xgboost}. Each regression is tuned over ten random configurations of the learning rate, tree depth, subsampling and regularization parameters. Every configuration is evaluated by ten-fold cross-validation with up to $1{,}000$ boosting rounds and early stopping after ten rounds without improvement. The best configuration is then refitted to the full sample of that regression. The outcome regression of the R-learner is cross-fitted over the same ten folds. Both learners otherwise run at the package defaults and receive the same propensity score as the other models. They return point estimates, so coverage and interval length are not reported for them. The learners are not fitted in the application.

\paragraph{Propensity Scores.}
In the forest simulations every model, including the loss-based ones,
receives the true propensity score, $\widehat e_i=e(\bx_i)$. In ACIC 2016 and in the application every model receives the same estimate, from a five-fold cross-fitted logistic lasso \citep{friedman2010glmnet} whose penalty minimizes the deviance of an inner five-fold cross-validation. The estimate is computed once per data set and truncated to $[10^{-4},1-10^{-4}]$.

\paragraph{Metrics.}
Let $\delta_i=\tau(\bx_i)-n^{-1}\sum_{i'=1}^n\tau(\bx_{i'})$ be the centered effect of unit $i$ in a data set of size $n$. For the Bayesian models its estimate $\widehat\delta_i$ is the posterior mean of the same quantity computed from each draw of $\tau$, and for the loss-based models it is the centered point estimate. The root mean squared error
\begin{equation*}
\mathrm{RMSE}=\sqrt{\frac1n\sum_{i=1}^n\bigl(\widehat\delta_i-\delta_i\bigr)^2}
\end{equation*}
is the precision in estimation of heterogeneous effects of \citet{hill2011causal}, applied to the centered effect. The 90\% interval for $\delta_i$ runs from the 5\% to the 95\% quantile of the draws pooled over the chains. Coverage is the proportion of units whose $\delta_i$ lies in its interval, and length is the average width of the intervals. In scenarios where some units have no effect, a unit is reported as having an effect when the 90\% interval for $\tau(\bx_i)$ excludes zero. The false positive rate (FPR) is the proportion of units with $\tau(\bx_i)=0$ that are reported, and the true positive rate (TPR) is the proportion of units with $\tau(\bx_i)\ne0$ that are reported. In the forest simulations, predictive accuracy is measured by the log predictive density (lpd) of $n_{\mathrm{test}}=500$ test units drawn from the same process \citep{vehtari2017practical},
\begin{equation*}
\mathrm{lpd}=\frac1{n_{\mathrm{test}}}\sum_{k=1}^{n_{\mathrm{test}}}
\log\Bigl[\frac1S\sum_{s=1}^S\Normal\bigl\{y_k;\,\mu^{(s)}(\bx_k,e_k)
+A_k\tau^{(s)}(\bx_k),\,\sigma^{(s)2}\bigr\}\Bigr],
\end{equation*}
where $s$ indexes the $S$ posterior draws. In the application, the expected log-pointwise predictive density (elpd) for a new patient is estimated by Pareto-smoothed importance-sampling leave-one-out cross-validation (PSIS-LOO),with its standard error \citep{vehtari2017practical}. Its pointwise likelihood is that of the censored model on the log-time scale, the normal density of $\log Y_i$ for a patient with an observed event and the normal survival function at $\log Y_i$ for a censored patient (Appendix~\ref{app:sampler-censoring}). In the figures, every uncertainty band for a mean over data sets is the mean plus or minus two standard errors. In the figures and tables, the standard error of such a mean is the standard deviation over data sets divided by the square root of their number, computed separately for each model. The ratios of Table~\ref{tab:acic} are treated in the same way on the log scale (Appendix~\ref{app:exp-acic}).

\paragraph{Software, Data and Computing.}
Table~\ref{tab:app-software} lists the software and data with their versions and licenses. Our sampler is written in R \citep{rcoreteam2026}, with C++ kernels compiled through Rcpp \citep{eddelbuettel2011rcpp}. The ACIC 2016 data sets are generated by the \texttt{aciccomp2016} package of \citet{dorie2019automated} from covariates of the Collaborative Perinatal Project. The pancreatic cohort is the TCGA PAAD object curated by \texttt{pdacR} \citep{torrehealy2023pdacr} from the data of \citet{tcga2017pdac}, which are open-access data of the Genomic Data Commons of the National Cancer Institute. The fits of the simulations and of ACIC 2016 ran on a shared server with two AMD EPYC 9655 processors (192 cores) and 2.3 TB of memory, under Ubuntu 24.04 and R 4.5.3. The fits of the application ran on that server and on a laptop (MacBook Pro, Apple M5 Pro, 24 GB of memory, macOS, R 4.6.0). The tables and figures were produced on the laptop. Each fit ran in its own R process, with single-threaded linear algebra.

\begin{table}[h]
\centering
\caption{Software and Data Used.}
\label{tab:app-software}
\small
\begin{tabular}{llll}
\toprule
Software or data & Version & License & Reference\\
\midrule
R & 4.5.3, 4.6.0 & GPL-2 or GPL-3 & \citet{rcoreteam2026}\\
Rcpp & 1.1.2 & GPL ($\ge2$) & \citet{eddelbuettel2011rcpp}\\
Our code & 0.1.0 & MIT & this paper\\
\texttt{bcf} & 2.0.2 & GPL-3 & \citet{hahn2020bcf}\\
\texttt{ShrinkageTrees} & 2.1.0 & MIT & \citet{jacobs2026horseshoe}\\
\texttt{rlearner} & 1.1.0 & GPL-3 & \citet{nie2021quasi}\\
\texttt{xgboost} & 1.7.8.1 & Apache 2.0 & \citet{chen2016xgboost}\\
\texttt{glmnet} & 5.0 & GPL-2 & \citet{friedman2010glmnet}\\
\texttt{survival} & 3.8-12 & LGPL ($\ge2$) & \citet{therneau2024survival}\\
\texttt{loo} & 2.9.0 & GPL ($\ge3$) & \citet{vehtari_loo_2023}\\
\texttt{aciccomp2016} (ACIC 2016 data) & 0.1-0 & GPL ($\ge2$) & \citet{dorie2019automated}\\
\texttt{pdacR} (TCGA PAAD data) & 0.1.2 & MIT; GDC open access & \citet{torrehealy2023pdacr}\\
\bottomrule
\end{tabular}
\end{table}

\subsection{Leaf Priors on a Fixed Partition}
\label{app:exp-partition}

This study separates the leaf prior from the tree search of the forest. Section~\ref{sec:exp-partition} reports its conditions at signal-to-noise ratio 3. This subsection gives its design and its complete results.

\paragraph{Data.}
The tree is the balanced binary tree with $L=2^D$ leaves, $D\in\{3,4,5,6\}$, held fixed. For each data set, $k$ leaves are drawn uniformly without replacement. Half of them receive $h_\ell=\Delta$ and half $h_\ell=-\Delta$. The other leaves have $h_\ell=0$, so the leaf values average to zero. Each leaf holds $m=16$ observations with residual standard deviation $\sigma=1$, and the effect size is $\Delta=\mathrm{SNR}\cdot\sigma/m^{1/2}$. The signal-to-noise ratio SNR is therefore $\Delta$ in units of the standard error of a leaf mean. The mean residual of leaf $\ell$ is $\bar R_\ell=h_\ell+\varepsilon_\ell$, with independent $\varepsilon_\ell\sim\Normal(0,\sigma^2/m)$. The sampler receives the count $m$ and the residual sum $m\bar R_\ell$ of each leaf. The 200 data sets of a condition depend only on $(L,k,\Delta)$ and their seeds, so every model and every value of $\varsigma$ is fitted to the same data sets. Besides the conditions of Figure~\ref{fig:partition}, the design varies the signal-to-noise ratio over $\{2,3,4\}$ at $k=2$, adds $k=32$ at $L=32$, where every leaf has an effect, and varies $\varsigma$ over $\{1/4,1/2,1\}$ at $L=32$, $k=2$ and SNR 3. In all, the design has 17 conditions.

\paragraph{Priors.}
Given its scales and shares, each leaf value has the prior $h_\ell\sim\Normal(0,V_\ell)$, independently across leaves. \method{} and \method{} ($\rho=1$) set $V_\ell=\sigma^2\omega\tsc^2\phi_\ell$ with $\bphi\sim\Dir(a\rho^D,\dots,a\rho^D)$ and $a=1/2$. Every leaf of a balanced tree has depth $D$, so both values of $\rho$ give a symmetric Dirichlet law, with concentration $a/L$ per leaf at $\rho=1/2$ and $a$ at $\rho=1$. Equal sets $V_\ell=\sigma^2\omega\tsc^2/L$. HS sets $V_\ell=\sigma^2\omega\tsc^2\lambda_\ell^2$ with independent standard half-Cauchy local scales $\lambda_\ell$. In these four models $\tsc^2\sim\BP(1/2,1/2,1)$, so $\tsc$ is standard half-Cauchy. BCF sets $V_\ell=\sigma^2\omega s^2$, with one standard half-normal scale $s$ shared by all leaves. Fixed sets $V_\ell=\sigma^2\omega$. The sampler runs the updates of Appendix~\ref{app:sampler} for the shares, the scales and the leaf values of one tree, without structural moves and with $\sigma$ known.

\paragraph{Prior Specification.}
For every model, every $L$ and every value of $\varsigma$, $\omega$ is chosen so that the prior median of $\{L^{-1}\sum_\ell(h_\ell-\bar h)^2\}^{1/2}$ equals $\varsigma$, where $\bar h$ is the mean of the leaf values. Because every leaf holds $m$ observations, this quantity is the across-unit standard deviation $S_\tau$ of Section~\ref{sec:hyperparameters}, up to its divisor. The median is estimated from 250{,}000 prior draws and checked on an independent set of 250{,}000. Every checked median lies within one percent of its target (Table~\ref{tab:app-partition-omega}). BCF (our version) is matched to the same $\varsigma$ here, so that the comparison concerns the form of its leaf prior. In the forest experiments it keeps its own prior width (Appendix~\ref{app:exp-setup}).

\paragraph{Losses.}
With $\widehat h_\ell$ the posterior mean of $h_\ell$, let $\mathrm{MSE}_1$ and $\mathrm{MSE}_0$ be the means of $(\widehat h_\ell - h_\ell)^2$ over the leaves with and without an effect in one data set. Figure~\ref{fig:partition} reports $\mathrm{MSE}_0^{1/2}$ and $\mathrm{MSE}_1^{1/2}$, averaged over data sets. Two combined losses summarize both kinds of leaves with different weights. The balanced RMSE $\{(\mathrm{MSE}_0+\mathrm{MSE}_1)/2\}^{1/2}$ gives the two kinds equal weight and was fixed before the experiment. The per-leaf RMSE $[\{k\,\mathrm{MSE}_1+(L-k)\mathrm{MSE}_0\}/L]^{1/2}$ gives every leaf equal weight and is therefore also the error per observation. It was defined after theresults were seen. Standard errors are over the 200 data sets, computed separately for each model.

\begin{figure}[!htbp]
\centering
\includegraphics[width=\textwidth]{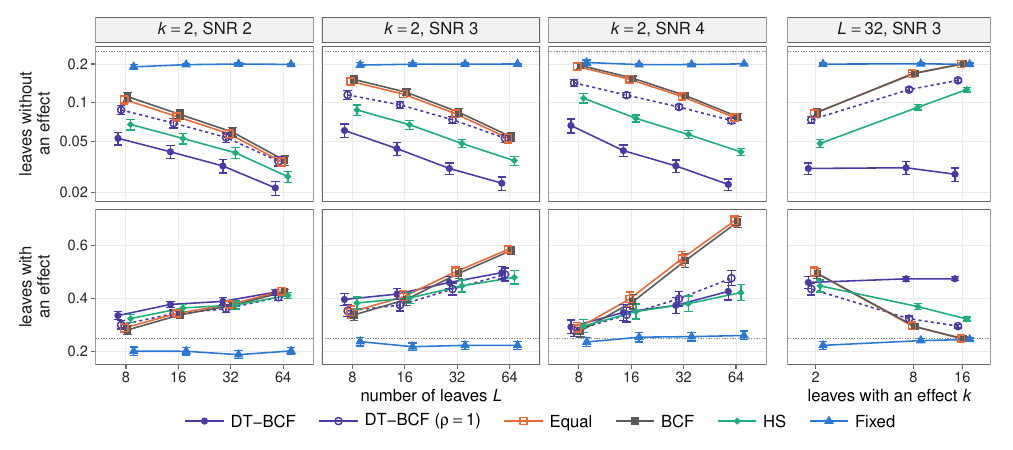}
\caption{Fixed Partition. RMSE of the posterior means of the leaf values on leaves without (top) and with (bottom) an effect, on a log scale, mean and two standard errors over 200 data sets. The first three columns vary $L$ at $k=2$ and signal-to-noise ratios 2, 3 and 4, and the last column varies $k$ at $L=32$ and SNR 3. Dotted, the RMSE $\sigma/m^{1/2}$ of the leaf means.}
\label{fig:app-partition}
\end{figure}

\paragraph{Results.}
Figure~\ref{fig:app-partition} and Table~\ref{tab:app-partition-levels} extend Figure~\ref{fig:partition} to signal-to-noise ratios 2 and 4. On leaves without an effect, \method{} has the smallest RMSE in every condition and HS the second smallest. The difference between the two exceeds two standard errors in every condition. On leaves with an effect, Fixed has the smallest RMSE whenever some leaves have no effect. At $\varsigma=1/2$ its matched leaf variance is four to five times $\sigma^2/m$ (Table~\ref{tab:app-partition-omega}), so its posterior mean keeps about four fifths of every leaf mean. The comparison of \method{} with BCF and Equal on these leaves depends on the signal. At SNR 2, BCF is more accurate than \method{} at $L\le16$, and the two are within two standard errors at $L\ge32$. At SNR 3 and 4, the RMSE of \method{} falls below those of BCF and Equal as $L$ grows, by more than two standard errors from $L=32$. BCF and Equal give every leaf a common variance. As $L$ grows, the leaves without an effect pull this variance toward zero, and the two leaves with an effect are shrunk with them. The shares of \method{} can instead place a large part of the tree's variance on the two leaves with large $|\bar R_\ell|$.

The combined losses order the models differently. The balanced RMSE is smallest for Fixed in every condition except $L=8$ at SNR 4, because $\mathrm{MSE}_1$ is much larger than $\mathrm{MSE}_0$ and Fixed has the smallest $\mathrm{MSE}_1$. The per-leaf RMSE is smallest for \method{}, or within 0.001 of the smallest, in every condition with $k=2$.

When more leaves carry an effect, the ordering on those leaves reverses
(Table~\ref{tab:app-partition-controls}). From $k=8$ at $L=32$, \method{} has the largest RMSE on leaves with an effect and the largest values of both combined losses. Its RMSE on leaves without an effect stays at or below its value at $k=2$. This is consistent with its prior, which at $a=1/2$ favors allocations in which a few leaves receive most of the tree's variance (Section~\ref{sec:hyperparameters}). Changing $\varsigma$ from $1/4$ to $1$ leaves the errors of \method{}, \method{} ($\rho=1$), Equal, BCF and HS nearly unchanged. Fixed instead trades error on the leaves with an effect against error on the others, since its leaf variance is $\omega$ itself.


\begin{table}[!htbp]
\centering
\footnotesize
\setlength{\tabcolsep}{3pt}
\caption{Fixed Partition. Here $k=2$ leaves with an effect and $\varsigma=1/2$. Mean over the 200 data sets of each error of the posterior means of the leaf values, with its standard error in parentheses. Bold, the smallest mean in each row. The balanced RMSE is $\{(\mathrm{MSE}_1+\mathrm{MSE}_0)/2\}^{1/2}$ and was fixed before the experiment. The per-leaf RMSE is $[\{k\,\mathrm{MSE}_1+(L-k)\,\mathrm{MSE}_0\}/L]^{1/2}$ and was defined after the results were seen.}
\label{tab:app-partition-levels}
\vspace{4pt}
\begin{tabular}{rr*{6}{c}}
\toprule
$L$ & SNR & \method{} & \method{} ($\rho=1$) & Equal & BCF & HS & Fixed\\
\midrule
\multicolumn{8}{l}{\emph{RMSE on leaves without an effect}}\\
8 & 2 & \textbf{0.053} (0.003) & 0.088 (0.004) & 0.106 (0.004) & 0.112 (0.004) & 0.068 (0.003) & 0.191 (0.004)\\
 & 3 & \textbf{0.061} (0.004) & 0.115 (0.004) & 0.147 (0.005) & 0.152 (0.005) & 0.088 (0.004) & 0.197 (0.005)\\
 & 4 & \textbf{0.066} (0.004) & 0.143 (0.005) & 0.192 (0.005) & 0.196 (0.005) & 0.109 (0.004) & 0.206 (0.005)\\
16 & 2 & \textbf{0.041} (0.002) & 0.069 (0.003) & 0.078 (0.003) & 0.081 (0.003) & 0.053 (0.003) & 0.198 (0.003)\\
 & 3 & \textbf{0.044} (0.003) & 0.096 (0.003) & 0.117 (0.004) & 0.121 (0.004) & 0.067 (0.003) & 0.200 (0.003)\\
 & 4 & \textbf{0.042} (0.002) & 0.115 (0.003) & 0.152 (0.003) & 0.155 (0.003) & 0.076 (0.002) & 0.198 (0.003)\\
32 & 2 & \textbf{0.032} (0.002) & 0.054 (0.002) & 0.057 (0.002) & 0.059 (0.002) & 0.041 (0.002) & 0.201 (0.002)\\
 & 3 & \textbf{0.031} (0.002) & 0.074 (0.002) & 0.082 (0.003) & 0.084 (0.003) & 0.048 (0.002) & 0.200 (0.002)\\
 & 4 & \textbf{0.032} (0.002) & 0.093 (0.002) & 0.112 (0.003) & 0.114 (0.003) & 0.057 (0.002) & 0.199 (0.002)\\
64 & 2 & \textbf{0.022} (0.001) & 0.035 (0.001) & 0.035 (0.001) & 0.035 (0.001) & 0.027 (0.001) & 0.199 (0.001)\\
 & 3 & \textbf{0.024} (0.001) & 0.053 (0.002) & 0.053 (0.002) & 0.054 (0.002) & 0.035 (0.001) & 0.201 (0.001)\\
 & 4 & \textbf{0.023} (0.001) & 0.073 (0.002) & 0.077 (0.002) & 0.078 (0.002) & 0.041 (0.001) & 0.201 (0.001)\\
\midrule
\multicolumn{8}{l}{\emph{RMSE on leaves with an effect}}\\
8 & 2 & 0.336 (0.007) & 0.301 (0.008) & 0.289 (0.008) & 0.282 (0.008) & 0.325 (0.008) & \textbf{0.202} (0.008)\\
 & 3 & 0.397 (0.011) & 0.354 (0.011) & 0.349 (0.011) & 0.338 (0.011) & 0.383 (0.011) & \textbf{0.238} (0.008)\\
 & 4 & 0.294 (0.013) & 0.279 (0.011) & 0.287 (0.011) & 0.278 (0.011) & 0.296 (0.012) & \textbf{0.236} (0.009)\\
16 & 2 & 0.378 (0.006) & 0.343 (0.007) & 0.344 (0.007) & 0.339 (0.007) & 0.365 (0.006) & \textbf{0.201} (0.007)\\
 & 3 & 0.417 (0.011) & 0.375 (0.011) & 0.406 (0.011) & 0.397 (0.011) & 0.402 (0.011) & \textbf{0.219} (0.007)\\
 & 4 & 0.348 (0.015) & 0.339 (0.014) & 0.398 (0.013) & 0.388 (0.013) & 0.351 (0.014) & \textbf{0.255} (0.009)\\
32 & 2 & 0.389 (0.007) & 0.361 (0.007) & 0.378 (0.006) & 0.374 (0.007) & 0.376 (0.007) & \textbf{0.189} (0.007)\\
 & 3 & 0.461 (0.011) & 0.436 (0.011) & 0.501 (0.009) & 0.495 (0.009) & 0.448 (0.011) & \textbf{0.224} (0.008)\\
 & 4 & 0.374 (0.015) & 0.399 (0.014) & 0.551 (0.013) & 0.543 (0.013) & 0.381 (0.014) & \textbf{0.257} (0.008)\\
64 & 2 & 0.424 (0.005) & 0.405 (0.006) & 0.427 (0.004) & 0.426 (0.004) & 0.411 (0.006) & \textbf{0.203} (0.007)\\
 & 3 & 0.498 (0.012) & 0.491 (0.012) & 0.585 (0.008) & 0.581 (0.008) & 0.480 (0.012) & \textbf{0.224} (0.008)\\
 & 4 & 0.427 (0.016) & 0.477 (0.014) & 0.693 (0.010) & 0.689 (0.010) & 0.423 (0.015) & \textbf{0.261} (0.008)\\
\midrule
\multicolumn{8}{l}{\emph{balanced RMSE}}\\
8 & 2 & 0.243 (0.005) & 0.228 (0.005) & 0.227 (0.005) & 0.225 (0.005) & 0.238 (0.005) & \textbf{0.205} (0.005)\\
 & 3 & 0.287 (0.008) & 0.271 (0.007) & 0.280 (0.006) & 0.275 (0.006) & 0.284 (0.007) & \textbf{0.227} (0.005)\\
 & 4 & \textbf{0.219} (0.009) & 0.233 (0.007) & 0.257 (0.007) & 0.253 (0.006) & 0.232 (0.008) & 0.231 (0.005)\\
16 & 2 & 0.270 (0.004) & 0.251 (0.004) & 0.255 (0.004) & 0.252 (0.004) & 0.262 (0.004) & \textbf{0.206} (0.004)\\
 & 3 & 0.298 (0.007) & 0.280 (0.007) & 0.307 (0.006) & 0.302 (0.006) & 0.292 (0.007) & \textbf{0.215} (0.004)\\
 & 4 & 0.250 (0.010) & 0.260 (0.009) & 0.310 (0.008) & 0.305 (0.008) & 0.258 (0.010) & \textbf{0.236} (0.005)\\
32 & 2 & 0.277 (0.005) & 0.261 (0.005) & 0.273 (0.004) & 0.271 (0.004) & 0.269 (0.005) & \textbf{0.202} (0.004)\\
 & 3 & 0.327 (0.008) & 0.315 (0.007) & 0.362 (0.006) & 0.358 (0.006) & 0.320 (0.008) & \textbf{0.219} (0.004)\\
 & 4 & 0.267 (0.011) & 0.295 (0.009) & 0.402 (0.008) & 0.397 (0.008) & 0.275 (0.010) & \textbf{0.236} (0.005)\\
64 & 2 & 0.301 (0.004) & 0.288 (0.004) & 0.304 (0.002) & 0.303 (0.003) & 0.292 (0.004) & \textbf{0.207} (0.003)\\
 & 3 & 0.353 (0.008) & 0.351 (0.008) & 0.417 (0.005) & 0.414 (0.005) & 0.341 (0.009) & \textbf{0.219} (0.004)\\
 & 4 & 0.303 (0.011) & 0.344 (0.010) & 0.494 (0.007) & 0.492 (0.007) & 0.302 (0.011) & \textbf{0.239} (0.005)\\
\midrule
\multicolumn{8}{l}{\emph{per-leaf RMSE}}\\
8 & 2 & 0.178 (0.004) & \textbf{0.177} (0.003) & 0.184 (0.003) & 0.184 (0.003) & 0.178 (0.003) & 0.200 (0.004)\\
 & 3 & \textbf{0.211} (0.006) & 0.214 (0.005) & 0.231 (0.004) & 0.230 (0.004) & 0.215 (0.005) & 0.215 (0.004)\\
 & 4 & \textbf{0.167} (0.006) & 0.199 (0.005) & 0.233 (0.005) & 0.232 (0.005) & 0.187 (0.006) & 0.222 (0.004)\\
16 & 2 & \textbf{0.142} (0.002) & 0.144 (0.002) & 0.151 (0.002) & 0.151 (0.002) & \textbf{0.142} (0.002) & 0.201 (0.003)\\
 & 3 & \textbf{0.157} (0.004) & 0.170 (0.003) & 0.192 (0.002) & 0.192 (0.002) & 0.162 (0.003) & 0.205 (0.003)\\
 & 4 & \textbf{0.134} (0.005) & 0.170 (0.004) & 0.212 (0.003) & 0.211 (0.003) & 0.150 (0.004) & 0.211 (0.003)\\
32 & 2 & \textbf{0.105} (0.002) & 0.110 (0.001) & 0.116 (0.001) & 0.116 (0.001) & 0.106 (0.002) & 0.202 (0.002)\\
 & 3 & \textbf{0.122} (0.003) & 0.137 (0.002) & 0.156 (0.001) & 0.156 (0.001) & 0.126 (0.002) & 0.204 (0.002)\\
 & 4 & \textbf{0.103} (0.004) & 0.142 (0.002) & 0.184 (0.002) & 0.184 (0.002) & 0.116 (0.003) & 0.205 (0.002)\\
64 & 2 & 0.080 (0.001) & 0.083 (0.001) & 0.086 ($<$0.001) & 0.086 ($<$0.001) & \textbf{0.079} (0.001) & 0.200 (0.001)\\
 & 3 & \textbf{0.093} (0.002) & 0.108 (0.001) & 0.121 (0.001) & 0.121 (0.001) & 0.095 (0.002) & 0.202 (0.001)\\
 & 4 & \textbf{0.082} (0.003) & 0.117 (0.002) & 0.149 (0.001) & 0.149 (0.001) & 0.090 (0.002) & 0.204 (0.001)\\
\bottomrule
\end{tabular}
\end{table}

\begin{table}[!htbp]
\centering
\footnotesize
\setlength{\tabcolsep}{3pt}
\caption{Fixed Partition. Here $L=32$ and SNR 3. The first four rows of each block vary the number $k$ of leaves with an effect at $\varsigma=1/2$, half of them at $+\Delta$ and half at $-\Delta$. The last two rows vary $\varsigma$ at $k=2$, whose row at $\varsigma=1/2$ is shared. At $k=32$ every leaf has an effect, so the error on leaves without an effect and the balanced RMSE are undefined, and the per-leaf RMSE equals the RMSE on leaves with an effect. Entries and bold as in Table~\ref{tab:app-partition-levels}.}
\label{tab:app-partition-controls}
\vspace{4pt}
\begin{tabular}{ll*{6}{c}}
\toprule
$k$ & $\varsigma$ & \method{} & \method{} ($\rho=1$) & Equal & BCF & HS & Fixed\\
\midrule
\multicolumn{8}{l}{\emph{RMSE on leaves without an effect}}\\
2 & 1/2 & \textbf{0.031} (0.002) & 0.074 (0.002) & 0.082 (0.003) & 0.084 (0.003) & 0.048 (0.002) & 0.200 (0.002)\\
8 & 1/2 & \textbf{0.031} (0.002) & 0.127 (0.002) & 0.168 (0.002) & 0.170 (0.002) & 0.092 (0.002) & 0.201 (0.002)\\
16 & 1/2 & \textbf{0.028} (0.002) & 0.150 (0.003) & 0.200 (0.003) & 0.201 (0.003) & 0.127 (0.003) & 0.199 (0.002)\\
32 & 1/2 & -- & -- & -- & -- & -- & --\\
2 & 1/4 & \textbf{0.028} (0.002) & 0.069 (0.002) & 0.077 (0.003) & 0.082 (0.003) & 0.042 (0.002) & 0.127 (0.001)\\
2 & 1 & \textbf{0.033} (0.002) & 0.076 (0.002) & 0.084 (0.003) & 0.085 (0.003) & 0.053 (0.002) & 0.234 (0.002)\\
\midrule
\multicolumn{8}{l}{\emph{RMSE on leaves with an effect}}\\
2 & 1/2 & 0.461 (0.011) & 0.436 (0.011) & 0.501 (0.009) & 0.495 (0.009) & 0.448 (0.011) & \textbf{0.224} (0.008)\\
8 & 1/2 & 0.474 (0.005) & 0.324 (0.005) & 0.297 (0.005) & 0.294 (0.005) & 0.370 (0.005) & \textbf{0.242} (0.004)\\
16 & 1/2 & 0.475 (0.004) & 0.296 (0.003) & 0.249 (0.003) & 0.248 (0.003) & 0.323 (0.004) & \textbf{0.246} (0.003)\\
32 & 1/2 & 0.473 (0.003) & 0.276 (0.002) & \textbf{0.235} (0.002) & \textbf{0.235} (0.002) & 0.285 (0.003) & 0.245 (0.002)\\
2 & 1/4 & 0.472 (0.011) & 0.450 (0.011) & 0.516 (0.009) & 0.502 (0.009) & 0.465 (0.011) & \textbf{0.376} (0.006)\\
2 & 1 & 0.454 (0.012) & 0.431 (0.011) & 0.495 (0.009) & 0.494 (0.009) & 0.437 (0.011) & \textbf{0.211} (0.009)\\
\midrule
\multicolumn{8}{l}{\emph{balanced RMSE}}\\
2 & 1/2 & 0.327 (0.008) & 0.315 (0.007) & 0.362 (0.006) & 0.358 (0.006) & 0.320 (0.008) & \textbf{0.219} (0.004)\\
8 & 1/2 & 0.337 (0.004) & 0.248 (0.003) & 0.245 (0.003) & 0.244 (0.003) & 0.271 (0.004) & \textbf{0.224} (0.002)\\
16 & 1/2 & 0.337 (0.003) & 0.237 (0.002) & 0.228 (0.002) & 0.228 (0.002) & 0.247 (0.002) & \textbf{0.226} (0.002)\\
32 & 1/2 & -- & -- & -- & -- & -- & --\\
2 & 1/4 & 0.335 (0.008) & 0.324 (0.007) & 0.372 (0.006) & 0.363 (0.006) & 0.331 (0.008) & \textbf{0.282} (0.004)\\
2 & 1 & 0.323 (0.008) & 0.312 (0.007) & 0.359 (0.006) & 0.358 (0.006) & 0.313 (0.008) & \textbf{0.232} (0.004)\\
\midrule
\multicolumn{8}{l}{\emph{per-leaf RMSE}}\\
2 & 1/2 & \textbf{0.122} (0.003) & 0.137 (0.002) & 0.156 (0.001) & 0.156 (0.001) & 0.126 (0.002) & 0.204 (0.002)\\
8 & 1/2 & 0.240 (0.003) & \textbf{0.199} (0.002) & 0.212 (0.002) & 0.212 (0.002) & 0.204 (0.002) & 0.214 (0.002)\\
16 & 1/2 & 0.337 (0.003) & 0.237 (0.002) & 0.228 (0.002) & 0.228 (0.002) & 0.247 (0.002) & \textbf{0.226} (0.002)\\
32 & 1/2 & 0.473 (0.003) & 0.276 (0.002) & \textbf{0.235} (0.002) & \textbf{0.235} (0.002) & 0.285 (0.003) & 0.245 (0.002)\\
2 & 1/4 & \textbf{0.123} (0.003) & 0.137 (0.002) & 0.156 (0.001) & 0.156 (0.001) & 0.127 (0.002) & 0.156 (0.001)\\
2 & 1 & \textbf{0.121} (0.003) & 0.137 (0.002) & 0.156 (0.001) & 0.156 (0.001) & 0.125 (0.002) & 0.234 (0.002)\\
\bottomrule
\end{tabular}
\end{table}

\begin{table}[!htbp]
\centering
\footnotesize
\setlength{\tabcolsep}{3pt}
\caption{Fixed Partition. The scaling constant $\omega$ of each model, chosen by prior-predictive simulation so that the median over prior draws of the standard deviation of the $L$ leaf values, with divisor $L$, equals $\varsigma$. Each value is solved from 250{,}000 prior draws and checked on an independent set of 250{,}000. The lower block gives the relative error of the checked median. No data enter the matching. For Fixed, $\omega$ is the leaf variance itself.}
\label{tab:app-partition-omega}
\vspace{4pt}
\begin{tabular}{rr*{6}{c}}
\toprule
$L$ & $\varsigma$ & \method{} & \method{} ($\rho=1$) & Equal & BCF & HS & Fixed\\
\midrule
\multicolumn{8}{l}{\emph{matched $\omega$}}\\
8 & 1/2 & 4.43 & 2.98 & 2.64 & 0.762 & $3.65\cdot10^{-2}$ & 0.315\\
16 & 1/2 & 8.10 & 4.99 & 4.59 & 0.634 & $1.55\cdot10^{-2}$ & 0.279\\
32 & 1/2 & 15.6 & 9.00 & 8.52 & 0.591 & $7.18\cdot10^{-3}$ & 0.264\\
64 & 1/2 & 30.5 & 16.9 & 16.5 & 0.572 & $3.48\cdot10^{-3}$ & 0.257\\
32 & 1/4 & 3.85 & 2.24 & 2.12 & 0.149 & $1.80\cdot10^{-3}$ & $6.59\cdot10^{-2}$\\
32 & 1 & 62.0 & 36.0 & 34.3 & 2.38 & $2.86\cdot10^{-2}$ & 1.05\\
\midrule
\multicolumn{8}{l}{\emph{relative error of the checked median (percent)}}\\
8 & 1/2 & $-$0.66 & $-$0.89 & $+$0.07 & $+$0.22 & $+$0.32 & $-$0.07\\
16 & 1/2 & $+$0.69 & $+$0.69 & $+$0.34 & $+$0.10 & $-$0.09 & $-$0.12\\
32 & 1/2 & $-$0.06 & 0.00 & $+$0.09 & $+$0.31 & $+$0.11 & $+$0.02\\
64 & 1/2 & $+$0.41 & $-$0.27 & $+$0.15 & $+$0.51 & $+$0.56 & $+$0.02\\
32 & 1/4 & $-$0.01 & $-$0.22 & $-$0.01 & $-$0.08 & $-$0.27 & $-$0.08\\
32 & 1 & $+$0.41 & $+$0.08 & $+$0.32 & $+$0.06 & $+$0.12 & $-$0.04\\
\bottomrule
\end{tabular}
\end{table}

\clearpage
\subsection{Forest Simulations}
\label{app:exp-forest}

This subsection gives the data-generating processes of
Section~\ref{sec:exp-forest} and the complete results of the tree-size and
dimension studies, for every model and metric.

\paragraph{Data.}
Each data set has $n=200$ training units and 500 test units drawn from the same process. The covariates are $\bx_i\sim\mathrm{U}(0,1)^p$, the treatment is $A_i\sim\mathrm{Bernoulli}\{e(\bx_i)\}$, and $y_i=\mu(\bx_i)+A_i\tau(\bx_i)+\varepsilon_i$ with $\varepsilon_i\sim\Normal(0,\sigma^2)$. In the two region scenarios, $\mu(\bx)=1.5+2x_1+x_2-1.5x_3+\sin(2\pi x_4)$, $e(\bx)=\operatorname{logit}^{-1}\{-0.4+0.5(1.3x_1-x_2)\}$ and $\sigma=1/2$. The rectangles are $R_1=\{x_1>0.6,\ x_2>0.5\}$, $R_2=\{x_1<0.3,\ x_3>0.7\}$, $R_3=\{x_2<0.25,\ x_4<0.35\}$ and $R_4=\{x_3>0.65,\ x_5<0.4\}$. The 2 Regions scenario sets $\tau(\bx)=2$ on $R_1\cup R_2$. The 4 Regions Graded scenario sets $\tau(\bx)=4\cdot0.6^{r-1}$ on the part of $R_r$ that lies outside $R_1,\dots,R_{r-1}$, for $r=1,\dots,4$. Both set $\tau(\bx)=0$ elsewhere. Smooth Nonlinear sets $\tau=\tau_0$ of Section~\ref{sec:exp-forest} and $\sigma=\sqrt3$. Its prognostic function is linear, $\mu(\bx)=\bx^\top\bm\beta$, with coefficients drawn for each data set that are zero with probability 0.9 and standard normal otherwise. Its propensity score is the probit $e(\bx)=\Phi(-0.5+0.4x_1-0.1x_3+0.3x_5)$, where $\Phi$ is the standard normal distribution function. Smooth Graded sets $\tau(\bx)=1.65^{-G(\bx)}\tau_0(\bx)$, with $G(\bx)=\ind(x_6>1/2)+2\,\ind(x_7>1/2)$, and $\sigma=\sqrt3$. Its $\mu$ and $e$ are those of the region scenarios. Each setting has 40 data sets with seeds 1 to 40. A scenario and a seed give the same data set at every $\eta$, so every model is fitted to the same data sets at every $\eta$. The dimension study at $p=50$ reuses the fits of the tree-size study at $\eta=2$.

\paragraph{Models.}
Both studies fit every model of Appendix~\ref{app:exp-setup}, including the DART versions of \method{}, Equal and HS. The tree-size exponent $\eta$ changes only the split probability of the treatment trees in our sampler. BCF keeps its own split probability $0.25(1+d)^{-3}$, and the loss-based models have no tree prior. Their fits therefore do not depend on $\eta$. BCF is fitted with its package in the tree-size study and at $p\in\{20,50\}$, and in our sampler at $p\ge200$.

\begin{figure}[!htbp]
\centering
\includegraphics[width=\textwidth]{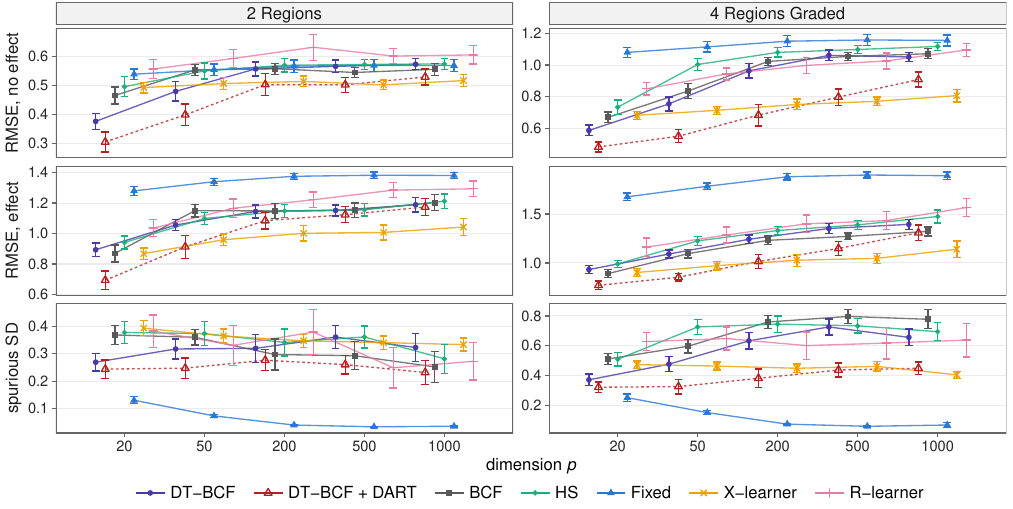}
\caption{Dimension Study. $n=200$ and $\eta=2$, region scenarios. RMSE of the
centered effect on units without (top) and with (middle) an effect, and the
spurious SD (bottom), against the number of covariates $p$, mean and two
standard errors over 40 data sets.}
\label{fig:app-dim-split}
\end{figure}

\paragraph{Errors on Units.}
Figure~\ref{fig:app-dim-split} splits the RMSE of the centered effect in the region scenarios into the units with and without an effect, as Figure~\ref{fig:partition} does for leaves. It also reports the spurious SD, the standard deviation of $\widehat\delta_i$ over the units without an effect. Because $\delta_i$ takes one value on these units, the spurious SD is zero for a model that estimates no heterogeneity among them.

\paragraph{Results.}
Tables~\ref{tab:app-tree-rmse} to~\ref{tab:app-tree-lpd} give the tree-size study, and Tables~\ref{tab:app-dim-rmse} to~\ref{tab:app-dim-lpd} the dimension study. They contain the values behind Table~\ref{tab:forest} and Figure~\ref{fig:experiments} and add the models and metrics that the main text does not show. On the region scenarios, Equal has a smaller mean RMSE than \method{} in all eight settings of the tree-size study, by more than two standard errors in four. Equal + DART has a smaller mean RMSE than \method{} + DART in all eight, by more than two standard errors in two. In the dimension study, Equal has a smaller mean RMSE than \method{} in all ten settings, by more than two standard errors in four. Equal + DART has a smaller mean RMSE than \method{} + DART in eight of the ten, in none by more than two standard errors. Equal + DART also has the smallest FPR in most region settings of the tree-size study. A DART version has the largest lpd in every region setting, Equal + DART in all of those of the tree-size study. In the dimension study, the X-learner has the smallest RMSE on 2 Regions from $p=200$ and on 4 Regions Graded from $p=500$.

On the smooth scenarios, BCF has the smallest RMSE in every setting except Smooth Nonlinear at $p=50$, where Equal + DART is smallest at every $\eta$. The coverage of BCF is also the closest to 0.90 in most of these settings. The lpd of the models other than Fixed spans at most 0.09 in any smooth setting.

Coverage lies below 0.90 for most models and settings. Fixed has the lowest coverage and the smallest lpd in every setting.

\paragraph{Equal Division on Region Scenarios.}
Equal is the limit $a\to\infty$ of the weighted allocation at $\rho=1$, so
Equal, \method{} ($\rho=1$) and \method{} belong to one family. Each of the three has a smaller mean RMSE than BCF and HS on both region scenarios at every $\eta$. The three give the tree-level total variance $V_j$ the same form and differ in how unequally a tree divides it among its leaves. The posterior mean of $(\sum_\ell\phi_{j\ell}^2)^{-1}$, averaged over the treatment trees and the data sets of a setting, is the realized counterpart of $A_{\mathrm{eff}}(T_j)$ in Section~\ref{sec:hyperparameters}. It lies between 1.27 and 1.38 for \method{}, between 1.61 and 2.42 for \method{} ($\rho=1$), and between 2.50 and 4.67 for Equal, for which it equals the number of leaves. Without DART, the mean RMSE on the region scenarios follows the reverse order of this quantity, largest for \method{} and smallest for Equal, in seven of the eight settings of the tree-size study.

We attribute this ordering to three features of the designs, the small treatment trees, the shape of the regions and the independence of the covariates. Each rectangle is defined by two of the independent uniform covariates, so both leaves of any single split contain units with and without an effect. The split at $x_1=0.6$ in 2 Regions, for instance, leaves an effect on half of the units on one side and on 15 percent of them on the other. Because the trees are this small, their leaves typically contain units of both kinds. Such leaves call for comparable shares, whereas under \method{} the shares of a two-leaf tree follow $\Beta(1/4,1/4)$ instead, a prior that favors giving nearly all of $V_j$ to one of the two leaves. The fixed partition of Appendix~\ref{app:exp-partition}, where every leaf either carries an effect or does not, shows when this concentration helps. There \method{} has the smallest RMSE on the leaves without an effect in every condition. With 8 or more of the 32 leaves carrying an effect, its RMSE on those leaves is the largest. When all 32 leaves carry an effect, Equal and BCF have the smallest (Table~\ref{tab:app-partition-controls}). The concentrated shares also affect detection. In all eight region settings of the tree-size study, \method{} has a larger mean TPR than Equal, by more than two standard errors in one, and a larger mean FPR. The forest fits cannot separate the shape of the regions from the independence of the covariates. With correlated covariates, or with regions defined by a single covariate, one split could separate most units with an effect from those without, as the leaves of the fixed partition do.

\begin{figure}[!htbp]
\centering
\includegraphics[width=\textwidth]{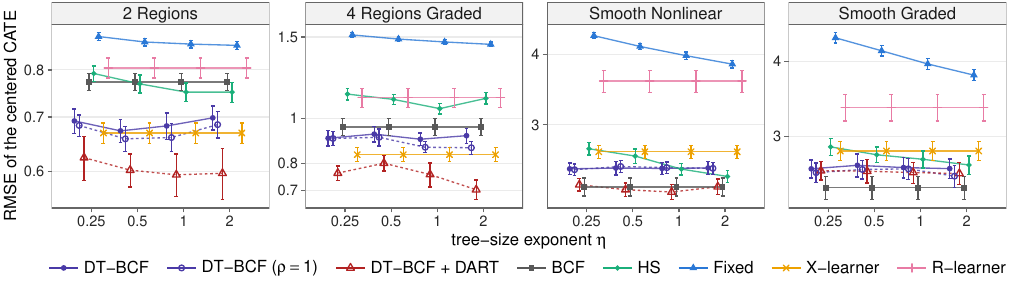}
\caption{Tree-size Study. $n=200$ and $p=50$. RMSE of the centered effect
against the tree-size exponent $\eta$, where smaller $\eta$ favors larger
treatment trees, mean and two standard errors over 40 data sets, on a log
scale. BCF and the loss-based models do not depend on $\eta$.}
\label{fig:app-tree-rmse}
\end{figure}

\begin{table}[!htbp]
\centering
\footnotesize
\setlength{\tabcolsep}{4pt}
\caption{Tree-size Study. $n=200$ and $p=50$. RMSE of the centered effect by scenario and the tree-size exponent $\eta$. Mean over 40 data sets with its standard error in parentheses. Bold, the smallest mean in each column of a scenario. BCF and the loss-based models do not depend on $\eta$, so their entries repeat the same fits.}
\label{tab:app-tree-rmse}
\vspace{4pt}
\begin{tabular}{l*{4}{c}}
\toprule
& \multicolumn{4}{c}{$\eta$}\\
\cmidrule(lr){2-5}
Model & 0.25 & 0.5 & 1 & 2\\
\midrule
\multicolumn{5}{l}{\emph{2 Regions}}\\
\method{} & 0.69 (0.01) & 0.67 (0.01) & 0.68 (0.01) & 0.70 (0.01)\\
\method{} ($\rho=1$) & 0.68 (0.01) & 0.66 (0.01) & 0.66 (0.01) & 0.69 (0.01)\\
\method{} + DART & 0.62 (0.02) & 0.60 (0.01) & 0.59 (0.02) & 0.60 (0.02)\\
Equal & 0.66 (0.01) & 0.65 (0.01) & 0.66 (0.01) & 0.66 (0.01)\\
Equal + DART & \textbf{0.55} (0.02) & \textbf{0.57} (0.02) & \textbf{0.53} (0.02) & \textbf{0.55} (0.02)\\
BCF & 0.77 (0.01) & 0.77 (0.01) & 0.77 (0.01) & 0.77 (0.01)\\
HS & 0.79 (0.01) & 0.77 (0.01) & 0.75 (0.01) & 0.75 (0.01)\\
HS + DART & 0.76 (0.01) & 0.73 (0.01) & 0.67 (0.02) & 0.64 (0.02)\\
Fixed & 0.88 ($<$0.01) & 0.87 ($<$0.01) & 0.86 ($<$0.01) & 0.86 ($<$0.01)\\
X-learner & 0.67 (0.01) & 0.67 (0.01) & 0.67 (0.01) & 0.67 (0.01)\\
R-learner & 0.80 (0.01) & 0.80 (0.01) & 0.80 (0.01) & 0.80 (0.01)\\
\midrule
\multicolumn{5}{l}{\emph{4 Regions Graded}}\\
\method{} & 0.91 (0.02) & 0.93 (0.02) & 0.90 (0.01) & 0.92 (0.02)\\
\method{} ($\rho=1$) & 0.91 (0.02) & 0.91 (0.02) & 0.87 (0.01) & 0.87 (0.01)\\
\method{} + DART & 0.76 (0.01) & 0.80 (0.02) & 0.76 (0.02) & 0.70 (0.02)\\
Equal & 0.88 (0.02) & 0.87 (0.02) & 0.88 (0.01) & 0.86 (0.01)\\
Equal + DART & \textbf{0.75} (0.02) & \textbf{0.76} (0.02) & \textbf{0.72} (0.01) & \textbf{0.69} (0.02)\\
BCF & 0.96 (0.02) & 0.96 (0.02) & 0.96 (0.02) & 0.96 (0.02)\\
HS & 1.13 (0.02) & 1.10 (0.01) & 1.05 (0.01) & 1.11 (0.02)\\
HS + DART & 0.93 (0.02) & 0.89 (0.02) & 0.83 (0.02) & 0.78 (0.02)\\
Fixed & 1.52 (0.01) & 1.48 (0.01) & 1.46 (0.01) & 1.45 (0.01)\\
X-learner & 0.84 (0.01) & 0.84 (0.01) & 0.84 (0.01) & 0.84 (0.01)\\
R-learner & 1.11 (0.03) & 1.11 (0.03) & 1.11 (0.03) & 1.11 (0.03)\\
\midrule
\multicolumn{5}{l}{\emph{Smooth Nonlinear}}\\
\method{} & 2.50 (0.03) & 2.51 (0.04) & 2.51 (0.03) & 2.51 (0.03)\\
\method{} ($\rho=1$) & 2.50 (0.03) & 2.53 (0.03) & 2.52 (0.03) & 2.51 (0.03)\\
\method{} + DART & 2.35 (0.03) & 2.30 (0.03) & 2.28 (0.04) & 2.33 (0.04)\\
Equal & 2.49 (0.03) & 2.50 (0.03) & 2.49 (0.03) & 2.53 (0.03)\\
Equal + DART & \textbf{2.24} (0.04) & \textbf{2.27} (0.04) & \textbf{2.24} (0.04) & \textbf{2.29} (0.03)\\
BCF & 2.33 (0.04) & 2.33 (0.04) & 2.33 (0.04) & 2.33 (0.04)\\
HS & 2.72 (0.03) & 2.64 (0.04) & 2.50 (0.03) & 2.43 (0.03)\\
HS + DART & 2.66 (0.04) & 2.51 (0.04) & 2.36 (0.03) & 2.33 (0.04)\\
Fixed & 4.32 (0.03) & 4.14 (0.03) & 3.98 (0.03) & 3.85 (0.03)\\
X-learner & 2.69 (0.04) & 2.69 (0.04) & 2.69 (0.04) & 2.69 (0.04)\\
R-learner & 3.59 (0.08) & 3.59 (0.08) & 3.59 (0.08) & 3.59 (0.08)\\
\midrule
\multicolumn{5}{l}{\emph{Smooth Graded}}\\
\method{} & 2.65 (0.05) & 2.69 (0.06) & 2.66 (0.05) & 2.65 (0.05)\\
\method{} ($\rho=1$) & 2.61 (0.05) & 2.64 (0.05) & 2.65 (0.05) & 2.58 (0.05)\\
\method{} + DART & 2.63 (0.05) & 2.63 (0.06) & 2.61 (0.05) & 2.60 (0.06)\\
Equal & 2.59 (0.05) & 2.64 (0.05) & 2.57 (0.05) & 2.58 (0.04)\\
Equal + DART & 2.60 (0.06) & 2.63 (0.06) & 2.62 (0.06) & 2.60 (0.05)\\
BCF & \textbf{2.46} (0.06) & \textbf{2.46} (0.06) & \textbf{2.46} (0.06) & \textbf{2.46} (0.06)\\
HS & 2.88 (0.05) & 2.80 (0.04) & 2.75 (0.05) & 2.69 (0.05)\\
HS + DART & 2.87 (0.05) & 2.75 (0.05) & 2.66 (0.04) & 2.64 (0.05)\\
Fixed & 4.38 (0.04) & 4.17 (0.04) & 3.96 (0.04) & 3.79 (0.04)\\
X-learner & 2.84 (0.05) & 2.84 (0.05) & 2.84 (0.05) & 2.84 (0.05)\\
R-learner & 3.35 (0.08) & 3.35 (0.08) & 3.35 (0.08) & 3.35 (0.08)\\
\bottomrule
\end{tabular}
\end{table}

\begin{table}[!htbp]
\centering
\footnotesize
\setlength{\tabcolsep}{4pt}
\caption{Tree-size Study. $n=200$ and $p=50$. Coverage of the 90\% intervals for the centered effect, averaged over units. Entries as in Table~\ref{tab:app-tree-rmse}. Bold, the coverage closest to 0.90 in each column of a scenario.}
\label{tab:app-tree-cov}
\vspace{4pt}
\begin{tabular}{l*{4}{c}}
\toprule
& \multicolumn{4}{c}{$\eta$}\\
\cmidrule(lr){2-5}
Model & 0.25 & 0.5 & 1 & 2\\
\midrule
\multicolumn{5}{l}{\emph{2 Regions}}\\
\method{} & 0.77 (0.03) & 0.80 (0.02) & 0.79 (0.03) & 0.75 (0.02)\\
\method{} ($\rho=1$) & 0.80 (0.02) & 0.84 (0.02) & 0.83 (0.02) & 0.77 (0.03)\\
\method{} + DART & 0.77 (0.04) & 0.84 (0.02) & 0.83 (0.02) & 0.78 (0.03)\\
Equal & 0.85 (0.01) & 0.85 (0.02) & 0.82 (0.02) & 0.80 (0.02)\\
Equal + DART & \textbf{0.90} (0.01) & \textbf{0.90} (0.01) & \textbf{0.90} (0.01) & \textbf{0.85} (0.02)\\
BCF & 0.52 (0.03) & 0.52 (0.03) & 0.52 (0.03) & 0.52 (0.03)\\
HS & 0.46 (0.03) & 0.58 (0.03) & 0.66 (0.03) & 0.68 (0.03)\\
HS + DART & 0.45 (0.05) & 0.61 (0.04) & 0.72 (0.03) & 0.76 (0.03)\\
Fixed & 0.00 ($<$0.01) & 0.07 (0.02) & 0.11 (0.02) & 0.12 (0.02)\\
\midrule
\multicolumn{5}{l}{\emph{4 Regions Graded}}\\
\method{} & 0.76 (0.02) & 0.79 (0.01) & 0.78 (0.02) & 0.73 (0.02)\\
\method{} ($\rho=1$) & 0.81 (0.01) & 0.83 (0.02) & 0.81 (0.01) & 0.77 (0.02)\\
\method{} + DART & 0.75 (0.02) & 0.79 (0.02) & 0.77 (0.02) & 0.72 (0.02)\\
Equal & \textbf{0.85} (0.01) & 0.84 (0.01) & \textbf{0.82} (0.01) & 0.77 (0.02)\\
Equal + DART & \textbf{0.85} (0.01) & \textbf{0.87} (0.01) & \textbf{0.82} (0.02) & \textbf{0.78} (0.02)\\
BCF & 0.73 (0.02) & 0.73 (0.02) & 0.73 (0.02) & 0.73 (0.02)\\
HS & 0.57 (0.02) & 0.66 (0.02) & 0.71 (0.02) & 0.65 (0.02)\\
HS + DART & 0.55 (0.03) & 0.63 (0.03) & 0.71 (0.02) & 0.69 (0.02)\\
Fixed & 0.14 (0.01) & 0.15 ($<$0.01) & 0.15 (0.01) & 0.16 (0.01)\\
\midrule
\multicolumn{5}{l}{\emph{Smooth Nonlinear}}\\
\method{} & 0.79 (0.01) & 0.81 (0.01) & 0.81 (0.01) & 0.78 (0.01)\\
\method{} ($\rho=1$) & 0.82 (0.01) & 0.83 (0.01) & 0.81 (0.01) & 0.78 (0.01)\\
\method{} + DART & 0.79 (0.01) & 0.82 (0.01) & 0.81 (0.01) & 0.77 (0.01)\\
Equal & 0.84 (0.01) & 0.84 (0.01) & 0.83 (0.01) & 0.77 (0.01)\\
Equal + DART & 0.84 (0.01) & 0.85 (0.01) & 0.83 (0.01) & 0.79 (0.01)\\
BCF & \textbf{0.87} (0.01) & \textbf{0.87} (0.01) & \textbf{0.87} (0.01) & \textbf{0.87} (0.01)\\
HS & 0.72 (0.01) & 0.76 (0.01) & 0.79 (0.01) & 0.78 (0.01)\\
HS + DART & 0.71 (0.01) & 0.77 (0.01) & 0.78 (0.01) & 0.76 (0.01)\\
Fixed & 0.30 ($<$0.01) & 0.37 (0.01) & 0.40 (0.01) & 0.40 (0.01)\\
\midrule
\multicolumn{5}{l}{\emph{Smooth Graded}}\\
\method{} & 0.83 (0.01) & 0.84 (0.01) & 0.84 (0.01) & 0.81 (0.01)\\
\method{} ($\rho=1$) & 0.85 (0.01) & 0.85 (0.01) & 0.85 (0.01) & 0.82 (0.01)\\
\method{} + DART & 0.79 (0.01) & 0.80 (0.01) & 0.80 (0.01) & 0.77 (0.01)\\
Equal & \textbf{0.86} (0.01) & \textbf{0.87} (0.01) & \textbf{0.86} (0.01) & 0.83 (0.01)\\
Equal + DART & 0.82 (0.01) & 0.84 (0.01) & 0.81 (0.01) & 0.78 (0.01)\\
BCF & \textbf{0.86} (0.01) & 0.86 (0.01) & \textbf{0.86} (0.01) & \textbf{0.86} (0.01)\\
HS & 0.74 (0.01) & 0.80 (0.01) & 0.80 (0.01) & 0.79 (0.01)\\
HS + DART & 0.71 (0.02) & 0.78 (0.01) & 0.78 (0.01) & 0.75 (0.02)\\
Fixed & 0.30 (0.01) & 0.40 (0.01) & 0.45 (0.01) & 0.48 (0.01)\\
\bottomrule
\end{tabular}
\end{table}

\begin{table}[!htbp]
\centering
\footnotesize
\setlength{\tabcolsep}{4pt}
\caption{Tree-size Study. $n=200$ and $p=50$. Average length of the 90\% intervals for the centered effect. Entries as in Table~\ref{tab:app-tree-rmse}. No entry is bold, because Fixed has the shortest intervals in every column together with the lowest coverage (Table~\ref{tab:app-tree-cov}), so length is read together with coverage.}
\label{tab:app-tree-len}
\vspace{4pt}
\begin{tabular}{l*{4}{c}}
\toprule
& \multicolumn{4}{c}{$\eta$}\\
\cmidrule(lr){2-5}
Model & 0.25 & 0.5 & 1 & 2\\
\midrule
\multicolumn{5}{l}{\emph{2 Regions}}\\
\method{} & 1.34 (0.03) & 1.39 (0.03) & 1.35 (0.03) & 1.30 (0.02)\\
\method{} ($\rho=1$) & 1.43 (0.03) & 1.50 (0.03) & 1.41 (0.03) & 1.29 (0.03)\\
\method{} + DART & 1.23 (0.04) & 1.35 (0.03) & 1.31 (0.03) & 1.16 (0.03)\\
Equal & 1.50 (0.03) & 1.49 (0.03) & 1.39 (0.03) & 1.28 (0.03)\\
Equal + DART & 1.36 (0.04) & 1.41 (0.03) & 1.34 (0.03) & 1.21 (0.03)\\
BCF & 1.11 (0.06) & 1.11 (0.06) & 1.11 (0.06) & 1.11 (0.06)\\
HS & 1.06 (0.04) & 1.22 (0.03) & 1.32 (0.04) & 1.33 (0.03)\\
HS + DART & 1.03 (0.05) & 1.23 (0.04) & 1.27 (0.04) & 1.25 (0.04)\\
Fixed & 0.69 (0.01) & 0.88 (0.01) & 0.90 (0.01) & 0.90 (0.01)\\
\midrule
\multicolumn{5}{l}{\emph{4 Regions Graded}}\\
\method{} & 2.17 (0.05) & 2.24 (0.06) & 2.19 (0.05) & 2.04 (0.05)\\
\method{} ($\rho=1$) & 2.29 (0.06) & 2.37 (0.06) & 2.16 (0.05) & 1.97 (0.04)\\
\method{} + DART & 1.65 (0.06) & 1.92 (0.06) & 1.71 (0.06) & 1.46 (0.06)\\
Equal & 2.31 (0.05) & 2.31 (0.06) & 2.23 (0.06) & 2.00 (0.04)\\
Equal + DART & 1.90 (0.07) & 2.02 (0.05) & 1.78 (0.05) & 1.51 (0.04)\\
BCF & 2.10 (0.06) & 2.10 (0.06) & 2.10 (0.06) & 2.10 (0.06)\\
HS & 2.06 (0.04) & 2.33 (0.05) & 2.39 (0.05) & 2.23 (0.04)\\
HS + DART & 1.84 (0.06) & 1.95 (0.06) & 1.86 (0.06) & 1.61 (0.06)\\
Fixed & 1.02 (0.01) & 1.26 (0.01) & 1.33 (0.02) & 1.33 (0.02)\\
\midrule
\multicolumn{5}{l}{\emph{Smooth Nonlinear}}\\
\method{} & 6.50 (0.11) & 6.77 (0.11) & 6.71 (0.10) & 6.27 (0.09)\\
\method{} ($\rho=1$) & 6.90 (0.12) & 7.07 (0.11) & 6.80 (0.11) & 6.30 (0.10)\\
\method{} + DART & 5.92 (0.08) & 6.28 (0.10) & 6.03 (0.08) & 5.63 (0.09)\\
Equal & 7.09 (0.11) & 7.10 (0.11) & 6.94 (0.10) & 6.23 (0.09)\\
Equal + DART & 6.34 (0.10) & 6.58 (0.09) & 6.27 (0.10) & 5.85 (0.09)\\
BCF & 7.04 (0.10) & 7.04 (0.10) & 7.04 (0.10) & 7.04 (0.10)\\
HS & 6.17 (0.10) & 6.61 (0.12) & 6.53 (0.11) & 6.26 (0.09)\\
HS + DART & 5.88 (0.10) & 6.20 (0.10) & 5.98 (0.10) & 5.64 (0.09)\\
Fixed & 3.52 (0.04) & 4.25 (0.04) & 4.41 (0.04) & 4.39 (0.04)\\
\midrule
\multicolumn{5}{l}{\emph{Smooth Graded}}\\
\method{} & 6.77 (0.15) & 6.96 (0.14) & 6.89 (0.15) & 6.38 (0.12)\\
\method{} ($\rho=1$) & 6.96 (0.14) & 7.03 (0.14) & 6.92 (0.14) & 6.42 (0.14)\\
\method{} + DART & 6.08 (0.15) & 6.34 (0.15) & 6.15 (0.15) & 5.66 (0.15)\\
Equal & 7.03 (0.13) & 7.30 (0.14) & 6.95 (0.13) & 6.43 (0.14)\\
Equal + DART & 6.40 (0.15) & 6.56 (0.14) & 6.15 (0.14) & 5.85 (0.16)\\
BCF & 6.99 (0.14) & 6.99 (0.14) & 6.99 (0.14) & 6.99 (0.14)\\
HS & 6.20 (0.15) & 6.86 (0.15) & 6.58 (0.16) & 6.37 (0.14)\\
HS + DART & 5.61 (0.19) & 6.24 (0.16) & 6.13 (0.14) & 5.59 (0.16)\\
Fixed & 3.79 (0.05) & 4.51 (0.06) & 4.64 (0.06) & 4.57 (0.06)\\
\bottomrule
\end{tabular}
\end{table}

\begin{table}[!htbp]
\centering
\footnotesize
\setlength{\tabcolsep}{4pt}
\caption{Tree-size Study. $n=200$ and $p=50$. False positive rate, the share of units without an effect whose 90\% interval for $\tau(\bx_i)$ excludes zero. Entries as in Table~\ref{tab:app-tree-rmse}. Bold, the smallest rate in each column of a scenario.}
\label{tab:app-tree-fpr}
\vspace{4pt}
\begin{tabular}{l*{4}{c}}
\toprule
& \multicolumn{4}{c}{$\eta$}\\
\cmidrule(lr){2-5}
Model & 0.25 & 0.5 & 1 & 2\\
\midrule
\multicolumn{5}{l}{\emph{2 Regions}}\\
\method{} & 0.15 (0.03) & 0.11 (0.02) & 0.12 (0.03) & 0.17 (0.03)\\
\method{} ($\rho=1$) & 0.12 (0.02) & 0.08 (0.02) & 0.10 (0.03) & 0.16 (0.03)\\
\method{} + DART & 0.15 (0.03) & 0.10 (0.03) & 0.11 (0.03) & 0.16 (0.04)\\
Equal & 0.08 (0.02) & 0.08 (0.02) & 0.09 (0.02) & 0.12 (0.03)\\
Equal + DART & \textbf{0.05} (0.01) & \textbf{0.05} (0.02) & \textbf{0.04} (0.01) & \textbf{0.09} (0.02)\\
BCF & 0.33 (0.03) & 0.33 (0.03) & 0.33 (0.03) & 0.33 (0.03)\\
HS & 0.34 (0.04) & 0.26 (0.04) & 0.21 (0.03) & 0.22 (0.03)\\
HS + DART & 0.37 (0.05) & 0.25 (0.05) & 0.19 (0.04) & 0.15 (0.03)\\
Fixed & 0.87 (0.05) & 0.63 (0.06) & 0.59 (0.06) & 0.60 (0.06)\\
\midrule
\multicolumn{5}{l}{\emph{4 Regions Graded}}\\
\method{} & 0.20 (0.03) & 0.15 (0.03) & 0.16 (0.03) & 0.24 (0.03)\\
\method{} ($\rho=1$) & 0.13 (0.03) & 0.10 (0.03) & \textbf{0.12} (0.02) & 0.17 (0.03)\\
\method{} + DART & 0.19 (0.04) & 0.16 (0.04) & 0.18 (0.04) & 0.21 (0.04)\\
Equal & \textbf{0.09} (0.02) & 0.10 (0.03) & \textbf{0.12} (0.03) & 0.17 (0.03)\\
Equal + DART & 0.10 (0.03) & \textbf{0.09} (0.02) & 0.13 (0.03) & \textbf{0.14} (0.03)\\
BCF & 0.23 (0.02) & 0.23 (0.02) & 0.23 (0.02) & 0.23 (0.02)\\
HS & 0.41 (0.03) & 0.30 (0.02) & 0.26 (0.02) & 0.32 (0.02)\\
HS + DART & 0.44 (0.05) & 0.32 (0.05) & 0.24 (0.04) & 0.23 (0.04)\\
Fixed & 1.00 ($<$0.01) & 0.98 (0.01) & 0.96 (0.02) & 0.94 (0.02)\\
\bottomrule
\end{tabular}
\end{table}

\begin{table}[!htbp]
\centering
\footnotesize
\setlength{\tabcolsep}{4pt}
\caption{Tree-size Study. $n=200$ and $p=50$. True positive rate, the share of units with an effect whose 90\% interval for $\tau(\bx_i)$ excludes zero. Entries as in Table~\ref{tab:app-tree-rmse}. No entry is bold, because Fixed has the largest rate in every column together with the largest FPR (Table~\ref{tab:app-tree-fpr}), so the TPR is read together with the FPR.}
\label{tab:app-tree-tpr}
\vspace{4pt}
\begin{tabular}{l*{4}{c}}
\toprule
& \multicolumn{4}{c}{$\eta$}\\
\cmidrule(lr){2-5}
Model & 0.25 & 0.5 & 1 & 2\\
\midrule
\multicolumn{5}{l}{\emph{2 Regions}}\\
\method{} & 0.69 (0.02) & 0.65 (0.02) & 0.66 (0.02) & 0.69 (0.02)\\
\method{} ($\rho=1$) & 0.63 (0.02) & 0.62 (0.02) & 0.64 (0.02) & 0.69 (0.02)\\
\method{} + DART & 0.73 (0.02) & 0.73 (0.02) & 0.73 (0.02) & 0.77 (0.02)\\
Equal & 0.63 (0.02) & 0.63 (0.02) & 0.65 (0.02) & 0.68 (0.02)\\
Equal + DART & 0.69 (0.02) & 0.67 (0.02) & 0.74 (0.02) & 0.77 (0.02)\\
BCF & 0.74 (0.01) & 0.74 (0.01) & 0.74 (0.01) & 0.74 (0.01)\\
HS & 0.74 (0.02) & 0.72 (0.02) & 0.69 (0.03) & 0.70 (0.02)\\
HS + DART & 0.82 (0.02) & 0.72 (0.02) & 0.74 (0.02) & 0.73 (0.02)\\
Fixed & 0.92 (0.03) & 0.82 (0.04) & 0.81 (0.04) & 0.83 (0.04)\\
\midrule
\multicolumn{5}{l}{\emph{4 Regions Graded}}\\
\method{} & 0.61 (0.02) & 0.57 (0.02) & 0.57 (0.02) & 0.63 (0.02)\\
\method{} ($\rho=1$) & 0.54 (0.02) & 0.52 (0.02) & 0.55 (0.02) & 0.59 (0.02)\\
\method{} + DART & 0.68 (0.02) & 0.62 (0.02) & 0.67 (0.02) & 0.71 (0.02)\\
Equal & 0.51 (0.02) & 0.52 (0.02) & 0.53 (0.02) & 0.59 (0.02)\\
Equal + DART & 0.60 (0.02) & 0.59 (0.02) & 0.63 (0.02) & 0.66 (0.02)\\
BCF & 0.67 (0.01) & 0.67 (0.01) & 0.67 (0.01) & 0.67 (0.01)\\
HS & 0.65 (0.02) & 0.62 (0.02) & 0.61 (0.02) & 0.63 (0.02)\\
HS + DART & 0.76 (0.03) & 0.70 (0.03) & 0.69 (0.02) & 0.70 (0.02)\\
Fixed & 1.00 ($<$0.01) & 1.00 ($<$0.01) & 0.99 (0.01) & 0.98 (0.01)\\
\bottomrule
\end{tabular}
\end{table}

\begin{table}[!htbp]
\centering
\footnotesize
\setlength{\tabcolsep}{4pt}
\caption{Tree-size Study. $n=200$ and $p=50$. Log predictive density of 500 test units, averaged over the units. Entries as in Table~\ref{tab:app-tree-rmse}. Bold, the largest mean in each column of a scenario.}
\label{tab:app-tree-lpd}
\vspace{4pt}
\begin{tabular}{l*{4}{c}}
\toprule
& \multicolumn{4}{c}{$\eta$}\\
\cmidrule(lr){2-5}
Model & 0.25 & 0.5 & 1 & 2\\
\midrule
\multicolumn{5}{l}{\emph{2 Regions}}\\
\method{} & $-$1.186 (0.007) & $-$1.175 (0.007) & $-$1.178 (0.007) & $-$1.188 (0.007)\\
\method{} ($\rho=1$) & $-$1.181 (0.008) & $-$1.169 (0.007) & $-$1.167 (0.008) & $-$1.179 (0.008)\\
\method{} + DART & $-$1.005 (0.011) & $-$0.994 (0.008) & $-$0.990 (0.009) & $-$0.994 (0.012)\\
Equal & $-$1.169 (0.008) & $-$1.166 (0.007) & $-$1.164 (0.007) & $-$1.166 (0.008)\\
Equal + DART & \textbf{$-$0.967} (0.010) & \textbf{$-$0.976} (0.009) & \textbf{$-$0.960} (0.010) & \textbf{$-$0.971} (0.010)\\
BCF & $-$1.203 (0.009) & $-$1.203 (0.009) & $-$1.203 (0.009) & $-$1.203 (0.009)\\
HS & $-$1.227 (0.007) & $-$1.216 (0.007) & $-$1.211 (0.007) & $-$1.212 (0.007)\\
HS + DART & $-$1.075 (0.010) & $-$1.058 (0.009) & $-$1.038 (0.011) & $-$1.015 (0.011)\\
Fixed & $-$1.268 (0.006) & $-$1.264 (0.006) & $-$1.263 (0.006) & $-$1.262 (0.006)\\
\midrule
\multicolumn{5}{l}{\emph{4 Regions Graded}}\\
\method{} & $-$1.293 (0.011) & $-$1.294 (0.010) & $-$1.291 (0.009) & $-$1.298 (0.009)\\
\method{} ($\rho=1$) & $-$1.274 (0.008) & $-$1.276 (0.008) & $-$1.261 (0.007) & $-$1.267 (0.007)\\
\method{} + DART & $-$1.070 (0.008) & $-$1.078 (0.007) & $-$1.058 (0.009) & $-$1.053 (0.009)\\
Equal & $-$1.267 (0.008) & $-$1.259 (0.006) & $-$1.263 (0.007) & $-$1.264 (0.007)\\
Equal + DART & \textbf{$-$1.051} (0.009) & \textbf{$-$1.062} (0.009) & \textbf{$-$1.045} (0.008) & \textbf{$-$1.043} (0.009)\\
BCF & $-$1.297 (0.010) & $-$1.297 (0.010) & $-$1.297 (0.010) & $-$1.297 (0.010)\\
HS & $-$1.404 (0.010) & $-$1.395 (0.010) & $-$1.367 (0.008) & $-$1.402 (0.009)\\
HS + DART & $-$1.156 (0.012) & $-$1.138 (0.013) & $-$1.105 (0.012) & $-$1.090 (0.012)\\
Fixed & $-$1.577 (0.006) & $-$1.570 (0.006) & $-$1.564 (0.006) & $-$1.559 (0.006)\\
\midrule
\multicolumn{5}{l}{\emph{Smooth Nonlinear}}\\
\method{} & $-$2.339 (0.006) & $-$2.337 (0.005) & $-$2.340 (0.006) & $-$2.338 (0.006)\\
\method{} ($\rho=1$) & $-$2.339 (0.005) & $-$2.342 (0.005) & $-$2.334 (0.005) & $-$2.337 (0.006)\\
\method{} + DART & $-$2.304 (0.006) & $-$2.297 (0.006) & $-$2.293 (0.005) & $-$2.301 (0.006)\\
Equal & $-$2.329 (0.006) & $-$2.335 (0.005) & $-$2.333 (0.006) & $-$2.338 (0.006)\\
Equal + DART & \textbf{$-$2.286} (0.007) & \textbf{$-$2.290} (0.005) & \textbf{$-$2.286} (0.006) & \textbf{$-$2.295} (0.006)\\
BCF & $-$2.304 (0.008) & $-$2.304 (0.008) & $-$2.304 (0.008) & $-$2.304 (0.008)\\
HS & $-$2.369 (0.007) & $-$2.357 (0.006) & $-$2.336 (0.006) & $-$2.325 (0.005)\\
HS + DART & $-$2.353 (0.007) & $-$2.327 (0.005) & $-$2.306 (0.005) & $-$2.297 (0.007)\\
Fixed & $-$2.584 (0.004) & $-$2.565 (0.005) & $-$2.547 (0.004) & $-$2.531 (0.005)\\
\midrule
\multicolumn{5}{l}{\emph{Smooth Graded}}\\
\method{} & $-$2.362 (0.008) & $-$2.364 (0.009) & $-$2.361 (0.007) & $-$2.366 (0.008)\\
\method{} ($\rho=1$) & $-$2.349 (0.008) & $-$2.344 (0.006) & $-$2.350 (0.009) & $-$2.350 (0.009)\\
\method{} + DART & $-$2.337 (0.008) & $-$2.333 (0.009) & $-$2.330 (0.010) & $-$2.334 (0.009)\\
Equal & $-$2.340 (0.008) & $-$2.342 (0.007) & $-$2.341 (0.007) & $-$2.343 (0.007)\\
Equal + DART & \textbf{$-$2.325} (0.010) & \textbf{$-$2.320} (0.009) & \textbf{$-$2.329} (0.009) & \textbf{$-$2.325} (0.009)\\
BCF & $-$2.346 (0.008) & $-$2.346 (0.008) & $-$2.346 (0.008) & $-$2.346 (0.008)\\
HS & $-$2.402 (0.009) & $-$2.389 (0.007) & $-$2.378 (0.007) & $-$2.375 (0.008)\\
HS + DART & $-$2.379 (0.010) & $-$2.359 (0.008) & $-$2.346 (0.009) & $-$2.347 (0.008)\\
Fixed & $-$2.629 (0.007) & $-$2.607 (0.007) & $-$2.583 (0.007) & $-$2.562 (0.007)\\
\bottomrule
\end{tabular}
\end{table}

\begin{table}[!htbp]
\centering
\footnotesize
\setlength{\tabcolsep}{4pt}
\caption{Dimension Study. $n=200$ and $\eta=2$. RMSE of the centered effect by scenario and the number of covariates $p$. Mean over 40 data sets with its standard error in parentheses. Bold, the smallest mean in each column of a scenario.}
\label{tab:app-dim-rmse}
\vspace{4pt}
\begin{tabular}{l*{5}{c}}
\toprule
& \multicolumn{5}{c}{$p$}\\
\cmidrule(lr){2-6}
Model & 20 & 50 & 200 & 500 & 1{,}000\\
\midrule
\multicolumn{6}{l}{\emph{2 Regions}}\\
\method{} & 0.58 (0.01) & 0.70 (0.01) & 0.78 (0.01) & 0.78 (0.01) & 0.80 (0.01)\\
\method{} ($\rho=1$) & 0.54 (0.02) & 0.69 (0.01) & 0.76 (0.01) & 0.79 (0.01) & 0.80 (0.01)\\
\method{} + DART & 0.46 (0.02) & 0.60 (0.02) & 0.72 (0.02) & 0.74 (0.01) & 0.77 (0.02)\\
Equal & 0.54 (0.01) & 0.66 (0.01) & 0.76 (0.01) & 0.77 (0.01) & 0.79 (0.01)\\
Equal + DART & \textbf{0.42} (0.02) & \textbf{0.55} (0.02) & \textbf{0.69} (0.02) & 0.74 (0.01) & 0.76 (0.01)\\
BCF & 0.61 (0.01) & 0.77 (0.01) & 0.78 (0.01) & 0.77 (0.01) & 0.80 (0.01)\\
HS & 0.66 (0.01) & 0.75 (0.01) & 0.78 (0.01) & 0.79 (0.01) & 0.81 (0.01)\\
HS + DART & 0.50 (0.02) & 0.64 (0.02) & 0.74 (0.02) & 0.75 (0.01) & 0.80 (0.01)\\
Fixed & 0.82 (0.01) & 0.86 ($<$0.01) & 0.88 (0.01) & 0.88 ($<$0.01) & 0.88 (0.01)\\
X-learner & 0.62 (0.01) & 0.67 (0.01) & \textbf{0.69} (0.01) & \textbf{0.69} (0.01) & \textbf{0.71} (0.01)\\
R-learner & 0.73 (0.01) & 0.80 (0.01) & 0.85 (0.02) & 0.86 (0.01) & 0.86 (0.01)\\
\midrule
\multicolumn{6}{l}{\emph{4 Regions Graded}}\\
\method{} & 0.76 (0.01) & 0.92 (0.02) & 1.09 (0.02) & 1.20 (0.02) & 1.21 (0.01)\\
\method{} ($\rho=1$) & 0.75 (0.01) & 0.87 (0.01) & 1.06 (0.02) & 1.18 (0.02) & 1.23 (0.01)\\
\method{} + DART & 0.63 (0.02) & 0.70 (0.02) & 0.85 (0.03) & 0.97 (0.03) & 1.10 (0.03)\\
Equal & 0.73 (0.01) & 0.86 (0.01) & 1.08 (0.02) & 1.14 (0.01) & 1.19 (0.02)\\
Equal + DART & \textbf{0.62} (0.02) & \textbf{0.69} (0.02) & \textbf{0.84} (0.03) & 1.00 (0.03) & 1.08 (0.03)\\
BCF & 0.78 (0.02) & 0.96 (0.02) & 1.12 (0.01) & 1.15 (0.01) & 1.19 (0.01)\\
HS & 0.86 (0.02) & 1.11 (0.02) & 1.20 (0.01) & 1.23 (0.01) & 1.29 (0.02)\\
HS + DART & 0.69 (0.01) & 0.78 (0.02) & 0.94 (0.03) & 1.07 (0.03) & 1.19 (0.03)\\
Fixed & 1.37 (0.01) & 1.45 (0.01) & 1.51 (0.01) & 1.53 (0.01) & 1.52 (0.01)\\
X-learner & 0.79 (0.01) & 0.84 (0.01) & 0.88 (0.02) & \textbf{0.90} (0.02) & \textbf{0.97} (0.03)\\
R-learner & 1.00 (0.03) & 1.11 (0.03) & 1.19 (0.03) & 1.22 (0.03) & 1.32 (0.03)\\
\midrule
\multicolumn{6}{l}{\emph{Smooth Nonlinear}}\\
\method{} & 2.32 (0.04) & 2.51 (0.03) & 2.80 (0.06) & 3.26 (0.06) & 3.65 (0.06)\\
\method{} ($\rho=1$) & 2.32 (0.04) & 2.51 (0.03) & 2.81 (0.05) & 3.27 (0.06) & 3.65 (0.07)\\
\method{} + DART & 2.30 (0.03) & 2.33 (0.04) & 2.70 (0.06) & 3.24 (0.07) & 3.69 (0.07)\\
Equal & 2.32 (0.04) & 2.53 (0.03) & 2.80 (0.05) & 3.23 (0.06) & 3.63 (0.06)\\
Equal + DART & 2.27 (0.03) & \textbf{2.29} (0.03) & 2.67 (0.07) & 3.24 (0.06) & 3.71 (0.07)\\
BCF & \textbf{2.10} (0.02) & 2.33 (0.04) & \textbf{2.61} (0.05) & \textbf{3.07} (0.05) & \textbf{3.53} (0.06)\\
HS & 2.34 (0.03) & 2.43 (0.03) & 2.77 (0.05) & 3.18 (0.06) & 3.69 (0.07)\\
HS + DART & 2.28 (0.04) & 2.33 (0.04) & 2.74 (0.05) & 3.28 (0.05) & 3.81 (0.06)\\
Fixed & 3.50 (0.02) & 3.85 (0.03) & 4.23 (0.03) & 4.46 (0.04) & 4.53 (0.03)\\
X-learner & 2.32 (0.03) & 2.69 (0.04) & 3.11 (0.04) & 3.56 (0.06) & 3.88 (0.05)\\
R-learner & 3.42 (0.10) & 3.59 (0.08) & 4.21 (0.13) & 4.45 (0.11) & 4.74 (0.12)\\
\midrule
\multicolumn{6}{l}{\emph{Smooth Graded}}\\
\method{} & 2.36 (0.04) & 2.65 (0.05) & 2.89 (0.04) & 3.12 (0.05) & 3.32 (0.06)\\
\method{} ($\rho=1$) & 2.37 (0.04) & 2.58 (0.05) & 2.86 (0.05) & 3.12 (0.05) & 3.38 (0.07)\\
\method{} + DART & 2.44 (0.04) & 2.60 (0.06) & 2.84 (0.06) & 3.03 (0.06) & 3.28 (0.06)\\
Equal & 2.37 (0.05) & 2.58 (0.04) & 2.90 (0.05) & 3.11 (0.05) & 3.32 (0.06)\\
Equal + DART & 2.39 (0.05) & 2.60 (0.05) & 2.76 (0.06) & 3.02 (0.07) & 3.28 (0.06)\\
BCF & \textbf{2.28} (0.03) & \textbf{2.46} (0.06) & \textbf{2.70} (0.04) & \textbf{2.84} (0.05) & \textbf{3.01} (0.05)\\
HS & 2.44 (0.04) & 2.69 (0.05) & 2.90 (0.04) & 3.18 (0.05) & 3.41 (0.06)\\
HS + DART & 2.48 (0.05) & 2.64 (0.05) & 2.85 (0.05) & 3.26 (0.07) & 3.42 (0.05)\\
Fixed & 3.55 (0.04) & 3.79 (0.04) & 4.05 (0.04) & 4.20 (0.05) & 4.25 (0.05)\\
X-learner & 2.58 (0.04) & 2.84 (0.05) & 3.12 (0.05) & 3.32 (0.06) & 3.47 (0.06)\\
R-learner & 3.10 (0.08) & 3.35 (0.08) & 3.68 (0.07) & 3.79 (0.08) & 3.82 (0.08)\\
\bottomrule
\end{tabular}
\end{table}

\begin{table}[!htbp]
\centering
\footnotesize
\setlength{\tabcolsep}{4pt}
\caption{Dimension Study. $n=200$ and $\eta=2$. Coverage of the 90\% intervals for the centered effect, averaged over units. Entries as in Table~\ref{tab:app-dim-rmse}. Bold, the coverage closest to 0.90 in each column of a scenario.}
\label{tab:app-dim-cov}
\vspace{4pt}
\begin{tabular}{l*{5}{c}}
\toprule
& \multicolumn{5}{c}{$p$}\\
\cmidrule(lr){2-6}
Model & 20 & 50 & 200 & 500 & 1{,}000\\
\midrule
\multicolumn{6}{l}{\emph{2 Regions}}\\
\method{} & 0.85 (0.01) & 0.75 (0.02) & 0.61 (0.04) & \textbf{0.71} (0.02) & 0.63 (0.03)\\
\method{} ($\rho=1$) & 0.88 (0.02) & 0.77 (0.03) & 0.64 (0.04) & 0.68 (0.03) & 0.62 (0.04)\\
\method{} + DART & 0.92 (0.01) & 0.78 (0.03) & 0.64 (0.03) & 0.68 (0.03) & 0.59 (0.04)\\
Equal & 0.88 (0.02) & 0.80 (0.02) & 0.64 (0.04) & 0.68 (0.03) & 0.64 (0.04)\\
Equal + DART & 0.92 (0.02) & \textbf{0.85} (0.02) & \textbf{0.70} (0.03) & 0.66 (0.03) & 0.64 (0.04)\\
BCF & 0.82 (0.02) & 0.52 (0.03) & 0.68 (0.03) & \textbf{0.71} (0.03) & \textbf{0.66} (0.04)\\
HS & 0.77 (0.02) & 0.68 (0.03) & 0.59 (0.04) & 0.68 (0.02) & 0.60 (0.03)\\
HS + DART & \textbf{0.89} (0.01) & 0.76 (0.03) & 0.65 (0.03) & 0.66 (0.03) & 0.57 (0.03)\\
Fixed & 0.23 (0.02) & 0.12 (0.02) & 0.04 (0.02) & 0.02 (0.01) & 0.08 (0.03)\\
\midrule
\multicolumn{6}{l}{\emph{4 Regions Graded}}\\
\method{} & 0.77 (0.02) & 0.73 (0.02) & 0.68 (0.02) & 0.70 (0.02) & 0.75 (0.02)\\
\method{} ($\rho=1$) & 0.79 (0.02) & 0.77 (0.02) & 0.74 (0.02) & 0.73 (0.01) & 0.75 (0.02)\\
\method{} + DART & 0.78 (0.01) & 0.72 (0.02) & 0.68 (0.02) & 0.68 (0.02) & 0.71 (0.03)\\
Equal & 0.81 (0.01) & 0.77 (0.02) & 0.74 (0.02) & 0.76 (0.01) & 0.79 (0.02)\\
Equal + DART & 0.81 (0.02) & \textbf{0.78} (0.02) & 0.73 (0.02) & 0.73 (0.02) & 0.73 (0.02)\\
BCF & \textbf{0.84} (0.01) & 0.73 (0.02) & \textbf{0.79} (0.01) & \textbf{0.81} (0.01) & \textbf{0.84} (0.02)\\
HS & 0.76 (0.01) & 0.65 (0.02) & 0.66 (0.01) & 0.71 (0.01) & 0.73 (0.02)\\
HS + DART & 0.74 (0.02) & 0.69 (0.02) & 0.66 (0.02) & 0.68 (0.02) & 0.68 (0.03)\\
Fixed & 0.18 (0.01) & 0.16 (0.01) & 0.15 ($<$0.01) & 0.16 (0.01) & 0.16 (0.01)\\
\midrule
\multicolumn{6}{l}{\emph{Smooth Nonlinear}}\\
\method{} & 0.78 (0.01) & 0.78 (0.01) & 0.76 (0.02) & 0.75 (0.01) & 0.73 (0.01)\\
\method{} ($\rho=1$) & 0.78 (0.01) & 0.78 (0.01) & 0.78 (0.01) & 0.76 (0.01) & 0.74 (0.01)\\
\method{} + DART & 0.75 (0.01) & 0.77 (0.01) & 0.73 (0.02) & 0.70 (0.01) & 0.68 (0.02)\\
Equal & 0.79 (0.01) & 0.77 (0.01) & 0.78 (0.01) & 0.77 (0.01) & 0.75 (0.01)\\
Equal + DART & 0.76 (0.01) & 0.79 (0.01) & 0.74 (0.02) & 0.72 (0.02) & 0.70 (0.02)\\
BCF & \textbf{0.87} (0.01) & \textbf{0.87} (0.01) & \textbf{0.87} (0.01) & \textbf{0.86} (0.01) & \textbf{0.84} (0.01)\\
HS & 0.77 (0.01) & 0.78 (0.01) & 0.77 (0.01) & 0.78 (0.01) & 0.76 (0.01)\\
HS + DART & 0.74 (0.01) & 0.76 (0.01) & 0.74 (0.01) & 0.73 (0.01) & 0.70 (0.02)\\
Fixed & 0.41 (0.01) & 0.40 (0.01) & 0.40 (0.01) & 0.40 (0.01) & 0.45 (0.01)\\
\midrule
\multicolumn{6}{l}{\emph{Smooth Graded}}\\
\method{} & 0.84 (0.01) & 0.81 (0.01) & 0.83 (0.01) & 0.84 (0.01) & 0.85 (0.01)\\
\method{} ($\rho=1$) & 0.85 (0.01) & 0.82 (0.01) & 0.85 (0.01) & 0.85 (0.01) & 0.87 (0.01)\\
\method{} + DART & 0.79 (0.01) & 0.77 (0.01) & 0.77 (0.01) & 0.79 (0.01) & 0.80 (0.01)\\
Equal & 0.84 (0.01) & 0.83 (0.01) & 0.84 (0.01) & 0.86 (0.01) & 0.86 (0.01)\\
Equal + DART & 0.81 (0.01) & 0.78 (0.01) & 0.80 (0.01) & 0.80 (0.01) & 0.82 (0.01)\\
BCF & \textbf{0.87} (0.01) & \textbf{0.86} (0.01) & \textbf{0.88} (0.01) & \textbf{0.90} (0.01) & \textbf{0.91} (0.01)\\
HS & 0.81 (0.01) & 0.79 (0.01) & 0.82 (0.01) & 0.84 (0.01) & 0.86 (0.01)\\
HS + DART & 0.79 (0.01) & 0.75 (0.02) & 0.78 (0.01) & 0.80 (0.01) & 0.81 (0.01)\\
Fixed & 0.50 (0.01) & 0.48 (0.01) & 0.47 (0.01) & 0.46 (0.01) & 0.47 (0.01)\\
\bottomrule
\end{tabular}
\end{table}

\begin{table}[!htbp]
\centering
\footnotesize
\setlength{\tabcolsep}{4pt}
\caption{Dimension Study. $n=200$ and $\eta=2$. Average length of the 90\% intervals for the centered effect. Entries as in Table~\ref{tab:app-dim-rmse}. No entry is bold, because Fixed has the shortest intervals in every column together with the lowest coverage (Table~\ref{tab:app-dim-cov}), so length is read together with coverage.}
\label{tab:app-dim-len}
\vspace{4pt}
\begin{tabular}{l*{5}{c}}
\toprule
& \multicolumn{5}{c}{$p$}\\
\cmidrule(lr){2-6}
Model & 20 & 50 & 200 & 500 & 1{,}000\\
\midrule
\multicolumn{6}{l}{\emph{2 Regions}}\\
\method{} & 1.24 (0.03) & 1.30 (0.02) & 1.34 (0.04) & 1.52 (0.03) & 1.51 (0.04)\\
\method{} ($\rho=1$) & 1.19 (0.03) & 1.29 (0.03) & 1.35 (0.05) & 1.48 (0.04) & 1.45 (0.05)\\
\method{} + DART & 1.09 (0.03) & 1.16 (0.03) & 1.27 (0.04) & 1.42 (0.05) & 1.40 (0.05)\\
Equal & 1.21 (0.03) & 1.28 (0.03) & 1.35 (0.04) & 1.48 (0.04) & 1.47 (0.05)\\
Equal + DART & 1.04 (0.03) & 1.21 (0.03) & 1.38 (0.05) & 1.42 (0.05) & 1.51 (0.06)\\
BCF & 1.48 (0.03) & 1.11 (0.06) & 1.62 (0.06) & 1.73 (0.06) & 1.67 (0.07)\\
HS & 1.36 (0.03) & 1.33 (0.03) & 1.39 (0.04) & 1.54 (0.03) & 1.48 (0.04)\\
HS + DART & 1.22 (0.03) & 1.25 (0.04) & 1.33 (0.04) & 1.40 (0.04) & 1.43 (0.04)\\
Fixed & 0.90 (0.01) & 0.90 (0.01) & 0.84 (0.01) & 0.83 (0.01) & 0.88 (0.02)\\
\midrule
\multicolumn{6}{l}{\emph{4 Regions Graded}}\\
\method{} & 1.72 (0.05) & 2.04 (0.05) & 2.36 (0.04) & 2.60 (0.05) & 2.96 (0.06)\\
\method{} ($\rho=1$) & 1.67 (0.04) & 1.97 (0.04) & 2.45 (0.05) & 2.71 (0.04) & 2.92 (0.05)\\
\method{} + DART & 1.36 (0.03) & 1.46 (0.06) & 1.73 (0.09) & 2.14 (0.07) & 2.54 (0.08)\\
Equal & 1.68 (0.05) & 2.00 (0.04) & 2.47 (0.05) & 2.78 (0.05) & 2.98 (0.06)\\
Equal + DART & 1.42 (0.03) & 1.51 (0.04) & 1.91 (0.08) & 2.25 (0.08) & 2.53 (0.07)\\
BCF & 2.05 (0.03) & 2.10 (0.06) & 2.85 (0.04) & 3.03 (0.04) & 3.34 (0.06)\\
HS & 2.07 (0.04) & 2.23 (0.04) & 2.45 (0.05) & 2.77 (0.05) & 2.90 (0.06)\\
HS + DART & 1.43 (0.04) & 1.61 (0.06) & 2.01 (0.08) & 2.39 (0.06) & 2.60 (0.06)\\
Fixed & 1.31 (0.01) & 1.33 (0.02) & 1.23 (0.02) & 1.22 (0.02) & 1.28 (0.03)\\
\midrule
\multicolumn{6}{l}{\emph{Smooth Nonlinear}}\\
\method{} & 5.81 (0.09) & 6.27 (0.09) & 6.82 (0.15) & 7.78 (0.16) & 8.38 (0.21)\\
\method{} ($\rho=1$) & 5.85 (0.09) & 6.30 (0.10) & 7.21 (0.16) & 8.05 (0.15) & 8.53 (0.20)\\
\method{} + DART & 5.35 (0.09) & 5.63 (0.09) & 6.16 (0.15) & 7.15 (0.15) & 7.86 (0.20)\\
Equal & 5.88 (0.10) & 6.23 (0.09) & 7.23 (0.14) & 8.03 (0.17) & 8.74 (0.20)\\
Equal + DART & 5.40 (0.10) & 5.85 (0.09) & 6.19 (0.15) & 7.43 (0.18) & 8.15 (0.22)\\
BCF & 6.29 (0.09) & 7.04 (0.10) & 7.99 (0.13) & 9.24 (0.15) & 10.08 (0.21)\\
HS & 5.67 (0.10) & 6.26 (0.09) & 7.09 (0.14) & 8.10 (0.15) & 9.01 (0.18)\\
HS + DART & 5.29 (0.09) & 5.64 (0.09) & 6.31 (0.15) & 7.64 (0.15) & 8.55 (0.20)\\
Fixed & 4.07 (0.03) & 4.39 (0.04) & 4.70 (0.05) & 4.99 (0.05) & 5.68 (0.06)\\
\midrule
\multicolumn{6}{l}{\emph{Smooth Graded}}\\
\method{} & 6.16 (0.11) & 6.38 (0.12) & 7.51 (0.15) & 8.41 (0.16) & 8.95 (0.19)\\
\method{} ($\rho=1$) & 6.32 (0.11) & 6.42 (0.14) & 7.53 (0.16) & 8.52 (0.20) & 9.40 (0.19)\\
\method{} + DART & 5.64 (0.12) & 5.66 (0.15) & 6.20 (0.15) & 7.19 (0.19) & 8.07 (0.21)\\
Equal & 6.20 (0.11) & 6.43 (0.14) & 7.64 (0.16) & 8.55 (0.14) & 9.32 (0.18)\\
Equal + DART & 5.77 (0.11) & 5.85 (0.16) & 6.51 (0.17) & 7.35 (0.18) & 8.57 (0.24)\\
BCF & 6.53 (0.10) & 6.99 (0.14) & 7.90 (0.13) & 8.88 (0.15) & 9.60 (0.16)\\
HS & 6.04 (0.11) & 6.37 (0.14) & 7.52 (0.17) & 8.67 (0.18) & 9.56 (0.21)\\
HS + DART & 5.82 (0.12) & 5.59 (0.16) & 6.48 (0.17) & 7.99 (0.21) & 8.65 (0.23)\\
Fixed & 4.29 (0.04) & 4.57 (0.06) & 4.93 (0.06) & 5.08 (0.07) & 5.24 (0.06)\\
\bottomrule
\end{tabular}
\end{table}

\begin{table}[!htbp]
\centering
\footnotesize
\setlength{\tabcolsep}{4pt}
\caption{Dimension Study. $n=200$ and $\eta=2$. False positive rate, the share of units without an effect whose 90\% interval for $\tau(\bx_i)$ excludes zero. Entries as in Table~\ref{tab:app-dim-rmse}. Bold, the smallest rate in each column of a scenario.}
\label{tab:app-dim-fpr}
\vspace{4pt}
\begin{tabular}{l*{5}{c}}
\toprule
& \multicolumn{5}{c}{$p$}\\
\cmidrule(lr){2-6}
Model & 20 & 50 & 200 & 500 & 1{,}000\\
\midrule
\multicolumn{6}{l}{\emph{2 Regions}}\\
\method{} & 0.07 (0.02) & 0.17 (0.03) & 0.28 (0.04) & 0.19 (0.02) & 0.26 (0.04)\\
\method{} ($\rho=1$) & 0.06 (0.02) & 0.16 (0.03) & 0.27 (0.05) & 0.21 (0.03) & 0.28 (0.05)\\
\method{} + DART & \textbf{0.05} (0.01) & 0.16 (0.04) & 0.24 (0.03) & 0.22 (0.04) & 0.28 (0.04)\\
Equal & 0.06 (0.02) & 0.12 (0.03) & 0.26 (0.05) & 0.17 (0.02) & 0.25 (0.04)\\
Equal + DART & \textbf{0.05} (0.01) & \textbf{0.09} (0.02) & \textbf{0.19} (0.03) & 0.21 (0.03) & \textbf{0.22} (0.04)\\
BCF & 0.11 (0.02) & 0.33 (0.03) & \textbf{0.19} (0.04) & \textbf{0.14} (0.02) & 0.24 (0.04)\\
HS & 0.15 (0.02) & 0.22 (0.03) & 0.30 (0.04) & 0.19 (0.02) & 0.28 (0.04)\\
HS + DART & 0.07 (0.01) & 0.15 (0.03) & 0.23 (0.03) & 0.22 (0.04) & 0.30 (0.04)\\
Fixed & 0.61 (0.04) & 0.60 (0.06) & 0.79 (0.06) & 0.77 (0.06) & 0.68 (0.07)\\
\midrule
\multicolumn{6}{l}{\emph{4 Regions Graded}}\\
\method{} & 0.17 (0.03) & 0.24 (0.03) & 0.28 (0.02) & 0.27 (0.02) & 0.21 (0.02)\\
\method{} ($\rho=1$) & 0.16 (0.03) & 0.17 (0.03) & 0.21 (0.02) & 0.22 (0.02) & 0.21 (0.02)\\
\method{} + DART & 0.18 (0.03) & 0.21 (0.04) & 0.28 (0.03) & 0.29 (0.04) & 0.30 (0.04)\\
Equal & \textbf{0.11} (0.02) & 0.17 (0.03) & 0.20 (0.02) & 0.20 (0.02) & 0.19 (0.02)\\
Equal + DART & 0.13 (0.02) & \textbf{0.14} (0.03) & 0.23 (0.03) & 0.22 (0.03) & 0.30 (0.04)\\
BCF & 0.12 (0.01) & 0.23 (0.02) & \textbf{0.17} (0.01) & \textbf{0.17} (0.02) & \textbf{0.13} (0.02)\\
HS & 0.21 (0.02) & 0.32 (0.02) & 0.31 (0.02) & 0.25 (0.02) & 0.21 (0.02)\\
HS + DART & 0.20 (0.02) & 0.23 (0.04) & 0.31 (0.03) & 0.30 (0.04) & 0.33 (0.04)\\
Fixed & 0.90 (0.02) & 0.94 (0.02) & 0.99 (0.01) & 0.99 (0.01) & 0.95 (0.03)\\
\bottomrule
\end{tabular}
\end{table}

\begin{table}[!htbp]
\centering
\footnotesize
\setlength{\tabcolsep}{4pt}
\caption{Dimension Study. $n=200$ and $\eta=2$. True positive rate, the share of units with an effect whose 90\% interval for $\tau(\bx_i)$ excludes zero. Entries as in Table~\ref{tab:app-dim-rmse}. No entry is bold, because Fixed has the largest rate in every column together with the largest FPR (Table~\ref{tab:app-dim-fpr}), so the TPR is read together with the FPR.}
\label{tab:app-dim-tpr}
\vspace{4pt}
\begin{tabular}{l*{5}{c}}
\toprule
& \multicolumn{5}{c}{$p$}\\
\cmidrule(lr){2-6}
Model & 20 & 50 & 200 & 500 & 1{,}000\\
\midrule
\multicolumn{6}{l}{\emph{2 Regions}}\\
\method{} & 0.74 (0.02) & 0.69 (0.02) & 0.72 (0.02) & 0.65 (0.03) & 0.65 (0.04)\\
\method{} ($\rho=1$) & 0.75 (0.02) & 0.69 (0.02) & 0.72 (0.02) & 0.63 (0.03) & 0.66 (0.04)\\
\method{} + DART & 0.82 (0.02) & 0.77 (0.02) & 0.73 (0.02) & 0.67 (0.03) & 0.68 (0.03)\\
Equal & 0.75 (0.02) & 0.68 (0.02) & 0.70 (0.03) & 0.63 (0.02) & 0.66 (0.03)\\
Equal + DART & 0.83 (0.02) & 0.77 (0.02) & 0.71 (0.03) & 0.67 (0.03) & 0.62 (0.03)\\
BCF & 0.79 (0.01) & 0.74 (0.01) & 0.67 (0.03) & 0.61 (0.03) & 0.60 (0.05)\\
HS & 0.76 (0.02) & 0.70 (0.02) & 0.72 (0.03) & 0.63 (0.03) & 0.63 (0.04)\\
HS + DART & 0.79 (0.02) & 0.73 (0.02) & 0.71 (0.02) & 0.69 (0.03) & 0.67 (0.03)\\
Fixed & 0.90 (0.02) & 0.83 (0.04) & 0.88 (0.04) & 0.86 (0.04) & 0.76 (0.06)\\
\midrule
\multicolumn{6}{l}{\emph{4 Regions Graded}}\\
\method{} & 0.70 (0.02) & 0.63 (0.02) & 0.59 (0.02) & 0.55 (0.02) & 0.50 (0.02)\\
\method{} ($\rho=1$) & 0.69 (0.02) & 0.59 (0.02) & 0.55 (0.02) & 0.52 (0.02) & 0.50 (0.02)\\
\method{} + DART & 0.75 (0.02) & 0.71 (0.02) & 0.69 (0.02) & 0.62 (0.03) & 0.59 (0.03)\\
Equal & 0.66 (0.02) & 0.59 (0.02) & 0.53 (0.02) & 0.51 (0.02) & 0.48 (0.03)\\
Equal + DART & 0.71 (0.02) & 0.66 (0.02) & 0.63 (0.02) & 0.54 (0.03) & 0.57 (0.03)\\
BCF & 0.70 (0.01) & 0.67 (0.01) & 0.56 (0.02) & 0.53 (0.01) & 0.47 (0.02)\\
HS & 0.67 (0.02) & 0.63 (0.02) & 0.60 (0.02) & 0.54 (0.02) & 0.50 (0.02)\\
HS + DART & 0.75 (0.02) & 0.70 (0.02) & 0.67 (0.02) & 0.63 (0.03) & 0.61 (0.03)\\
Fixed & 0.96 (0.01) & 0.98 (0.01) & 1.00 ($<$0.01) & 0.99 (0.01) & 0.97 (0.02)\\
\bottomrule
\end{tabular}
\end{table}

\begin{table}[!htbp]
\centering
\footnotesize
\setlength{\tabcolsep}{4pt}
\caption{Dimension Study. $n=200$ and $\eta=2$. Log predictive density of 500 test units, averaged over the units. Entries as in Table~\ref{tab:app-dim-rmse}. Bold, the largest mean in each column of a scenario.}
\label{tab:app-dim-lpd}
\vspace{4pt}
\begin{tabular}{l*{5}{c}}
\toprule
& \multicolumn{5}{c}{$p$}\\
\cmidrule(lr){2-6}
Model & 20 & 50 & 200 & 500 & 1{,}000\\
\midrule
\multicolumn{6}{l}{\emph{2 Regions}}\\
\method{} & $-$1.072 (0.008) & $-$1.188 (0.007) & $-$1.280 (0.008) & $-$1.306 (0.006) & $-$1.355 (0.008)\\
\method{} ($\rho=1$) & $-$1.055 (0.009) & $-$1.179 (0.008) & $-$1.272 (0.009) & $-$1.302 (0.005) & $-$1.348 (0.007)\\
\method{} + DART & $-$0.917 (0.011) & $-$0.994 (0.012) & $-$1.060 (0.012) & $-$1.092 (0.010) & $-$1.150 (0.011)\\
Equal & $-$1.057 (0.008) & $-$1.166 (0.008) & $-$1.274 (0.009) & $-$1.295 (0.006) & $-$1.351 (0.007)\\
Equal + DART & \textbf{$-$0.903} (0.012) & \textbf{$-$0.971} (0.010) & \textbf{$-$1.056} (0.012) & $-$1.096 (0.011) & $-$1.159 (0.013)\\
BCF & $-$1.095 (0.010) & $-$1.203 (0.009) & $-$1.287 (0.008) & $-$1.307 (0.005) & $-$1.356 (0.007)\\
HS & $-$1.112 (0.008) & $-$1.212 (0.007) & $-$1.280 (0.008) & $-$1.297 (0.006) & $-$1.344 (0.007)\\
HS + DART & $-$0.939 (0.010) & $-$1.015 (0.011) & $-$1.076 (0.011) & \textbf{$-$1.085} (0.010) & \textbf{$-$1.147} (0.011)\\
Fixed & $-$1.189 (0.006) & $-$1.262 (0.006) & $-$1.321 (0.007) & $-$1.334 (0.006) & $-$1.376 (0.008)\\
\midrule
\multicolumn{6}{l}{\emph{4 Regions Graded}}\\
\method{} & $-$1.166 (0.007) & $-$1.298 (0.009) & $-$1.438 (0.010) & $-$1.518 (0.010) & $-$1.574 (0.009)\\
\method{} ($\rho=1$) & $-$1.152 (0.007) & $-$1.267 (0.007) & $-$1.421 (0.010) & $-$1.509 (0.008) & $-$1.578 (0.008)\\
\method{} + DART & $-$1.006 (0.010) & $-$1.053 (0.009) & $-$1.126 (0.017) & \textbf{$-$1.204} (0.014) & $-$1.350 (0.016)\\
Equal & $-$1.139 (0.007) & $-$1.264 (0.007) & $-$1.421 (0.011) & $-$1.480 (0.008) & $-$1.558 (0.010)\\
Equal + DART & \textbf{$-$0.998} (0.010) & \textbf{$-$1.043} (0.009) & \textbf{$-$1.114} (0.014) & $-$1.218 (0.019) & \textbf{$-$1.322} (0.016)\\
BCF & $-$1.171 (0.010) & $-$1.297 (0.010) & $-$1.472 (0.008) & $-$1.510 (0.006) & $-$1.572 (0.007)\\
HS & $-$1.220 (0.009) & $-$1.402 (0.009) & $-$1.489 (0.010) & $-$1.535 (0.007) & $-$1.591 (0.009)\\
HS + DART & $-$1.029 (0.010) & $-$1.090 (0.012) & $-$1.179 (0.019) & $-$1.272 (0.019) & $-$1.367 (0.016)\\
Fixed & $-$1.487 (0.006) & $-$1.559 (0.006) & $-$1.625 (0.007) & $-$1.639 (0.006) & $-$1.674 (0.008)\\
\midrule
\multicolumn{6}{l}{\emph{Smooth Nonlinear}}\\
\method{} & $-$2.292 (0.008) & $-$2.338 (0.006) & $-$2.458 (0.010) & $-$2.631 (0.010) & $-$2.827 (0.011)\\
\method{} ($\rho=1$) & $-$2.289 (0.008) & $-$2.337 (0.006) & $-$2.459 (0.009) & $-$2.629 (0.009) & $-$2.829 (0.011)\\
\method{} + DART & $-$2.278 (0.008) & $-$2.301 (0.006) & $-$2.441 (0.012) & $-$2.628 (0.011) & $-$2.838 (0.012)\\
Equal & $-$2.288 (0.008) & $-$2.338 (0.006) & $-$2.455 (0.009) & $-$2.624 (0.010) & $-$2.826 (0.011)\\
Equal + DART & $-$2.273 (0.008) & \textbf{$-$2.295} (0.006) & $-$2.440 (0.013) & $-$2.637 (0.011) & $-$2.839 (0.011)\\
BCF & \textbf{$-$2.251} (0.009) & $-$2.304 (0.008) & \textbf{$-$2.426} (0.009) & \textbf{$-$2.606} (0.010) & \textbf{$-$2.813} (0.011)\\
HS & $-$2.291 (0.008) & $-$2.325 (0.005) & $-$2.449 (0.009) & $-$2.619 (0.009) & $-$2.829 (0.011)\\
HS + DART & $-$2.273 (0.010) & $-$2.297 (0.007) & $-$2.448 (0.012) & $-$2.639 (0.010) & $-$2.849 (0.011)\\
Fixed & $-$2.442 (0.006) & $-$2.531 (0.005) & $-$2.644 (0.008) & $-$2.770 (0.007) & $-$2.908 (0.009)\\
\midrule
\multicolumn{6}{l}{\emph{Smooth Graded}}\\
\method{} & $-$2.288 (0.008) & $-$2.366 (0.008) & $-$2.409 (0.008) & $-$2.454 (0.009) & $-$2.480 (0.008)\\
\method{} ($\rho=1$) & $-$2.302 (0.010) & $-$2.350 (0.009) & $-$2.403 (0.008) & $-$2.453 (0.009) & $-$2.482 (0.008)\\
\method{} + DART & $-$2.291 (0.009) & $-$2.334 (0.009) & $-$2.391 (0.013) & $-$2.427 (0.013) & $-$2.472 (0.011)\\
Equal & $-$2.293 (0.010) & $-$2.343 (0.007) & $-$2.414 (0.010) & $-$2.449 (0.008) & $-$2.483 (0.008)\\
Equal + DART & \textbf{$-$2.279} (0.009) & \textbf{$-$2.325} (0.009) & \textbf{$-$2.373} (0.010) & $-$2.429 (0.012) & $-$2.477 (0.010)\\
BCF & $-$2.290 (0.009) & $-$2.346 (0.008) & $-$2.382 (0.008) & \textbf{$-$2.414} (0.008) & \textbf{$-$2.439} (0.007)\\
HS & $-$2.312 (0.010) & $-$2.375 (0.008) & $-$2.417 (0.008) & $-$2.468 (0.009) & $-$2.505 (0.010)\\
HS + DART & $-$2.305 (0.009) & $-$2.347 (0.008) & $-$2.394 (0.009) & $-$2.468 (0.013) & $-$2.498 (0.010)\\
Fixed & $-$2.488 (0.009) & $-$2.562 (0.007) & $-$2.632 (0.007) & $-$2.664 (0.009) & $-$2.663 (0.008)\\
\bottomrule
\end{tabular}
\end{table}

\clearpage
\subsection{ACIC 2016 Benchmark}
\label{app:exp-acic}

The simulated designs of Appendix~\ref{app:exp-forest} draw independent uniform covariates. To assess the models on an empirical covariate distribution with known unit-level effects, Section~\ref{sec:exp-acic} uses the data of the 2016 Atlantic Causal Inference Conference competition \citep{dorie2019automated}. This subsection gives its design and its complete results.

\paragraph{Data.}
The competition fixes a design matrix of $n=4{,}802$ units and 58 covariates from the Collaborative Perinatal Project, of which 3 are categorical, 5 binary, 27 counts and 23 continuous \citep{dorie2019automated}. We code the three categorical covariates, with 6, 16 and 5 levels, as 25 indicator columns, one for every level of the first and for every level but one of the other two, which gives 80 columns. On this design the competition simulates the treatment assignment and the outcome under 77 settings with 100 replicates each. A setting is defined by six parameters, the functional form of the assignment model (linear, polynomial or step), a parameter of the baseline treatment probability (0.35 or 0.65), the overlap (full, or restricted by one term), the functional form of the response model (linear, exponential or step), the alignment of the confounders between the two models (0, 0.25 or 0.75), and the heterogeneity of the treatment effect (none, medium or high). For every unit the competition supplies the conditional means $\mu_0(\bx_i)$ and $\mu_1(\bx_i)$ of the outcome under control and treatment, so the effect $\tau(\bx_i)=\mu_1(\bx_i)-\mu_0(\bx_i)$ and the centered effect $\delta_i$ are known. We take $\sigma$ to be the standard deviation of $y_i-\mu_{A_i}(\bx_i)$ over the units of a data set.

\paragraph{Setting Selection.}
We define the active fraction
\[
F_\epsilon=\frac1n\#\{i:|\delta_i|>\epsilon\sigma\},\qquad \epsilon=0.2,
\]
as the share of units whose centered effect exceeds a fifth of the residual standard deviation in absolute value. Before any model was fitted, the 77 settings were ordered by $F_\epsilon$, averaged over replicates 1 to 3, and divided into 20 groups of consecutive settings of nearly equal size. One setting was drawn from each group with a recorded seed. Table~\ref{tab:app-acic-settings} lists the 20 settings with their parameters. Two of the 77 settings have a constant effect and $F_\epsilon=0$, and one of them, setting 3, was drawn. The other 19 drawn settings have $F_\epsilon$ between 0.889 and 0.979, and the replicates of one setting differ in $F_\epsilon$ by up to 0.16. The selected settings therefore vary little in this fraction. They vary more in the spread of the effect, whose standard deviation across units ranges from 2.6 to 6.1 times $\sigma$ in replicates 1 to 3. We fit replicates 1 to 10 of every selected setting, 200 data sets in all.

\paragraph{Preprocessing and Models.}
In each data set, the outcome is centered and scaled to unit standard
deviation before fitting, and $\tau$, $\delta_i$ and $\sigma$ are scaled with it. Every error below is therefore in units of the standard deviation of the outcome of its data set. On this scale $\sigma$ lies between 0.13 and 0.26. Every model receives the same cross-fitted estimate of the propensity score (Appendix~\ref{app:exp-setup}), and the true assignment probabilities are not
used. BCF is fitted with its package at its defaults. In our sampler we fit \method{}, \method{} ($\rho=1$), the DART version of each, HS and Fixed, with $\omega$ chosen on the 80 columns of the competition design before any outcome was read. On an independent set of 1{,}000 prior draws, the median of $S_\tau/\sigma$ lies between 0.46 and 0.52 for every model. Equal and the DART version of HS were not fitted in this study.

\paragraph{Losses.}
The RMSE of the centered effect ranges over more than an order of magnitude across data sets, so Section~\ref{sec:exp-acic} compares the models through its ratio to that of BCF on the same data set. With $\bar\ell$ and $s$ the mean and standard deviation of the log ratio over the $N$ data sets, the reported ratio is $\exp(\bar\ell)$ with the interval $\exp(\bar\ell\pm2s/N^{1/2})$. The BCF package failed on two data sets, replicate 2 of setting 4 and replicate 9 of setting 48, so $N=188$. A setting counts as below one when the geometric mean of the ratio over its replicates is below one.

\paragraph{Results.}
Table~\ref{tab:app-acic-summary} extends Table~\ref{tab:acic} with
\method{} ($\rho=1$) and its DART version, whose ratios are lower than those of \method{} and \method{} + DART by 0.007 and 0.011. The ratio of \method{} to BCF is below one in 10 of the 19 settings (Table~\ref{tab:app-acic-by-setting}), so the comparison between the two changes sign across settings. The ratio of \method{} + DART is below one in 18. Because every setting with a non-constant effect has $F_\epsilon$ between 0.889 and 0.979, the benchmark does not show how the comparison depends on $F_\epsilon$. On setting 3, where the effect is constant, the models in our sampler have an RMSE of the centered effect between 0.011 and 0.021, and their intervals contain the centered effect of almost every unit (Table~\ref{tab:app-acic-constant}). BCF has an RMSE of 0.022 and a coverage of 0.91, and the X-learner and the R-learner have RMSEs of 0.060 and 0.146.

t
\begin{table}[!htbp]
\centering
\footnotesize
\setlength{\tabcolsep}{2.5pt}
\caption{ACIC 2016 Settings. The 20 settings drawn before any model was fitted, in increasing order of the active fraction $F_\epsilon$ ($\epsilon=0.2$, mean over replicates 1 to 3, as used for the selection). The next six columns are the competition's parameters of the setting \citep{dorie2019automated}. The range of $F_\epsilon$, the number of treated units and the ratio of the standard deviation of the centered effect to the residual standard deviation $\sigma$ are computed over replicates 1 to 3, the last as a mean. Every data set has $n=4{,}802$ units.}
\label{tab:app-acic-settings}
\vspace{4pt}
\begin{tabular}{rrcclclcclcc}
\toprule
& & & & \multicolumn{3}{c}{treatment assignment} & \multicolumn{3}{c}{response} & \multicolumn{2}{c}{replicates 1 to 3} \\
\cmidrule(lr){5-7}\cmidrule(lr){8-10}\cmidrule(lr){11-12}
rank & setting & $F_\epsilon$ & range & model & root & overlap & model & alignment & heterogeneity & treated & sd$(\delta)/\sigma$ \\
\midrule
1 & 3 & 0.000 & 0.00--0.00 & linear & 0.35 & one-term & linear & 0.75 & none & 1{,}399--1{,}455 & 0.0 \\
2 & 68 & 0.889 & 0.79--0.95 & step & 0.65 & one-term & exponential & 0.75 & med & 1{,}832--3{,}001 & 3.3 \\
3 & 8 & 0.901 & 0.85--0.95 & polynomial & 0.35 & one-term & exponential & 0.00 & high & 1{,}237--1{,}535 & 5.9 \\
4 & 24 & 0.912 & 0.86--0.96 & polynomial & 0.35 & one-term & step & 0.75 & high & 1{,}314--1{,}455 & 3.0 \\
5 & 29 & 0.918 & 0.88--0.95 & polynomial & 0.35 & full & step & 0.75 & med & 1{,}707--1{,}790 & 2.6 \\
6 & 36 & 0.923 & 0.84--0.98 & polynomial & 0.65 & one-term & step & 0.75 & med & 1{,}708--2{,}731 & 3.2 \\
7 & 35 & 0.926 & 0.86--1.00 & polynomial & 0.65 & one-term & step & 0.25 & high & 2{,}207--2{,}779 & 4.1 \\
8 & 37 & 0.935 & 0.91--0.97 & polynomial & 0.65 & one-term & step & 0.75 & high & 2{,}289--2{,}692 & 3.3 \\
9 & 48 & 0.935 & 0.90--0.96 & step & 0.35 & one-term & step & 0.25 & med & 1{,}504--1{,}551 & 3.0 \\
10 & 4 & 0.944 & 0.87--0.99 & polynomial & 0.35 & full & exponential & 0.75 & high & 1{,}530--1{,}766 & 5.2 \\
11 & 61 & 0.945 & 0.89--0.99 & step & 0.35 & full & exponential & 0.75 & med & 1{,}597--1{,}643 & 4.5 \\
12 & 18 & 0.948 & 0.91--0.98 & linear & 0.65 & one-term & exponential & 0.25 & high & 2{,}660--2{,}773 & 4.9 \\
13 & 17 & 0.950 & 0.89--1.00 & step & 0.65 & one-term & step & 0.75 & high & 2{,}468--2{,}699 & 4.1 \\
14 & 47 & 0.952 & 0.95--0.96 & polynomial & 0.65 & full & exponential & 0.75 & high & 3{,}141--3{,}181 & 3.7 \\
15 & 27 & 0.959 & 0.92--1.00 & polynomial & 0.35 & full & step & 0.25 & med & 1{,}667--1{,}773 & 3.8 \\
16 & 49 & 0.967 & 0.95--0.98 & step & 0.35 & one-term & step & 0.25 & high & 990--1{,}263 & 5.2 \\
17 & 57 & 0.968 & 0.94--0.98 & step & 0.35 & full & step & 0.75 & med & 1{,}662--1{,}792 & 3.9 \\
18 & 28 & 0.969 & 0.97--0.97 & polynomial & 0.35 & full & step & 0.25 & high & 1{,}700--1{,}804 & 5.5 \\
19 & 42 & 0.974 & 0.96--0.99 & polynomial & 0.65 & full & step & 0.75 & med & 3{,}032--3{,}198 & 6.1 \\
20 & 64 & 0.979 & 0.95--1.00 & step & 0.65 & one-term & step & 0.25 & high & 2{,}206--2{,}569 & 3.3 \\
\bottomrule
\end{tabular}
\end{table}

\begin{table}[!htbp]
\centering
\footnotesize
\setlength{\tabcolsep}{4pt}
\caption{ACIC 2016. The 19 settings with a non-constant effect. The RMSE ratio is the geometric mean over the 188 data sets with a BCF fit of the RMSE of the centered effect divided by that of BCF on the same data set, with its two-standard-error interval. Settings is the number of settings whose geometric-mean ratio is below one. Coverage and length are those of the 90\% intervals for the centered effect, averaged over the 190 data sets, or 188 for BCF. Bold, the smallest ratio, the largest number of settings and the coverage closest to 0.90.}
\label{tab:app-acic-summary}
\vspace{4pt}
\begin{tabular}{lcccc}
\toprule
Model & RMSE ratio & Settings & Coverage & Length\\
\midrule
\method{} & 0.99 (0.95, 1.04) & 10 & 0.80 & 0.20\\
\method{} ($\rho=1$) & 0.99 (0.94, 1.03) & 9 & 0.80 & 0.20\\
\method{} + DART & 0.88 (0.83, 0.92) & \textbf{18} & 0.80 & 0.17\\
\method{} ($\rho=1$) + DART & \textbf{0.86} (0.82, 0.91) & \textbf{18} & 0.80 & 0.17\\
BCF & 1 &  & 0.77 & 0.21\\
HS & 0.99 (0.95, 1.04) & 10 & 0.79 & 0.20\\
Fixed & 1.06 (0.99, 1.13) & 5 & \textbf{0.83} & 0.19\\
X-learner & 1.25 (1.19, 1.30) & 0 &  & \\
R-learner & 1.90 (1.74, 2.08) & 0 &  & \\
\bottomrule
\end{tabular}
\end{table}

\begin{table}[!htbp]
\centering
\footnotesize
\setlength{\tabcolsep}{3pt}
\caption{ACIC 2016. Settings, in increasing order of $F_\epsilon$. The third column gives the mean RMSE of the centered effect of BCF on the standardized outcome, and the others the geometric mean over the replicates of a setting of the RMSE of each model divided by that of BCF on the same data set. Every setting has 10 replicates, except settings 4 and 48 with 9. Bold, the smallest ratio in each row when it is below one.}
\label{tab:app-acic-by-setting}
\vspace{4pt}
\begin{tabular}{rcccccccccc}
\toprule
Setting & $F_\epsilon$ & BCF RMSE & \method{} & \shortstack{\method{}\\($\rho=1$)} & \shortstack{\method{}\\+ DART} & \shortstack{\method{} ($\rho=1$)\\+ DART} & HS & Fixed & X-learner & R-learner\\
\midrule
68 & 0.889 & 0.138 & 1.02 & 1.01 & 0.93 & \textbf{0.92} & 1.00 & 1.13 & 1.17 & 2.39\\
8 & 0.901 & 0.410 & 0.97 & 0.93 & 0.89 & \textbf{0.88} & 0.92 & 1.10 & 1.12 & 1.32\\
24 & 0.912 & 0.149 & 1.12 & 1.11 & \textbf{0.97} & \textbf{0.97} & 1.12 & 1.21 & 1.25 & 2.02\\
29 & 0.918 & 0.128 & 1.05 & 1.05 & 0.96 & 0.94 & 1.07 & \textbf{0.78} & 1.26 & 1.10\\
36 & 0.923 & 0.132 & 0.74 & 0.76 & 0.68 & \textbf{0.67} & 0.77 & 0.77 & 1.28 & 2.30\\
35 & 0.926 & 0.214 & 1.00 & 0.90 & 0.88 & \textbf{0.82} & 0.95 & 1.18 & 1.19 & 2.02\\
37 & 0.935 & 0.163 & 0.96 & 0.95 & 0.81 & \textbf{0.79} & 0.97 & 1.05 & 1.26 & 2.10\\
48 & 0.935 & 0.092 & 0.96 & 0.97 & 0.87 & \textbf{0.84} & 0.90 & 0.87 & 1.27 & 1.54\\
4 & 0.944 & 0.294 & 1.02 & 1.01 & \textbf{0.92} & \textbf{0.92} & 0.94 & 1.28 & 1.27 & 1.51\\
61 & 0.945 & 0.145 & 1.05 & 1.03 & 0.93 & \textbf{0.92} & 1.07 & 1.23 & 1.32 & 1.83\\
18 & 0.948 & 0.286 & 1.05 & 1.05 & \textbf{0.98} & 0.99 & 1.06 & 1.17 & 1.17 & 1.80\\
17 & 0.950 & 0.175 & 1.00 & 0.98 & 0.87 & \textbf{0.86} & 1.02 & 1.08 & 1.32 & 3.00\\
47 & 0.952 & 0.215 & 0.95 & 0.96 & 0.83 & \textbf{0.80} & 0.94 & 1.16 & 1.11 & 1.48\\
27 & 0.959 & 0.107 & 0.97 & 1.01 & 0.80 & \textbf{0.78} & 1.11 & 0.87 & 1.16 & 1.28\\
49 & 0.967 & 0.083 & 1.09 & 1.12 & 1.01 & 1.03 & 1.10 & 1.23 & 1.58 & 3.58\\
57 & 0.968 & 0.062 & 1.13 & 1.11 & 0.94 & \textbf{0.93} & 1.11 & 1.00 & 1.31 & 1.73\\
28 & 0.969 & 0.108 & 0.99 & 1.00 & 0.86 & \textbf{0.85} & 1.03 & 1.33 & 1.30 & 2.02\\
42 & 0.974 & 0.158 & 0.88 & 0.85 & \textbf{0.73} & 0.74 & 0.86 & 0.87 & 1.09 & 1.36\\
64 & 0.979 & 0.071 & 0.98 & 0.98 & 0.87 & \textbf{0.86} & 0.98 & 1.02 & 1.32 & 3.80\\
\bottomrule
\end{tabular}
\end{table}

\begin{table}[!htbp]
\centering
\footnotesize
\setlength{\tabcolsep}{4pt}
\caption{ACIC 2016. Constant setting, 10 replicates. RMSE of the centered effect on the standardized outcome, mean over data sets with its standard error in parentheses, and the coverage of the 90\% intervals for the centered effect. Because the centered effect is zero for every unit, the RMSE measures spurious heterogeneity. Bold, the smallest RMSE.}
\label{tab:app-acic-constant}
\vspace{4pt}
\begin{tabular}{lcc}
\toprule
Model & RMSE & Coverage\\
\midrule
\method{} & 0.013 (0.004) & 1.00\\
\method{} ($\rho=1$) & 0.013 (0.004) & 1.00\\
\method{} + DART & 0.013 (0.004) & 1.00\\
\method{} ($\rho=1$) + DART & 0.013 (0.004) & 1.00\\
BCF & 0.022 (0.012) & 0.91\\
HS & \textbf{0.011} (0.004) & 1.00\\
Fixed & 0.021 (0.001) & 1.00\\
X-learner & 0.060 (0.007) & \\
R-learner & 0.146 (0.042) & \\
\bottomrule
\end{tabular}
\end{table}

\clearpage
\subsection{Pancreatic Cohort}
\label{app:exp-pdac}

This subsection describes the cohort, the covariates and the models of Section~\ref{sec:application}, gives the results of every model fitted to the cohort, and reports the diagnostics of the leave-one-out estimates.

\paragraph{Cohort.}
The data are the TCGA PAAD object of \texttt{pdacR} \citep{torrehealy2023pdacr}, which holds clinical records and gene expression for 181 tumors of the pancreatic ductal adenocarcinoma study of \citet{tcga2017pdac}. Following \citet{jacobs2026horseshoe}, we exclude the 51 patients with a missing survival time, vital status or radiotherapy status, which leaves 130 patients. The treatment is adjuvant radiotherapy, received by 36 patients, and the outcome is overall survival in days. Of the 130 patients, 71 died during follow-up and 59 are censored. The median follow-up, estimated by the reverse Kaplan--Meier method, is 23.2 months, against 23.3 months in \citet{jacobs2026horseshoe}. Our censoring rate of 45 percent is below their 47 percent, so the two analyses differ in the vital status of at least two patients. TCGA required every contributing site to document that its institutional review board had approved the use of the specimens and data or did not regard it as human-subjects research. Its open-access tier holds only de-identified clinical data \citep{nci2014tcgapolicies}. We use only these open-access data (Appendix~\ref{app:exp-setup}).

\paragraph{Covariates.}
The clinical block contains sex, grade, pathological T and N stage, residual tumor, history of chronic pancreatitis, KRAS mutation status, the number of lymph nodes, the surgical procedure, race and histological type. Missing categories are coded as an additional level, and missing numeric values are imputed by the median with an indicator of missingness. The clinical block thus has 27 columns. The expression block applies $\log_2(x+1)$ to the expression values, keeps the 3000 genes with the largest median absolute deviation, as in \citet{jacobs2026horseshoe}, and centers and scales each gene. \citet{jacobs2026horseshoe} also include
genes selected from the literature, which we omit because their list was not available to us. With $p=3{,}027$ covariates and $n=130$, BCF is fitted in our sampler (Appendix~\ref{app:exp-setup}). The propensity score is the cross-fitted logistic lasso of Appendix~\ref{app:exp-setup} and ranges from 0.19 to 0.42.

\paragraph{Models.}
Every model uses the accelerated failure time model of Appendix~\ref{app:aft} for the log survival time. In our sampler, the log time of a censored patient is a latent variable, drawn in each iteration from its normal full conditional truncated below at the log censoring time (Appendix~\ref{app:sampler-censoring}). Given the latent times, the model is the BCF model with a normal outcome. Besides the models of Table~\ref{tab:pdac}, we fit \method{} ($\rho=1$), its DART version. In this subsection, HS denotes the \texttt{CausalHorseForest} fit of Table~\ref{tab:pdac}. The priors of the models in our sampler are those of Appendix~\ref{app:exp-setup}, with $\omega$ matched on the 130 patients. Each of them runs 4 chains of 40000 iterations, of which the first 10000 are warmup, and keeps every third draw. A fit took 25 to 39 minutes with one core per chain. HS runs 4 chains of 10000 iterations, of which the first 5000 are warmup.

\begin{figure}[!htbp]
\centering
\includegraphics{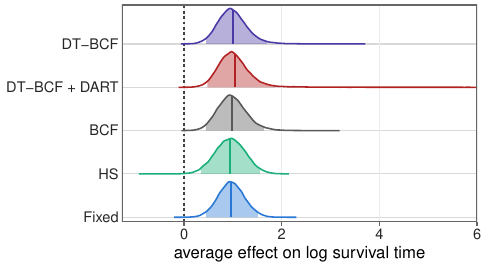}
\caption{Pancreatic Cohort. Posterior densities of the average effect on log survival time under the models of Table~\ref{tab:pdac}. The shaded region is the 95\% interval, the vertical line the posterior mean, and the dashed line marks zero. The axis ends at 6, beyond which lie 8 of the 40{,}000 draws of \method{}
+ DART.}
\label{fig:app-pdac-ate}
\end{figure}

\paragraph{Results.}
Table~\ref{tab:app-pdac-results} extends Table~\ref{tab:pdac} with the two additional models. The posterior means of the average effect lie between 0.95 and 1.15, and every 95\% interval lies above zero
(Figure~\ref{fig:app-pdac-ate}). \method{} ($\rho=1$) has the longest upper tails, with 95\% intervals that reach 4.42 for the average effect and 5.21 for the SD across patients. Its DART version reaches 2.26 and 3.07, and \method{} reaches 1.70 and 1.87. Under every matched prior, the posterior median of the SD across patients lies below its prior median. For no model and no patient does the 95\% interval of the departure $\tau(\bx_i)-\bar\tau$ exclude zero. Under each model of Table~\ref{tab:pdac}, the 95\% intervals for $\tau(\bx_i)$ of all 130 patients overlap in a common range of values (Figure~\ref{fig:app-pdac-cate}).

The survival curves of Figure~\ref{fig:app-pdac-km} average over all 130
patients, while each Kaplan--Meier curve describes only the patients who
received that treatment. The Kaplan--Meier curve of the 94 untreated patients lies within the 95\% bands of the models at nearly every time shown. That of the 36 treated patients lies above the bands until shortly after the first treated death at 7.7 months, and below them around 25 and 36 months. At these times few treated patients remain at risk, 8 at 24 months and 3 at 36 months, so each death is a large step.

\begin{table}[!htbp]
\centering
\footnotesize
\setlength{\tabcolsep}{4pt}
\caption{Pancreatic Cohort. Reported Models. ATE, posterior mean of the
average effect on log survival time with its 95\% interval. SD across patients, the standard deviation of $\tau(\bx_i)$ over the 130 patients, with its prior median at $\sigma=1$ and its posterior median with its 95\% interval. The prior median comes from 4{,}000 prior draws that grow the treatment trees as the sampler does, on the treated patients with cut points drawn uniformly and, for the DART versions, with DART split probabilities. $\sigma$, posterior median of the residual standard deviation on the log time scale. HS is the \texttt{CausalHorseForest} fit of Table~\ref{tab:pdac}, whose prior is not matched. No entry is bold, because the cohort has no known effect against which the models could be ranked.}
\label{tab:app-pdac-results}
\vspace{4pt}
\begin{tabular}{lcccc}
\toprule
& & \multicolumn{2}{c}{SD across patients} & \\
\cmidrule(lr){3-4}
Model & ATE & prior & posterior & $\sigma$\\
\midrule
\method{} & 1.00 (0.45, 1.70) & 0.49 & 0.31 (0.10, 1.87) & 0.90\\
\method{} ($\rho=1$) & 1.15 (0.44, 4.42) & 0.47 & 0.31 (0.10, 5.21) & 0.87\\
\method{} + DART & 1.04 (0.47, 1.87) & 0.44 & 0.31 (0.10, 1.95) & 0.99\\
\method{} ($\rho=1$) + DART & 1.08 (0.48, 2.26) & 0.47 & 0.33 (0.10, 3.07) & 1.00\\
BCF & 0.99 (0.45, 1.64) & 0.96 & 0.34 (0.02, 1.17) & 0.86\\
HS & 0.95 (0.34, 1.56) & -- & 0.78 (0.46, 1.24) & 1.01\\
Fixed & 0.97 (0.44, 1.51) & 0.47 & 0.42 (0.25, 0.61) & 0.90\\
\bottomrule
\end{tabular}
\end{table}

\paragraph{Leave-one-out Estimates.}
The elpd of Table~\ref{tab:pdac} is estimated by PSIS-LOO \citep{vehtari2017practical}. For each patient, the posterior draws
$\theta^{(s)}$ are reweighted by importance ratios proportional to $1/p(y_i\mid\theta^{(s)})$, and the largest ratios are smoothed by a fitted generalized Pareto distribution. The estimated shape $\hat k_i$ of that distribution measures how reliable the reweighting is. Above 0.7, the estimate of the term of patient $i$ can have a large error \citep{vehtari2024pareto}. Table~\ref{tab:app-pdac-loo} reports the diagnostics. The number of patients above 0.7 ranges from one for each DART version to 24 for Fixed and 36 for \method{}.

The patients flagged under \method{} and under Fixed are largely the same, nearly all of them untreated, and most of them died during follow-up. They differ from the observations that usually produce a large $\hat k_i$, which are fitted poorly or have a large influence on the fit. Under \method{}, the posterior mean of $\mu(\bx_i,\widehat e_i)+A_i\tau(\bx_i)$ lies close to their observed log times, within a median of 0.19 residual standard deviations against 0.82 for the other patients. Their contributions $\mathrm{lpd}_i-\mathrm{elpd}_{\mathrm{loo},i}$ to the effective number of parameters $p_{\mathrm{loo}}$ are small, with a median of 0.11 against 0.20, and together they account for 9.3 of the 35.0 effective parameters of \method{}. The patients flagged under HS follow the usual pattern instead, lying a median of 1.31 residual standard deviations from their fitted values and contributing a median of 0.95 to $p_{\mathrm{loo}}$.

We are nonetheless conservative in the use of these estimates and read them only for the ordering of the models. The elpd of the models in our sampler lie within 4.5 of each other, less than half of one standard error, and that of HS lies between 23 and 28 below them. The ordering is unchanged on the 85 patients whose $\hat k_i$ is at most 0.7 under \method{}, BCF, HS and Fixed, where the sums of the pointwise estimates are $-78.4$, $-78.0$, $-92.0$ and $-78.9$. Exact refits that leave out each flagged patient would replace the approximation where it is least reliable.


\begin{table}[!htbp]
\centering
\footnotesize
\setlength{\tabcolsep}{4pt}
\caption{Pancreatic Cohort. PSIS-LOO estimates and their Pareto diagnostics. elpd, expected log pointwise predictive density on the log time scale with its standard error, where the Jacobian of the log transformation is common to all models and omitted. $p_{\mathrm{loo}}$, the estimated effective number of parameters. The last three columns give the number of the 130 patients with $\hat k_i>0.5$ and $\hat k_i>0.7$, and the largest $\hat k_i$. No entry is bold, because the elpd of the models in our sampler differ by less than half of one
standard error.}
\label{tab:app-pdac-loo}
\vspace{4pt}
\begin{tabular}{lccccc}
\toprule
Model & elpd & $p_{\mathrm{loo}}$ & $\hat k_i>0.5$ & $\hat k_i>0.7$ & $\max_i\hat k_i$\\
\midrule
\method{} & $-140.7$ (9.9) & 35.0 & 112 & 36 & 0.86\\
\method{} ($\rho=1$) & $-139.4$ (9.9) & 38.3 & 104 & 13 & 0.82\\
\method{} + DART & $-142.8$ (10.2) & 17.9 & 15 & 1 & 0.71\\
\method{} ($\rho=1$) + DART & $-143.8$ (10.6) & 16.0 & 17 & 1 & 0.81\\
BCF & $-139.7$ (10.0) & 39.1 & 104 & 6 & 0.84\\
HS & $-167.0$ (12.8) & 43.0 & 52 & 9 & 1.15\\
Fixed & $-141.2$ (10.0) & 33.5 & 107 & 24 & 0.80\\
\bottomrule
\end{tabular}
\end{table}

\begin{figure}[!htbp]
\centering
\includegraphics{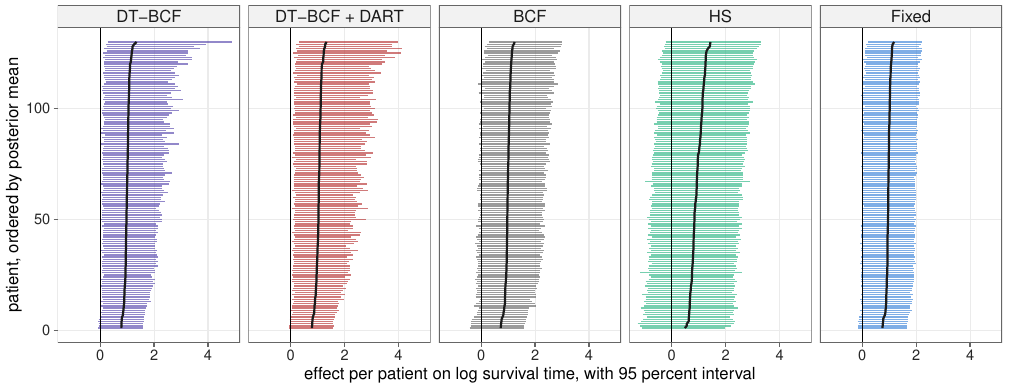}
\caption{Pancreatic Cohort. Effect $\tau(\bx_i)$ of adjuvant radiotherapy on log survival time for each patient under the models of Table~\ref{tab:pdac}. The horizontal lines are the 95\% intervals and the black line connects the posterior means, with the patients ordered by posterior mean within each model. The vertical line marks zero.}
\label{fig:app-pdac-cate}
\end{figure}

\begin{figure}[!htbp]
\centering
\includegraphics{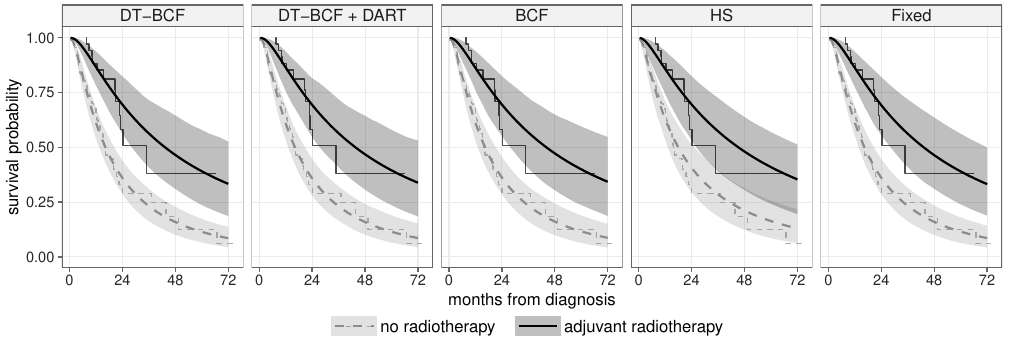}
\caption{Pancreatic cohort. Predicted survival of the 130 patients under each model of Table~\ref{tab:pdac} if every patient received adjuvant radiotherapy (solid) or none did (dashed). The bands show pointwise 95\% credible intervals and the steps show Kaplan--Meier curves of the 36 treated and the 94 untreated patients.}
\label{fig:app-pdac-km}
\end{figure}

\end{document}